\RequirePackage{ifpdf}
\documentclass[hyper,letterpaper]{JHEP3}
\usepackage{amsmath,amssymb,amsfonts,bm,amscd}
\usepackage{cite}
\usepackage{graphicx}
\usepackage{multirow}
\usepackage{verbatim}
\usepackage{appendix}
\usepackage{fancybox}

\usepackage{url}
\usepackage{float}

\newcommand{\bea}{\begin{eqnarray}}
\newcommand{\eea}{\end{eqnarray}}
\newcommand{\be}{\begin{equation}}
\newcommand{\ee}{\end{equation}}

\newcommand{\unknot}{{\,\raisebox{-.08cm}{\includegraphics[width=.8cm]{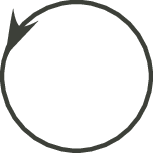}}\,}}

\newcommand{\Z}{{\mathbb Z}}
\newcommand{\R}{{\mathbb R}}
\newcommand{\C}{{\mathbb C}}

\def\Tr{{\rm Tr \,}}

\def\s{\sigma}

\def\frak{\mathfrak}

\def\tilde{\widetilde}
\def\hat{\widehat}
\def\bar{\overline}

\def\CA{{\mathcal A}}
\def\CB{{\mathcal B}}

\def\CE{{\mathcal E}}

\def\CG{{\mathcal G}}
\def\CH{{\mathcal H}}
\def\CI{{\mathcal I}}

\def\CL{{\mathcal L}}
\def\CM{{\mathcal M}}
\def\CN{{\mathcal N}}
\def\CO{{\mathcal O}}

\def\CQ{{\mathcal Q}}
\def\CR{{\mathcal R}}
\def\CS{{\mathcal S}}

\def\CV{{\mathcal V}}
\def\CW{{\mathcal W}}

\newcommand{\cp}{{\mathbb{C}}{\mathbf{P}}}

\renewcommand{\bar}{\overline}
\renewcommand{\hat}{\widehat}

\renewcommand{\d}{\partial}
\newcommand{\VW}{{\text{VW}}}

\newcommand{\basic}{\mathfrak{B}}
\newcommand{\sw}{\Delta}
\newcommand{\tGamma}{\widetilde{\Gamma}}
\DeclareRobustCommand{\Op}[1]{{\raisebox{-0.7ex}{$\overset{#1}{\bullet}$}}}

\title{Fivebranes and 4-manifolds}

\author{Abhijit Gadde$^{1}$, Sergei Gukov$^{1,2}$, and Pavel Putrov$^{1}$
\\
$^1$ California Institute of Technology, Pasadena, CA 91125, USA \\
$^2$ Max-Planck-Institut f\"ur Mathematik, Vivatsgasse 7, D-53111 Bonn, Germany}

\abstract{We describe rules for building 2d theories labeled by 4-manifolds.
Using the proposed dictionary between building blocks of 4-manifolds and 2d $\CN=(0,2)$ theories,
we obtain a number of results, which include
new 3d $\CN=2$ theories $T[M_3]$ associated with rational homology spheres
and new results for Vafa-Witten partition functions on 4-manifolds.
In particular, we point out that the gluing measure for the latter
is precisely the superconformal index of 2d $(0,2)$ vector multiplet
and relate the basic building blocks with coset branching functions.
We also offer a new look at the fusion of defect lines / walls,
and a physical interpretation of the 4d and 3d Kirby calculus
as dualities of 2d $\CN=(0,2)$ theories and 3d $\CN=2$ theories, respectively.
\\
\\
\\
\\
\\
\\
\\
{\tt CALT 68-2904}}

\begin{document}
\cornersize{1}


\section{Introduction}
\label{sec:intro}

We study a class of 2d $\CN=(0,2)$ theories $T[M_4]$ labeled by 4-manifolds (with boundary)
that enjoys all the standard operations on 4-manifolds, such as cutting, gluing, and the Kirby moves \cite{GompfS}.
Since the world-sheet SCFT of a heterotic string is a prominent member of this class
of 2d $\CN=(0,2)$ theories we shall call it ``class $\CH$'' in what follows.
By analogy with theories of class $\CS$ and class $\CR$ that can be thought of as
compactifications of six-dimensional $(2,0)$ theory on 2-manifolds \cite{Gaiotto:2008cd,Gaiotto:2009we,Alday:2009aq}
and 3-manifolds \cite{DGH,DGG,CCV}, respectively,
a theory $T[M_4]$ of class $\CH$ can be viewed as the effective two-dimensional theory
describing the physics of fivebranes wrapped on a 4-manifold $M_4$.

If 2d theories $T[M_4]$ are labeled by 4-manifolds, then what are 4-manifolds labeled by?
Unlike the classification of 2-manifolds and 3-manifolds that was of great help in taming
the zoo of theories $T[M_2]$ and $T[M_3]$, the world of 4-manifolds is much richer and less understood.
In particular, the answer to the above question is not known at present
if by a 4-manifold one means a {\it smooth} 4-manifold.
And, not surprisingly, there will be many points in our journey where this richness of
the world of 4-manifolds will translate into rich physics of 2d $\CN=(0,2)$ theories $T[M_4]$.
We hope that exploring the duality between 4-manifolds and theories $T[M_4]$ sufficiently far
will provide insights into classification of smooth structures in dimension four.

In dimensions $\le 6$, every combinatorial manifold --- a.k.a. simplicial complex or a manifold with piecewise linear (PL) structure --- admits
a unique compatible smooth (DIFF) structure. However, not every topological 4-manifold admits a smooth structure:
\be
\text{DIFF} \; = \; \text{PL} \; \subset \; \text{TOP}
\ee
and, furthermore, the smooth structure on a given topological 4-manifold may not be unique
(in fact, $M_4$ can admit infinitely many smooth structures).
When developing a dictionary between $M_4$ and $T[M_4]$, we will use various tools from string theory
and quantum field theory which directly or indirectly involve derivatives of various fields on $M_4$.
Therefore, in our duality between $M_4$ and $T[M_4]$ all 4-manifolds are assumed to be smooth, but not necessarily compact.
In particular, it makes sense to ask what the choice of smooth or PL structure on $M_4$ means for the 2d theory $T[M_4]$,
when the 4-manifold admits multiple smooth structures.

Returning to the above question, the basic topological invariants of a (compact) 4-manifold $M_4$
are the Betti numbers $b_i (M_4)$ or combinations thereof, such as the Euler characteristic and the signature:
\bea
b_2 & = & b_2^+ + b_2^- \nonumber \\
\sigma & = & b_2^+ - b_2^- = \frac{1}{3} \int_{M_4} p_1 \label{bchifirst} \\
\chi & = & 2 - 2b_1 + b_2^+ + b_2^- \nonumber
\eea
At least in this paper, we will aim to understand fivebranes on simply-connected 4-manifolds.
In particular, all compact 4-manifolds considered below will have $b_1 (M_4) = 0$.
We will be forced, however, to deviate from this assumption (in a minimal way)
when discussing cutting and gluing, where non-trivial fundamental groups $|\pi_1 (M_4)| < \infty$ will show up.

As long as $b_1 =0$, there are only two non-trivial integer invariants in \eqref{bchifirst},
which sometimes are replaced by the following topological invariants
\bea
\chi_h (M_4) & = & \frac{\chi (M_4) + \s (M_4)}{4} \\
c (M_4) & = & 2 \chi (M_4) + 3 \s (M_4) \quad (= c_1^2 ~~\text{when}~M_4~\text{is a complex surface}) \nonumber
\eea
also used in the literature on 4-manifolds.
These two integer invariants (or, simply $b_2$ and $\sigma$) determine the rank and the signature of
the bilinear intersection form
\be
Q_{M_4} : \Gamma \otimes \Gamma \; \to \; \Z
\label{intmatdef}
\ee
on the homology lattice
\be
\Gamma \; = \; H_2 (M_4; \Z) / \text{Tors}
\ee
The intersection pairing $Q_{M_4}$ (or, simply, $Q$) is a nondegenerate symmetric bilinear integer-valued form,
whose basic characteristics include the rank, the signature, and the parity (or type).
While the first two are determined by $b_2 (M_4)$ and $\sigma (M_4)$, the type is defined as follows.
The form $Q$ is called even if all diagonal entries in its matrix are even; otherwise it is odd.
We also define
\be
\Gamma^* = H^2 (M_4; \Z) / \text{Tors}
\ee
The relation between the two lattices $\Gamma$ and $\Gamma^*$ will play an important role in construction of theories $T[M_4]$
and will be discussed in section \ref{sec:generalities}.

For example, the intersection form for the K\"ummer surface has a matrix representation
\be
E_8 \oplus E_8 \oplus 3
\begin{pmatrix}
0 & 1 \\
1 & 0
\end{pmatrix}
\ee
where $\bigl( \begin{smallmatrix}
0 & 1 \\
1 & 0
\end{smallmatrix} \bigr)$
is the intersection form for $S^2 \times S^2$ and $E_8$ is minus the Cartan matrix for the exceptional Lie algebra by the same name.
A form $Q$ is called positive (resp. negative) definite if $\sigma (Q) = \text{rank} (Q)$ (resp. $\sigma (Q) = - \text{rank} (Q)$)
or, equivalently, if $Q(\gamma,\gamma)>0$ (resp. $Q(\gamma,\gamma)<0$) for all non-zero $\gamma \in \Gamma$.
There are finitely many unimodular\footnote{that is $\det Q = \pm 1$} definite forms of a fixed rank.
Thus, in the above example the intersection form for $S^2 \times S^2$ is indefinite and odd,
whereas $E_8$ is the unique unimodular negative definite even form of rank $8$.

If $M_4$ is a closed simply-connected oriented 4-manifold, its homeomorphism type is completely determined by $Q$.
To be a little more precise, according to the famous theorem of Michael Freedman \cite{Freedman},
compact simply-connected topological 4-manifolds are
completely characterized by an integral unimodular
symmetric bilinear form $Q$ and the Kirby-Siebenmann
triangulation obstruction invariant $\alpha (M_4) \in H^4 (M_4; \Z_2) \cong \Z_2$, such that $\frac{\sigma}{8} \equiv \alpha$ mod 2
if $Q$ is even.
In particular, there is a unique topological 4-manifold with the intersection pairing $E_8$.
This manifold, however, does not admit a smooth structure.
Indeed, by Rokhlin's theorem, if a simply-connected smooth 4-manifold has an even intersection form $Q$,
then $\sigma (M_4)$ is divisible by 16.
There is, however, a {\it non-compact} smooth manifold with $E_8$ intersection form that will be one of our examples below:
it corresponds to a nice 2d theory $T[E_8]$,
which for a single fivebrane we propose to be a realization of level-1 $E_8$ current algebra used in the world-sheet SCFT
of a heterotic string \cite[sec.6]{Green:1987sp} or in the construction of $E$-strings \cite{Estrings}:
\be
T[E_8] \; = \; \text{(bosonization of) 8 Fermi multiplets}
\label{TE8U1}
\ee

In the case of compact smooth 4-manifolds, the story is a lot more complicated and the complete classification is not known at present.
One major result that will be important to us in what follows is the Donaldson's theorem \cite{Donaldson},
which states that the intersection form $Q$ of a smooth simply-connected positive (resp. negative) definite 4-manifold
is equivalent over integers to the standard diagonal form $\text{diag} (1,1, \ldots, 1)$
or $\text{diag} (-1,-1, \ldots, -1)$, respectively.
(This result applies to compact $M_4$ and admits a generalization to 4-manifolds bounded by homology spheres,
which we will also need in the study of 2d theories $T[M_4]$.)
In particular, since $E_8 \oplus E_8$ is not diagonalizable over integers,
the unique topological 4-manifold with this intersection form does not admit a smooth structure.\footnote{Note,
this can not be deduced from the Rokhlin's theorem as in the case of the $E_8$ manifold.}
Curiously, this, in turn, implies that $\R^4$ does not have a unique differentiable structure.

We conclude this brief introduction to the wild world of 4-manifolds by noting
that any non-compact topological 4-manifold admits a smooth structure \cite{FQuinn}.
In fact, an interesting feature of non-compact 4-manifolds considered in this paper --- that
can be viewed either as a good news or as a bad news --- is that they all
admit {\it uncountably} many smooth structures. 

\begin{table}[htb]
\centering
\renewcommand{\arraystretch}{1.3}
\begin{tabular}{|@{\quad}c@{\quad}|@{\quad}c@{\quad}| }
\hline  {\bf 4-manifold} $M_4$ & {\bf 2d} $(0,2)$ {\bf theory} $T[M_4]$
\\
\hline
\hline handle slides & dualities of $T[M_4]$ \\
\hline boundary conditions & vacua of $T[M_3]$ \\
\hline 3d Kirby calculus & dualities of $T[M_3]$ \\
\hline cobordism & domain wall (interface) \\
from $M_3^-$ to $M_3^+$ & between $T[M_3^-]$ and $T[M_3^+]$ \\
\hline gluing & fusion \\
\hline Vafa-Witten & flavored (equivariant) \\
partition function & elliptic genus \\
\hline $Z_{\VW} (\text{cobordism})$ & branching function \\
\hline instanton number & $L_0$ \\
\hline embedded surfaces & chiral operators \\
\hline Donaldson polynomials & chiral ring relations \\
\hline
\end{tabular}
\caption{The dictionary between geometry and physics.}
\label{tab:dict}
\end{table}

In order to preserve supersymmetry in two remaining dimensions, the 6d theory must be partially ``twisted'' along the $M_4$.
The standard way to achieve this is to combine the Euclidean $Spin(4)$ symmetry of the 4-manifold with (part of) the R-symmetry.
Then, different choices
--- labeled by homomorphisms from $Spin(4)$ to the R-symmetry group, briefly summarized in appendix \ref{sec:survey} ---
lead to qualitatively different theories $T[M_4]$, with different amount of supersymmetry in two dimensions, {\it etc.}
The choice we are going to consider in this paper is essentially (the 6d lift of) the topological twist
introduced by Vafa and Witten \cite{VafaWitten}, which leads to $(0,2)$ supersymmetry in two dimensions.
In fact, the partition function of the Vafa-Witten TQFT that, under certain conditions,
computes Euler characteristics of instanton moduli spaces also plays an important role
in the dictionary beteen 4-manifolds and the corresponding 2d $\CN=(0,2)$ theories $T[M_4]$.

The basic ``protected quantity'' of any two-dimensional theory with at least $\CN=(0,1)$ supersymmetry
is the elliptic genus \cite{Witten} defined as a partition function on a 2-torus $T^2$
with periodic (Ramond) boundary conditions for fermions.
In the present case, it carries information about all left-moving states of the 2d $\CN=(0,2)$ theory $T[M_4]$
coupled to the supersymmetric Ramond ground states from the right.
To be more precise, we shall consider the ``flavored'' version of the elliptic genus
(studied in this context {\it e.g.} in \cite{Gadde:2013wq,Benini:2013nda}),
\be
\CI_{T[M_4]} (q,x) \; := \; \mbox{Tr}_{\CH} (-1)^{F} q^{L_0} x^{f} \,,
\label{indexdef}
\ee
that follows the standard definition of the superconformal index in radial quantization
and carries extra information about the flavor symmetry charges $f$.
In general, the flavor symmetry group of $T[M_4]$ is $U(1)^{b_2} \times G_{3d}$, where the second factor
is associated with the boundary $M_3 = \partial M_4$ and is gauged upon gluing operations.
Defined as a supersymmetric partition function on a torus $T^2$ with a modular parameter $\tau$
(where, as usual, $q = e^{2\pi i \tau}$), the index $\CI_{T[M_4]} (q;x)$ has a nice interpretation
as an invariant of the 4-manifold computed by the topological theory on $M_4$.

Indeed, since the theory $T[M_4]$ was obtained by compactification from six dimensions on a 4-manifold,
its supersymmetric partition function on a torus can be identified with the partition function
of the 6d $(2,0)$ theory on $T^2 \times M_4$.
As usual, by exchanging the order of compactificaion, we obtain two perspecties on this fivebrane partition function
$$
 \begin{array}{ccccc} \; & \; & \text{6d $(2,0)$ theory} & \; & \; \\
 \; & \; & \text{on $T^2 \times M_4$} & \; & \; \\ \; & \swarrow & \; & \searrow & \; \\
 \text{$\CN=4$ super-Yang-Mills} & \; & \; & \; & \text{2d $(0,2)$ theory $T[M_4]$} \\
 \text{on $M_4$} & \; & \; & \; & \text{on $T^2$}
 \end{array} 
$$
that are expected to produce the same result.
If we compactify first on $M_4$, we obtain a 2d theory $T[M_4]$,
whose partition function on $T^2$ is precisely the flavored elliptic genus \eqref{indexdef}.
On the other hand, if we first compactify on $T^2$, we get $\CN=4$ super-Yang-Mills\footnote{Sometimes,
to avoid clutter, we suppress the choice of the gauge group, $G$, which in most of our applications will
be either $G=U(N)$ or $G=SU(N)$ for some $N \ge 1$. It would be interesting to see if generalization to $G$
of Cartan type $D$ or $E$ leads to new phenomena. We will not aim to do this analysis here.}
with the Vafa-Witten twist on $M_4$ and coupling constant $\tau$.
This suggests the following natural relation
\be
Z_{\VW}^G [M_4] (q,x) \; = \; {\cal I}_{T[M_4;G]} (q,x)
\label{ZVWvsindex}
\ee
that will be one of our main tools in matching 4-manifolds with 2d $\CN=(0,2)$ theories $T[M_4]$.
Note, this in particular requires $M_4$ to be a smooth 4-manifold.
Both sides of \eqref{ZVWvsindex} are known to exhibit nice modular properties under certain
favorable assumptions \cite{VafaWitten,Witten} that we illustrate in numerous examples below.

In this paper, we approach the correspondence between 4-manifolds and 2d $\CN=(0,2)$ theories $T[M_4]$
mainly from the viewpoint of cutting and gluing.
For this reason, not only 4-manifolds with boundary are unavoidable, they also are the main subject of interest.
As a result, interesting new phenomena, such as a generalization of the Freed-Witten anomaly \cite{Freed:1999vc}
to manifolds with boundary, come into play.
It also affects the relation \eqref{ZVWvsindex}, where the left-hand side naturally becomes a function
of boundary conditions, and leads to one interesting novelty discussed in section \ref{sec:contbasis}.
Namely, in order to interpret the Vafa-Witten partition function on a non-compact 4-manifold as
the index \eqref{indexdef}, it is convenient to make a certain transformation
--- somewhat akin to a change of basis familiar in the literature on the superconformal index \cite{Gadde:2011ik} ---
changing {\it discrete} labels associated with boundary conditions to {\it continuous} variables.

The type of the topological twist that leads to 2d $(0,2)$ theory $T[M_4]$, namely the Vafa-Witten twist,
can be realized on the world-volume of fivebranes wrapped on a coassociative submanifold $M_4$ inside
a seven-dimensional manifold with $G_2$ holonomy \cite{Bershadsky:1995qy,Blau:1996bx}.
Locally, in the vicinity of $M_4$, this 7-dimensional manifold always looks like
the bundle of self-dual 2-forms over $M_4$ (see {\it e.g.} \cite{Acharya:2004qe} for a pedagogical review).
This realization of the 6d $(2,0)$ theory on the world-volume of M-theory fivebranes embedded in
11d space-time can provide some useful clues about the 2d superconformal theory $T[M_4]$,
especially when the number of fivebranes is large, $N \gg 1$, and the system admits a holographic
dual supergravity description ({\it cf.} appendix \ref{sec:survey} for a brief survey).

In the case of fivebranes on coassociative 4-manifolds, the existence of the holographic dual
supergravity solution \cite{Gauntlett:2000ng,Gauntlett:2001jj,Benini:2013cda}
requires $M_4$ to admit a conformally half-flat structure, {\it i.e.} metric with anti-self-dual Weyl tensor.
Since the signature of the 4-manifold can be expressed as the integral
\be
\sigma (M_4) \; = \; \frac{1}{12 \pi^2} \int_{M_4} \left( |W_+|^2 - |W_-|^2 \right)
\ee
where $W_{\pm}$ are the self-dual and anti-self-dual components of the Weyl tensor,
it suggests to focus on 2d $\CN=(0,2)$ superconformal theories $T[M_4]$ associated with negative definite $M_4$.
As we explained earlier, negative definite 4-manifolds are very simple in the smooth category
and, curiously, $W_+ = 0$ also happens to be the condition under which instantons on $M_4$
admit a description \cite{AtiyahWard} that involves holomorphic vector bundles (on the twistor space of $M_4$),
monads, and other standard tools from $(0,2)$ model building.

The holographic dual and the anomaly of the fivebrane system also allow to express left and right moving
central charges of the 2d $\CN=(0,2)$ superconformal theory $T[M_4]$ via basic topological invariants \eqref{bchifirst} of the 4-manifold.
Thus, in the case of the 6d $(2,0)$ theory of type $G$ one finds \cite{Benini:2013cda,Alday:2009qq}:
\bea
c_R & = & \frac{3}{2} (\chi + \s) r_G + (2 \chi + 3 \s) d_G h_G \\
c_L & = & \chi r_G + (2 \chi + 3 \s) d_G h_G \nonumber
\eea
where $r_G = \text{rank} (G)$, $d_G = \text{dim} (G)$, and $h_G$ is the Coxeter number.
In particular, for a single fivebrane ($r_G = 1$ and $d_G h_G = 0$) these expressions give
$c_L = \chi$ and $c_R = 3 + 3 b_2^+$, suggesting that $b_2^-$ is the number of Fermi
multiplets\footnote{Recall, that a free Fermi multiplet contributes to the central charge $(c_L , c_R) = (1,0)$.}
in the 2d $\CN=(0,2)$ theory $T[M_4; U(1)]$.
This conclusion agrees with the direct counting of bosonic and fermionic Kaluza-Klein modes \cite{Ganor:1996xg}
and confirms \eqref{TE8U1}.
As we shall see in the rest of this paper, the basic building blocks of 2d theories $T[M_4]$
are indeed very simple and, in many cases, can be reduced to Fermi multiplets
charged under global flavor symmetries (that are gauged in gluing operations).
However, the most interesting part of the story is about operations on 2d $(0,2)$ theories that correspond to {\it gluing}.\\

The paper is organized as follows.
In section \ref{sec:generalities} we describe the general ideas relating 4-manifolds and the corresponding
theories $T[M_4]$, fleshing out the basic elements of the dictionary in Table \ref{tab:dict}.
Then, we study the proposed rules in more detail and present various tests as well as new predictions
for Vafa-Witten partition functions on 4-manifolds (in section \ref{sec:VW})
and for 2d walls and boundaries in 3d $\CN=2$ theories (in section \ref{sec:2dtheory}).

The relation between Donaldson invariants of $M_4$ and $\bar \CQ_+$-cohomology of the corresponding 2d $(0,2)$
theory $T[M_4]$ will be discussed elsewhere.
More generally, and as we already remarked earlier, it would be interesting to study to what extent $T[M_4]$,
viewed as an invariant of 4-manifolds, can detect smooth structures. In particular, it would be interesting
to explore the relation between $T[M_4]$ and other invariants of smooth 4-manifolds originating from physics,
such as the celebrated Seiberg-Witten invariants \cite{Seiberg:1994rs,Witten:1994cg}
or various attempts based on gravity \cite{Rohm:1988yz,Asselmeyer:1996bh,Pfeiffer:2004pe,Sladkowski:2009dc}.


\section{2d theories labeled by 4-manifolds}
\label{sec:generalities}

Building theories $T[M_4]$ in many ways follows the same set of rules and tricks as building 4-manifolds.
Here, we describe some of the basic operations in the world of 4-manifolds and propose their realization
in the world of supersymmetric gauge theories.
While the emphasis is certainly on explaining the general rules,
we supplement each part with concrete examples and/or new calculations.
More examples, with further details, and new predictions based on the proposed relations in Table \ref{tab:dict}
will be discussed in sections \ref{sec:VW} and \ref{sec:2dtheory}.

\subsection{Kirby diagrams and plumbing}
\label{sec:plumbing}

We start by reviewing the standard construction of 4-manifolds, based on a handle decomposition,
mostly following \cite{GompfS} (see also \cite{Akbulut}).
Thus, if $M_4$ is connected, we take a single 0-handle ($\cong D^4$)
and successively attach to it $k$-handles ($\cong D^k \times D^{4-k}$) with $k=1,2,3$.
Then, depending on the application in mind, we can either stop at this stage
(if we are interesting in constructing non-compact 4-manifolds)
or cap it off with a 4-handle ($\cong D^4$) if the goal is to build a compact 4-manifold.

The data associated with this process is usually depicted in the form of a {\it Kirby diagram},
on which every $k$-handle ($\cong D^k \times D^{4-k}$) is represented
by its attaching region, $S^{k-1} \times D^{4-k}$, or by its attaching sphere, $S^{k-1}$.
To be a little more precise, a Kirby diagram of a smooth connected 4-manifold $M_4$
usually shows only 1-handles and 2-handles because 3-handles and 4-handles attach essentially in a unique way \cite{LaudenbachP}.
Moreover, in our applications we typically will not see 1-handles either
(due to our intention to work with simply-connected 4-manifolds).
Indeed, regarding a handle decomposition of $M_4$ as a cell complex, its $k$-th homology group
becomes an easy computation in which $k$-handles gives rise to generators and $(k+1)$-handles give rise to relations.
The same interpretation of the handlebody as a cell complex can be also used for the computation
of the fundamental group, where 1-handles correspond to generators and 2-handles lead to relations.
Therefore, the easiest way to ensure that $M_4$ is simply-connected is to avoid using 1-handles at all.

\begin{figure}[ht]
\centering
\includegraphics[width=5.5in]{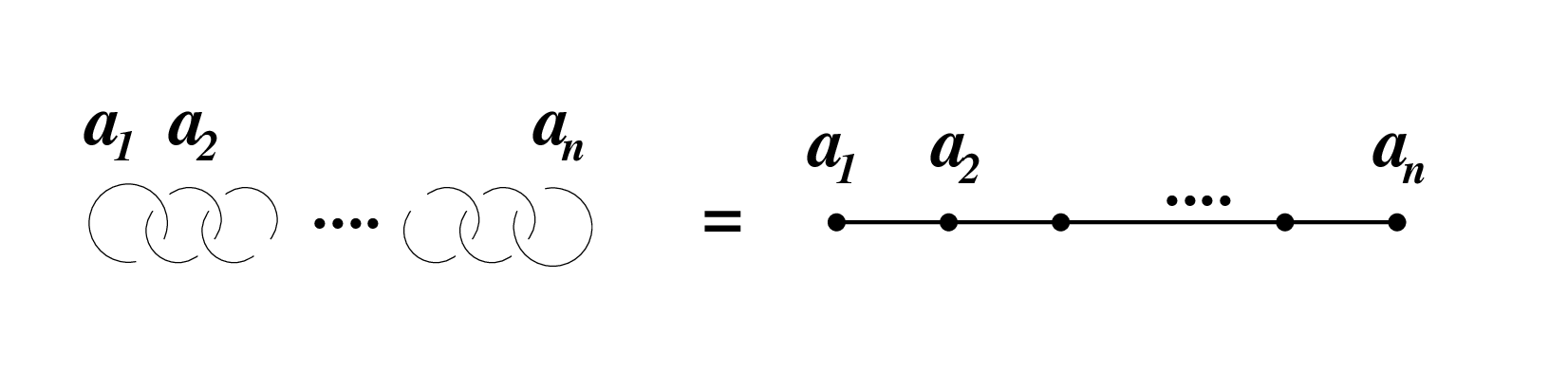}
\caption{A Kirby diagram and the corresponding plumbing graph
for the plumbing 4-manifold associated to the string $(a_1, a_2, \ldots, a_n)$.}
\label{fig:Anplumbing}
\end{figure}

\begin{figure}[ht]
\centering
\includegraphics[width=5.5in]{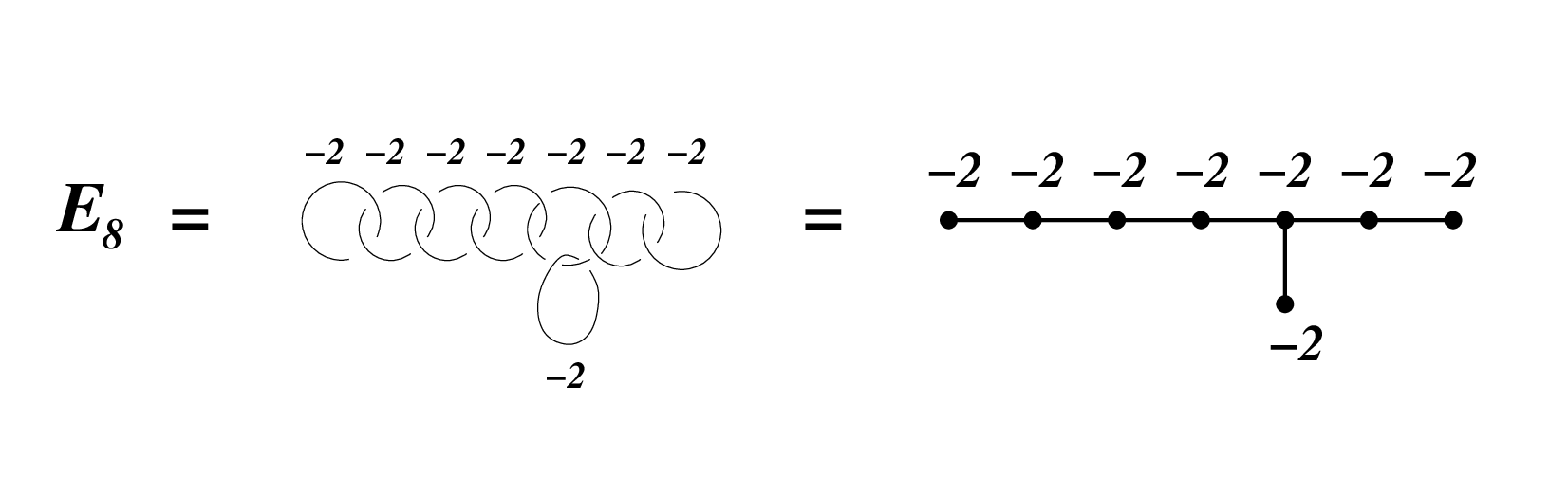}
\caption{A Kirby diagram and the corresponding plumbing graph
for the $E_8$ manifold with $b_2 = - \sigma = 8$ and $\partial E_8 \approx \Sigma (2,3,5)$.}
\label{fig:E8}
\end{figure}

\begin{figure}[ht]
\centering
\includegraphics[width=1.8in]{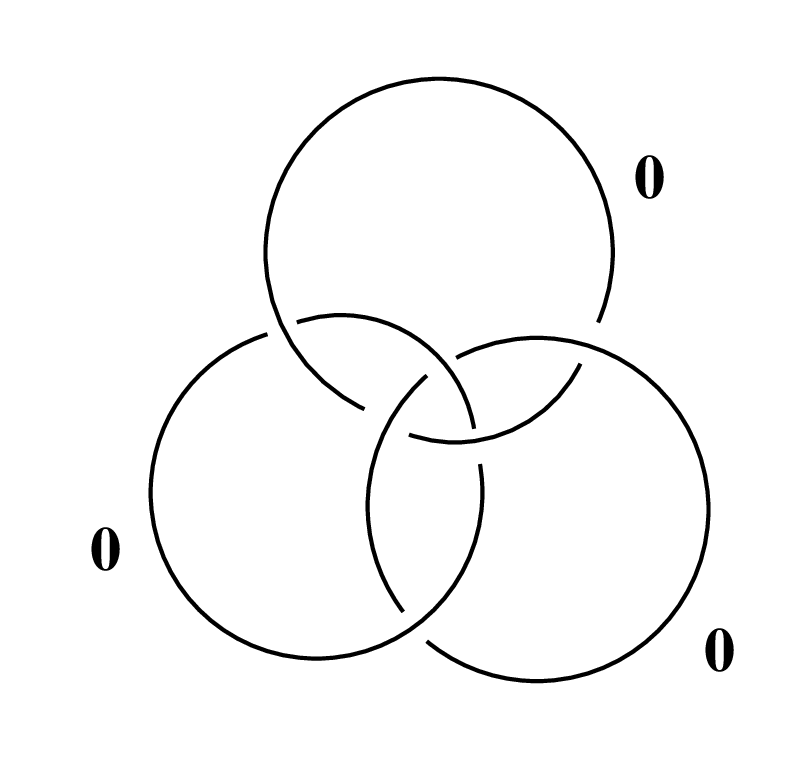}
\caption{Kirby diagram of a 4-manifold bounded by a 3-torus $T^3$.}
\label{fig:Borromean}
\end{figure}

Then, for this class of 4-manifolds, Kirby diagrams only contain framed circles,
{\it i.e.} attaching spheres of 2-handles, that can be knotted and linked inside $S^3$ (= boundary of the 0-handle).
To summarize, we shall mostly work with 4-manifolds labeled by framed links in a 3-sphere,
\be
M_4 ~:~ \quad K_1^{a_1} \, K_2^{a_2} \, \ldots \, K_n^{a_n}
\label{M4KKK}
\ee
where $K_i$ denotes the $i$-th component of the link and $a_i \in \Z$ is the corresponding framing coefficient.
Examples of Kirby diagrams for simple 4-manifolds are shown in Figures \ref{fig:Anplumbing},
\ref{fig:E8}, and \ref{fig:Borromean}.

At this stage, it is important to emphasize that
Kirby diagrams are not quite unique: there are certain moves which relate
different presentations of the same 4-manifold.
We refer the reader to excellent monographs \cite{GompfS,Akbulut} on Kirby calculus,
of which most relevant to us is the basic tool called 2-handle slide.
Indeed, since our assumptions led us to consider 4-manifolds built out of
2-handles,\footnote{Another nice property of such 4-manifolds is that they
admit an {\it achiral} Lefschetz fibration over the disk~\cite{Harer}.}
occasionally we will encounter the operation of sliding a 2-handle $i$ over a 2-handle $j$.
It changes the Kirby diagram and, in particular, the framing coefficients:
\bea
a_j & \mapsto & a_i + a_j \pm 2\text{lk} (K_i, K_j) \label{aaslide} \\
a_i & \mapsto & a_i \nonumber
\eea
where the sign depends on the choice of orientation
(``$+$'' for handle addition and ``$-$'' for handle subtraction)
and $\text{lk} (K_i, K_j)$ denotes the linking number.
We will see in what follows that this operation corresponds to changing the basis of flavor charges.

In the class of non-compact simply-connected 4-manifolds \eqref{M4KKK} labeled by framed links,
the simplest examples clearly correspond to Kirby diagrams where all $K_i$ are copies of the unknot.
Many\footnote{but not all! See Figure \ref{fig:Borromean} for an instructive (counter)example.\label{foot:limitations}}
such 4-manifolds can be equivalently represented by graphs with integer ``weights'' assigned to the vertices,
somewhat similar to quiver diagrams that conveniently encode the spectrum of fields and interactions
in a large class of gauge theories.
The 4-manifolds in question are constructed by gluing together $n$ copies of disk bundles
over 2-spheres, $D^2_i \to S^2_i$, each labeled by an integer Euler class $a_i \in \Z$.
Switching the role of the base and the fiber in the gluing process, one builds a simply-connected
4-manifold $M_4$, called {\it plumbing}, whose handle decomposition involves $n$ two-handles
(besides the ``universal'' 0-handle at the bottom).
As usual, we represent such 4-manifolds by Kirby diagrams drawing
the attaching framed circles $K_i$ of 2-handles inside $S^3$.

The simplest non-trivial plumbing manifold corresponds to the Kirby diagram:
\be
{-p \atop \unknot}
\label{single2handle}
\ee
In other words, its handlebody decomposition contains only one 2-handle with framing $-p$,
and the resulting manifold $M_4$ is a twisted $D^2$ bundle over $S^2$ or, as a complex manifold,
the total space of the $\CO (-p)$ bundle over $\cp^1$,
\be
M_4 ~:~ \quad \CO(-p) \to \cp^1
\ee
For $p>0$, which we are going to assume in what follows,
$M_4$ is a negative definite plumbing manifold bounded by the Lens space $L(p,1)$.

\bigskip

Another, equivalent way to encode the same data is by a plumbing graph $\Upsilon$.
In this presentation, each attaching circle $K_i$ of a 2-handle is replaced by
a vertex with an integer label $a_i$, and an edge between two vertices $i$ and $j$
indicates that the corresponding attaching circles $K_i$ and $K_j$ are linked.
Implicit in the plumbing graph is the orientation of edges, which, unless noted otherwise,
is assumed to be such that all linking numbers are $+1$.
More generally, one can consider plumbings of twisted $D^2$ bundles over higher-genus
Riemann surfaces, see {\it e.g.} \cite[sec. 2.1]{Akbulut},
in which case vertices of the corresponding plumbing graphs are labeled by Riemann surfaces (not necessarily orientable)
in addition to the integer labels $a_i$.
However, such 4-manifolds typically have non-trivial fundamental group
and we will not consider these generalizations here, focusing mainly on plumbings of 2-spheres.

The topology of a 4-manifold $M_4$ constructed via plumbing of 2-spheres
is easy to read off from its Kirby diagram or the corresponding plumbing graph.
Specifically, $M_4$ is a non-compact simply-connected 4-manifold,
and one can think of $K_i$ as generators of $\Gamma = H_2 (M_4; \Z)$
with the intersection pairing
\be
Q_{ij} =
\begin{cases}
\text{lk}(K_i, K_j), & \text{if } i \ne j \\
a_i, & \text{if }i=j
\end{cases}
\label{intformlink}
\ee
For example, the Kirby diagram in Figure \ref{fig:Anplumbing} corresponds to
\be
Q \; = \;
\begin{pmatrix}
a_1 & 1   & 0 & \cdots  & 0 \\
1   & a_2 & 1        &   & \vdots \\
0 & 1 &  & \ddots  &   0 \\
\vdots  & & \ddots & ~\ddots~  &  ~1~  \\
0 &  \cdots  & 0 & 1 & a_n
\end{pmatrix}
\label{Qmatrix}
\ee
A further specialization to $(a_1, a_2, \ldots, a_n) = (-2, -2, \ldots, -2)$
for obvious reasons is usually referred to as $A_{n}$,
whereas that in Figure \ref{fig:E8} is called $E_8$.

\bigskip

Similarly, given a weighted graph $\Upsilon$, one can plumb disk bundles with Euler numbers $a_i$
over 2-spheres together to produce a 4-manifold $M_4 (\Upsilon)$ with boundary
$M_3 (\Upsilon) = \partial M_4 (\Upsilon)$, such that
\begin{subequations}\label{plumbgraphBetti}
\be
b_1 (M_4) \; = \; b_1 (\Upsilon)
\ee
\be
b_2 (M_4) \; = \; \# \{ \text{vertices of } \Upsilon \}
\ee
\end{subequations}
In particular, aiming to construct simply-connected 4-manifolds, we will avoid plumbing graphs that have loops
or self-plumbing constructions.
Therefore, in what follows we typically assume that $\Upsilon$ is a tree, relegating generalizations to future work.
Besides the basic topological invariants \eqref{plumbgraphBetti}, the plumbing tree $\Upsilon$ also gives
a nice visual presentation of the intersection matrix $Q (\Upsilon) = (Q_{ij})$,
which in the natural basis of $H_2 (M_4 ; \Z)$ has entries
\be
Q_{ij} \; = \;
\begin{cases}
a_i, & \text{if } i=j \\
1, & \text{if } i \text{ is connected to } j \text{ by an edge} \\
0, & \text{otherwise}
\end{cases}
\label{Qplumbing}
\ee
The eigenvalues and the determinant of the intersection form $Q$ can be also easily extracted from $\Upsilon$
by using the algorithm described below in \eqref{plumbtreeQ} and illustrated in Figure~\ref{fig:detQrules}.

Note, this construction of non-compact 4-manifolds
admits vast generalizations that do not spoil any of our assumptions (including the simple connectivity of $M_4$).
Thus, in a Kirby diagram of an arbitrary plumbing tree,
we can replace every framed unknot (= attaching circle of a 2-handle) by a framed knot, with a framing coefficient $a_i$.
This does not change the homotopy type of the 4-manifold, but does affect the boundary $M_3 = \partial M_4$.
Put differently, all the interesting information about the knot can only be seen at the boundary.

Another important remark is that, although the description of 4-manifolds via plumbing graphs is very nice and simple,
it has certain limitations that were already mentioned in the footnote \ref{foot:limitations}.
Indeed, if the 4-manifold has self-plumbings or $\Upsilon$ has loops, it may not be possible to
consistently convert the Kirby diagram into a plumbing graph without introducing additional labels.
An example of such Kirby diagram is shown in Figure~\ref{fig:Borromean},
where each pair of the attaching circles $K_i$ with framing $a_i=0$ has linking number zero.
The corresponding 4-manifold, however, is different from that associated to three unlinked copies
of the unknot (with plumbing graph that has three vertices and no edges) and the same values of framing coefficients.

\bigskip

Finally, we point out that,
since all 4-manifolds constructed in this section have a boundary $M_3 = \partial M_4$,
the corresponding 2d $\CN=(0,2)$ theory $T[M_4]$ that will be described below
should properly be viewed as a boundary condition for the 3d $\CN=2$ theory $T[M_3]$.
For example, the plumbing on $A_n$ has the Lens space boundary $M_3 = L(n+1,n)$,
while the plumbing on $E_8$ has the Poincar\'e sphere boundary $M_3 = \Sigma (2,3,5)$,
where
\be
\Sigma (a,b,c) := S^5 \cap \{ (x,y,z) \in \C^3 \; \vert \; x^a + y^b + z^c = 0 \}
\label{Brieskorn}
\ee
is the standard notation for a family of Brieskorn spheres.
This remark naturally leads us to the study of boundaries $M_3$ and the corresponding theories $T[M_3]$
for more general sphere plumbings and 4-manifolds \eqref{M4KKK} labeled by framed links.

\subsection{$T{[} M_4 {]}$ as a boundary condition}
\label{sec:bdrycond}

Since we want to build 4-manifolds as well as the corresponding theories $T[M_4]$ by gluing basic pieces,
it is important to develop the physics-geometry dictionary for manifolds with boundary,
which will play a key role in gluing and other operations.

\subsection*{Vacua of the 3d $\CN=2$ theory $T{[} M_3 {]}$}

Our first goal is to describe supersymmetric vacua of the 3d $\CN=2$ theory $T[M_3]$ associated
to the boundary\footnote{Depending on the context, sometimes $M_3$ will refer to a single component of the boundary.}
of the 4-manifold $M_4$,
\be
M_3 \; = \; \partial M_4
\label{M3asbdryofM4}
\ee
This relation between $M_3$ and $M_4$ translates into the statement that 2d $\CN=(0,2)$ theory $T[M_4]$
is a boundary theory for the 3d $\CN=2$ theory $T[M_3]$ on a half-space $\R_+ \times \R^2$.
In order to see this, it is convenient to recall that both theories $T[M_3]$ and $T[M_4]$ can be defined
as fivebrane configurations (or, compactifications of 6d $(2,0)$ theory) on the corresponding manifolds, $M_3$ and $M_4$.
This gives a {\it coupled} system of 2d-3d theories $T[M_4]$ and $T[M_3]$ since both originate from the same configuration
in six dimensions, which looks like $M_3 \times \R_+ \times \R^2$ near the boundary and $M_4 \times \R^2$ away from the boundary.
In other words, a 4-manifold $M_4$ with a boundary $M_3$ defines a half-BPS (B-type) boundary condition in a 3d $\CN=2$ theory $T[M_3]$.

Therefore, in order to understand a 2d theory $T[M_4]$ we need to identify a 3d theory $T[M_3]$ or, at least, its necessary
elements.\footnote{\label{footnt:abelian}While this problem has been successfully solved for a large class
of 3-manifolds \cite{DGG,CCV,DGGindex}, unfortunately it will not be enough for our purposes here
and we need to resort to matching $M_3$ with $T[M_3]$ based on identification of vacua, as was originally proposed in \cite{DGH}.
One reason is that the methods of {\it loc. cit.} work best for 3-manifolds with sufficiently large boundary
and/or fundamental group, whereas in our present context $M_3$ is itself a boundary and, in many cases, is a rational homology sphere.
As we shall see below, 3d $\CN=2$ theories $T[M_3]$ seem to be qualitatively different in these two cases;
typically, they are (deformations of) superconformal theories in the former case and massive 3d $\CN=2$ theories in the latter.
Another, more serious issue is that 3d theories $T[M_3]$ constructed in \cite{DGG} do not account for {\it all} flat connections on $M_3$,
which will be crucial in our applications below.
This second issue can be avoided by considering larger 3d theories $T^{(\text{ref})} [M_3]$
that have to do with refinement/categorification and mix all branches of flat connections \cite{FGS,FGP}.
Pursuing this approach should lead to new relations with rich algebraic structure and functoriality of knot homologies.}
One important characteristic of a 3d $\CN=2$ theory $T[M_3]$ is the space of its supersymmetric vacua,
either in flat space-time $\R^3$, or on a circle, {\it i.e.} in space-time $S^1 \times \R^2$.
This will be the subject of our discussion here.

Specifically, when 3d $\CN=2$ theory $T[M_3;G]$ is considered on a circle,
its supersymmetric ground states are in one-to-one correspondence with gauge equivalence classes
of flat $G_{\C}$ connections on $M_3$ \cite{DGH}:
\be
d \CA + \CA \wedge \CA \; = \; 0
\label{Aflat}
\ee
This follows from the duality between fivebranes on $S^1$ and D4-branes combined with
the fact that D4-brane theory is partially twisted along the 3-manifold $M_3$.
The partial twist in the directions of $M_3$ is the dimensional reduction of the Vafa-Witten twist \cite{VafaWitten}
as well as the GL twist \cite{KW} of the $\CN=4$ super-Yang-Mills in four dimensions.
The resulting $\CN_T = 4$ three-dimensional topological gauge theory on $M_3$
is the equivariant version of the Blau-Thompson theory \cite{Blau:1996bx,Blau:1997pp}
that localizes on solutions of \eqref{Aflat}, where $\CA = A + i B$ is the $\text{Lie} (G_{\C})$-valued connection.

{}From the viewpoint of the topological Vafa-Witten theory on $M_4$, solutions to the equation \eqref{Aflat} provide
boundary conditions for PDEs in four dimensions.
To summarize,
$$
\boxed{{\text{boundary conditions} \atop \text{on}~M_4}}
\quad \longleftrightarrow \quad
\boxed{{\text{complex flat} \atop \text{connections on}~M_3}}
\quad \longleftrightarrow \quad
\boxed{\phantom{\int} \text{vacua of}~T[M_3] \phantom{\int}}
$$

In general, complex flat connections on $M_3$ are labeled by representations of the fundamental group $\pi_1 (M_3)$
into $G_{\C}$, modulo conjugation,
\be
\CV_{T[M_3;G]} \; = \; \text{Rep} \left( \pi_1 (M_3) \to G_{\C} \right) / \text{conj.}
\label{repVar}
\ee
In particular, in the basic case of abelian theory ({\it i.e.} a single fivebrane),
the vacua of the 3d $\CN=2$ theory $T[M_3]$ are simply abelian representations of $\pi_1 (M_3)$,
{\it i.e.} elements of $H_1 (M_3)$.
In the non-abelian case, $G_{\C}$ flat connection on $M_3$ are described by nice algebraic equations,
which play an important role in complex Chern-Simons theory and its relation to quantum group invariants~\cite{Apol}.

As will become clear shortly, for many simply-connected 4-manifolds \eqref{M4KKK} built from
2-handles --- such as sphere plumbings represented by trees ({\it i.e.} graphs without loops) ---
the boundary $M_3$ is a rational homology sphere ($b_1 (M_3) = 0$)
in which case the theory $T[M_3; U(1)]$ has finitely many isolated vacua,
\be
\# \{ \text{vacua of}~T[M_3; U(1)] \} \; = \; |H_1 (M_3; \Z)|
\label{H1vac}
\ee
Therefore, the basic piece of data that characterizes $M_3 = \partial M_4$ and the corresponding 3d theory $T[M_3]$
is the first homology group $H_1 (M_3; \Z)$.
Equivalently, when $H_1 (M_3; \Z)$ is torsion, by the Universal Coefficient Theorem we can label the vacua of $T[M_3; U(1)]$
by elements of $H^2 (M_3;\Z)$. Indeed, given a 1-cycle $\mu$ in $M_3$, the Poincar\'e dual class $[\mu] \in H^2 (M_3;\Z)$
can be interpreted as the first Chern class $c_1 (\CL) = [\mu]$ of a complex line bundle $\CL$,
which admits a flat connection whenever the first Chern class is torsion.
The (co)homology groups of the boundary 3-manifold $M_3$ --- that, according to \eqref{H1vac}, determine the vacua of $T[M_3]$ --- are
usually easy to read off from the Kirby diagram of $M_4$.

Now, once we explained the physical role of the boundary $M_3 = \partial M_4$,
we need to discuss its topology in more detail that will allow us to describe complex flat connections on $M_3$ and,
therefore, determine the vacua of the 3d $\CN=2$ theory $T[M_3]$.
In general, the boundary of a simply-connected 4-manifold \eqref{M4KKK} labeled by a framed link
is an integral surgery on that link in $S^3$.
This operation consists of removing the tubular neighborhood $N(K_i) \cong S^1 \times D^2$ of each link component
and then gluing it back in a different way, labeled by a non-trivial self-diffeomorphism $\phi : T^2 \to T^2$
of the boundary torus $\partial N(K_i) \cong T^2$.

This description of the boundary 3-manifold $M_3$ is also very convenient for describing complex flat connections.
Namely, from the viewpoint of $T^2$ that divides $M_3$ into two parts, complex flat connections on $M_3$
are those which can be simultaneously extended from the boundary torus to $M_3 \setminus K_i$ and $N(K_i) \cong S^1 \times D^2$,
equivalently, the intersection points
\be
\CV_{T[M_3]} \; = \; \CV_{T[M_3 \setminus K]} \cap \phi \left( \CV_{T[S^1 \times D^2]} \right)
\label{vacspaceintersect}
\ee
Here, the representation varieties of the knot complement and the solid torus can be interpreted as $(A,B,A)$ branes
in the moduli space of $G$ Higgs bundles on $T^2$.
In this interpretation, $\phi$ acts as an autoequivalence on the category of branes,
see {\it e.g.} \cite{GukovRTN} for some explicit examples and the computation of \eqref{vacspaceintersect} in the case $G_{\C} = SL(2,\C)$.

Coming back to the vacua \eqref{H1vac},
the cohomology group $H^2 (M_3;\Z)$ can be easily deduced from
the long exact sequence for the pair $(M_4,M_3)$ with integer coefficients:
\be
\begin{array}{ccccccccccccc}
0 & \to & H^2 (M_4 , M_3) & \to & H^2 (M_4) & \to & H^2 (M_3)
 & \to & H^3 (M_4 , M_3) & \to & H^3 (M_4) & \to & 0 \\
 &  & \parallel & & \| & & & & \| & & \| & & \\
 &  & \Z^{b_2} \oplus T_2 & & \Z^{b_2} \oplus T_1 & & & & T_1 & & T_2 & &
\end{array}
\label{TTTlong}
\ee
where $T_1$ and $T_2$ are torsion groups. Since $T_2 \to T_1$ is injective, one can introduce $t = |T_1| / |T_2|$.
Then,
\be
|H_1 (M_3; \Z)| \; = \; t^2 |\det Q|
\ee
In particular, when both torsion groups $T_1$ and $T_2$ are trivial, we simply have a short exact sequence
\be
0 \; \longrightarrow \; \Gamma \; \xrightarrow{~Q~} \; \Gamma^* \; \longrightarrow \; H^2 (M_3) \; \longrightarrow \; 0
\label{HHHseq}
\ee
so that $H_1 (M_3) \cong H^2 (M_3)$ is isomorphic to $\Z^{b_2} / Q (\Z^{b_2})$, generated by the meridians $\mu_i$
of the link components $K_i$, modulo relations imposed by the intersection form $Q$ of the 4-manifold \eqref{M4KKK}:
\be
H_1 (M_3; \Z) \; = \;  \Z [ \mu_1 , \ldots , \mu_n ] / \text{im} Q
\label{H1kerQ}
\ee
It follows that, in the case of $G=U(1)$ ({\it i.e.} a single fivebrane), the representation variety \eqref{repVar}
is parametrized by the eigenvalues $x_i \in \C^*$ of the $G_{\C}$-valued holonomies along the 1-cycles $\mu_i$,
subject to the relations in \eqref{H1kerQ}:
\be
\prod_{i=1}^n x_i^{Q_{ij}} \; = \; 1 \qquad \forall j = 1, \ldots, n
\label{xxvacQ}
\ee
There is a similar description of $\CV_{T[M_3;G]}$ for non-abelian groups as well~\cite{Apol}.
One important consequence of this calculation is that $H_1 (M_3; \Z)$ is finite and, therefore,
the 3d $\CN=2$ theory $T[M_3]$ has finitely many vacua if and only if all eigenvalues of the intersection form $Q_{M_4}$ are non-zero.
If $Q$ has zero eigenvalues, then $H_1 (M_3; \Z)$ contains free factors.
This happens, for example, for knots with zero framing coefficients, $a=0$.
Every such Kirby diagram leads to a boundary 3-manifold $M_3$,
whose first homology group is generated by the meridian $\mu$ of the knot $K$ with no relations.
This clarifies, for instance, why the boundary of a 4-manifold shown in Figure \ref{fig:Borromean}
has $H_1 (M_3; \Z) \cong \Z^3$.

\bigskip
\begin{figure}[ht]
\centering
\includegraphics[width=5.0in]{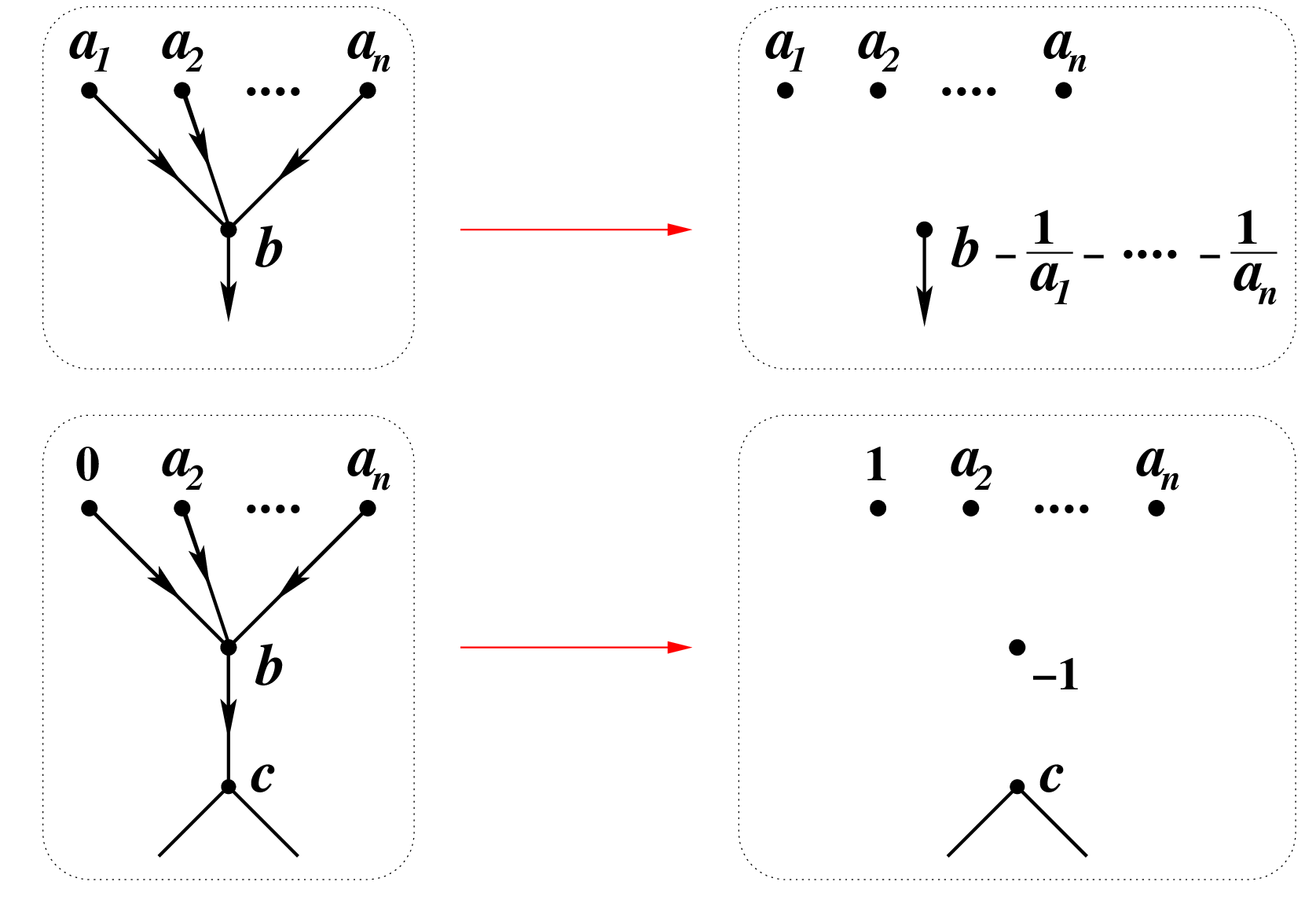}
\caption{For a plumbing tree, the eigenvalues (and, therefore, the determinant) of the intersection form $Q$
can be computed by orienting the edges toward a single vertex and then successively eliminating them using the two rules shown here.}
\label{fig:detQrules}
\end{figure}

If $M_4$ is a sphere plumbing represented by a plumbing tree $\Upsilon$,
then the eigenvalues of $Q$ can be obtained using a version of the Gauss algorithm
that consists of the following two simple steps (see {\it e.g.} \cite{Saveliev}):

\begin{enumerate}

\item
Pick any vertex in $\Upsilon$ and orient all edges toward it. Since $\Upsilon$ is a tree, this is always possible.

\item
Recursively applying the rules in Figure \ref{fig:detQrules} remove the edges, replacing the integer
weights $a_i$ (= framing coefficients of the original Kirby diagram) by rational weights.

\end{enumerate}

In the end of this process, when there are no more edges left, the rational weights $r_i$
are precisely the eigenvalues of the intersection form $Q$ and
\begin{subequations}\label{plumbtreeQ}
\be
\det (Q) \; = \; \prod_i r_i
\ee
\be
\text{sign} (Q) \; = \; \# \{ i | r_i > 0 \} - \# \{ i | r_i < 0 \}
\ee
\end{subequations}

\begin{figure}[ht]
\centering
\includegraphics[width=3.5in]{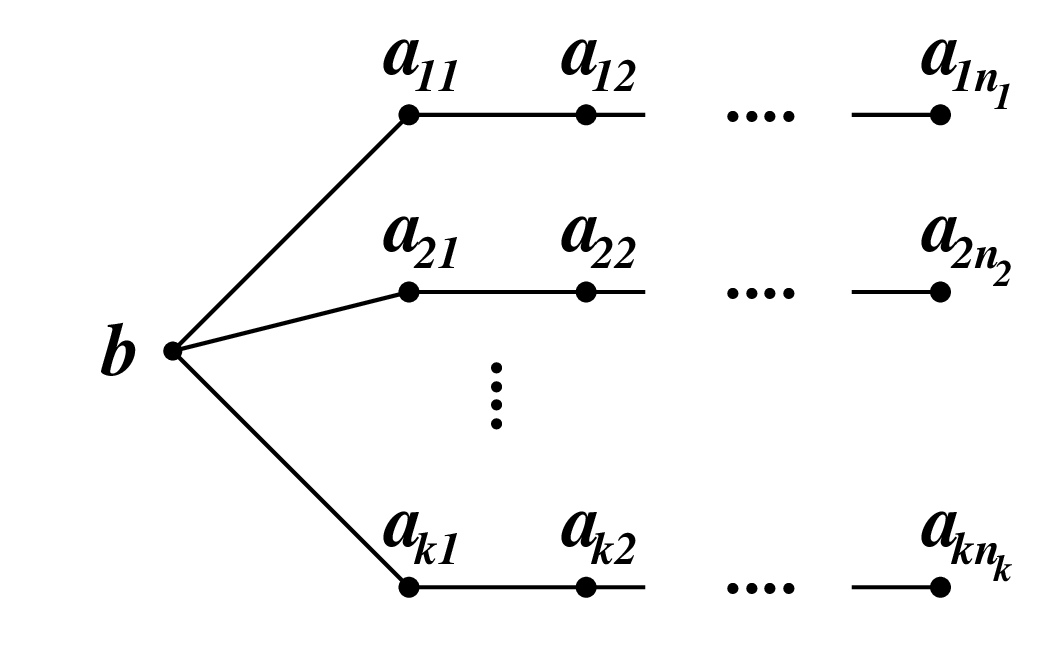}
\caption{Plumbing tree of a 4-manifold bounded by a Seifert fibration. We assume $b \le -1$ and $a_{ij} \le -2$.}
\label{fig:Seifert}
\end{figure}

For example, applying this algorithm to
the plumbing tree in Figure~\ref{fig:Seifert} we get
\be
\det (Q) \; = \; \left( b + \sum_{i=1}^k \frac{q_i}{p_i} \right) \cdot \prod_{i=1}^k p_i
\ee
where $-\frac{p_i}{q_i} = [a_{i1}, \ldots, a_{in_i}]$ are given by the continued fractions
\be
- \frac{p_i}{q_i} \; = \; a_{i1} - \cfrac{1}{a_{i2} - \cfrac{1}{\ddots - \cfrac{1}{a_{in_i}}}}
\ee
The boundary 3-manifold in this case is the Seifert fibered homology 3-sphere
$M_3 (b; (p_1,q_1),$ $\ldots, (p_k,q_k))$ with singular fibers of orders $p_i \ge 1$.
It is known that any Seifert fibred rational homology sphere bounds at least one definite form.
In our applications here, we are mostly interested in the choice of orientation, such that
a Seifert manifold $M_3$ bounds a plumbed 4-manifold with negative definite intersection form.
Then, $M_3$ is the link of a complex surface singularity.

\bigskip

\subsection*{Quiver Chern-Simons theory}

We already mentioned a striking similarity between plumbing graphs and quivers.
The latter are often used to communicate quickly and conveniently the field content of gauge theories,
in a way that each node of the quiver diagram represents a simple Lie group and every edge corresponds to a bifundamental matter.
Here, we take this hint a little bit more seriously and, with a slight modification of the standard rules,
associate a 3d $\CN=2$ gauge theory to a plumbing graph $\Upsilon$, which will turn out to be an example
of the sought-after theory $T[M_3]$.

Much as in the familiar quiver gauge theories, to every vertex of $\Upsilon$ we are going to
associate a gauge group factor. Usually, the integer label of the vertex represents the rank.
In our present example, however, we assign to each vertex a gauge group $U(1)$
with pure $\CN=2$ Chern-Simons action at level $k$ determined by the integer weight (= the framing coefficient) of that vertex:
\bea
S & = & \frac{k}{4\pi} \int d^3 x d^4 \theta \; V \Sigma \label{SUSYCSaction} \\
& = & \frac{k}{4\pi} \int (A \wedge dA - \bar \lambda \lambda + 2 D \sigma) \nonumber
\eea
Here, $V = (A_{\mu},\lambda,\sigma,D)$ is the three-dimensional $\CN=2$ vector superfield
and $\Sigma = \bar D^{\alpha} D_{\alpha} V$ is the field strength superfield.

Similarly, to every edge of $\Upsilon$ that connects a vertex ``$i$'' with a vertex ``$j$''
we associate 3d $\CN=2$ Chern-Simons coupling between the corresponding vector superfields $V_i$ and $V_j$:
\be
S = \frac{1}{2\pi} \int d^3 x d^4 \theta \; V_i \Sigma_j
\ee
Both of these basic building blocks can be combined together with the help of the symmetric bilinear form \eqref{Qplumbing}.
As a result, to a plumbing graph $\Upsilon$ we associate the following 3d $\CN=2$ theory:
\be
T[M_3; U(1)] \; = \;
\left \{
\begin{array}{l}
U(1)^n ~\text{quiver Chern-Simons theory with Lagrangian} \\[.2cm]
\quad \CL = \displaystyle \sum_{i,j=1}^n \int d^4 \theta \, \frac{Q_{ij}}{4\pi} \, V_i \Sigma_j
\, = \, \frac{1}{4\pi} \int Q (A,dA) + \ldots
\end{array}
\right .
\label{quiverCS}
\ee
where $n = \text{rank} (Q)$ and the ellipses represent $\CN=2$ supersymmetric completion of the bosonic Chern-Simons action.
Note, since the gauge group is abelian, the fermions in the $\CN=2$ supersymmetric completion of this Lagrangian decouple.
As for the bosonic part, quantum-mechanically it only depends on the discriminant group of the lattice $(\Gamma,Q)$,
\be
\frak{D} \; = \; H_1 (M_3; \Z)
\ee
and a $\mathbb{Q} / \Z$-valued quadratic form $\frak q$ on $\frak{D}$ \cite{Kapustin:2010hk}.

We claim that the quiver Chern-Simons theory \eqref{quiverCS} provides
a Lagrangian description of the 3d $\CN=2$ theory $T[M_3; U(1)]$ for {\it any} boundary 3-manifold $M_3$.
Indeed, by a theorem of Rokhlin, every closed oriented 3-manifold $M_3$ bounds a 4-manifold of the form \eqref{M4KKK}
and can be realized as an integral surgery on some link in $S^3$.
Denoting by $Q$ the intersection form (resp. the linking matrix) of the corresponding 4-manifold (resp. its Kirby diagram),
we propose 3d $\CN=2$ theory \eqref{quiverCS} with Chern-Simons coefficients $Q_{ij}$ to be a Lagrangian
description of the boundary theory $T[M_3;U(1)]$.

To justify this proposal, we note that supersymmetric vacua of the theory \eqref{quiverCS} on $S^1 \times \R^2$
are in one-to-one correspondence with solutions to \eqref{xxvacQ}.
Indeed, upon reduction on a circle, each 3d $\CN=2$ vector multiplet
becomes a twisted chiral multiplet, whose complex scalar component we denote $\sigma_i = \log x_i$.
The Chern-Simons coupling \eqref{quiverCS} becomes the twisted chiral superpotential, see {\it e.g.} \cite{DGG,FGP}:
\be
\tilde \CW \; = \; \sum_{i,j=1}^n \frac{Q_{ij}}{2} \, \log x_i \cdot \log x_j
\label{WquiverCS}
\ee
Extremizing the twisted superpotential with respect to the dynamical fields $\sigma_i = \log x_i$
gives equations for supersymmetric vacua:
\be
\exp \left( \frac{\partial \tilde \CW}{\partial \log x_i} \right) \; = \; 1
\label{Wcritptseqs}
\ee
which reproduce \eqref{xxvacQ}.

\subsection*{The Lens space theory}

Of particular importance to the construction of two-dimensional theories $T[M_4]$
are special cases that correspond to 4-manifolds bounded by Lens spaces $L(p,q)$.
We remind that the Lens space $L(p,q)$ is defined as the quotient
of $S^3 = \{ (z_1,z_2) \in \C^2 \; \vert \; |z_1|^2 + |z_2|^2 = 1 \}$
by a $\Z_p$-action generated by
\be
(z_1 , z_2) \sim (e^{2 \pi i/p} z_1 , e^{2 \pi i q/p} z_2)
\ee
We assume $p$ and $q$ to be coprime integers in order to ensure that $\Z_p$-action is free and the quotient is smooth.
Two Lens spaces $L(p,q_1)$ and $L(p,q_2)$ are homotopy equivalent if and only if $q_1 q_2 \equiv \pm n^2 \mod p$
for some $n \in \mathbb{N}$, and homeomorphic if and only if $q_1 \equiv \pm q_2^{\pm 1} \mod p$.
Reversing orientation means $L(p,-q) = - L(p,q)$.
Note, supersymmetry (of the cone built on the Lens space) requires $q+1 \equiv 0 \mod p$.

In the previous discussion we already encountered several examples of 4-manifolds bounded by Lens spaces.
These include the disk bundle over $S^2$ with the Kirby diagram \eqref{single2handle}
and the linear plumbing on $A_{p-1}$, which are bounded by $L(p,1)$ and $L(p,-1)$, respectively.
In particular, for future reference we write
\be
\partial A_p \; = \; L(p+1,p)
\ee
In fact, a more general linear plumbing of oriented circle bundles over spheres
with Euler numbers $a_1, a_2, \ldots, a_n$ (see Figure~\ref{fig:Anplumbing})
is bounded by a Lens space $L(p,q)$, such that $[a_1, a_2, \ldots, a_n]$
is a continued fraction expansion for $-\frac{p}{q}$,
\be
- \frac{p}{q} \; = \; a_1 - \cfrac{1}{a_2 - \cfrac{1}{\ddots - \cfrac{1}{a_n}}}
\label{contfract}
\ee
When $p>q>0$ we may restrict the continued fraction coefficients to be integers
$a_i \le -2$, for all $i = 1, \ldots, n$,
so that $L(p,q)$ is the oriented boundary of the negative definite plumbing
associated to the string $(a_1, a_2, \ldots, a_n)$.
With these orientation conventions,
the Lens space $L(p,q)$ is defined by a $(-\frac{p}{q})$-surgery on an unknot in $S^3$.
We also point out that any lens space $L(p,q)$ bounds both positive and negative definite forms $Q$.
(Note, according to the Donaldson's theorem \cite{Donaldson},
the only definite forms that $S^3$ bounds are the diagonal unimodular forms.)

Next, let us discuss 3d $\CN=2$ theory $T[M_3;G]$ for $M_3 = L(p,q)$ and $G=U(N)$.
First, since $H_1 (M_3) = \Z_p$ we immediately obtain the number of vacua on $S^1 \times \R^2$, {\it cf.} \eqref{H1vac}:
\be
\# \{ \text{vacua of}~T[L(p,q); U(N)] \} \; = \; \frac{(N+p-1)!}{N! (p-1)!}
\label{Lpqvac}
\ee
which, according to \eqref{repVar}, is obtained by counting $U(N)$ flat connections on $S^3 / \Z_p$.
Incidentally, this also equals the number of $SU(p)$ representations at level $N$,
which is crucial for identifying Vafa-Witten partition functions on ALE spaces with WZW characters \cite{Nakajima,VafaWitten}.

There are several ways to approach the theory $T[L(p,q); U(N)]$, in particular, to give a Lagrangian description,
that we illustrate starting with the simple case of $N=1$ and $q=1$.
For example, one approach is to make use of the Hopf fibration structure on the Lens space $L(p,1) = S^3 / \Z_p$
and to reduce the M-theory setup with a fivebrane on the $S^1$ fiber.
This reduction was very effective {\it e.g.} in analyzing a similar system of fivebranes on Lens spaces
with half as much supersymmetry \cite{Acharya:2001dz}.
It yields type IIA string theory with a D4-brane wrapped on the base $S^2$ of the Hopf
fibration with $-p$ units of Ramond-Ramond 2-form flux through the $S^2$.
The effective theory on the D4-brane is 3d $\CN=2$ theory with $U(1)$ gauge group
and supersymmetric Chern-Simons coupling at level $-p$ induced by the RR 2-form flux,
thus, motivating the following proposal:
\be
T[L(p,1); U(1)] \; = \; U(1) \text{ SUSY Chern-Simons theory at level } -p
\label{TLensU1}
\ee
To be more precise, this theory as well as quiver Chern-Simons theories \eqref{quiverCS}
labeled by plumbing graphs in addition include free chiral multiplets, one for each vertex in the plumbing graph.
Since in the abelian, $G = U(1)$ case these chiral multiplets decouple and do not affect the counting
of $G_{\C}$ flat connections, we tacitly omit them in our present discussion.
However, they play an important role and need to be included in the case of $G = U(N)$.

\bigskip
\begin{figure}[ht]
\centering
\includegraphics[width=4.5in]{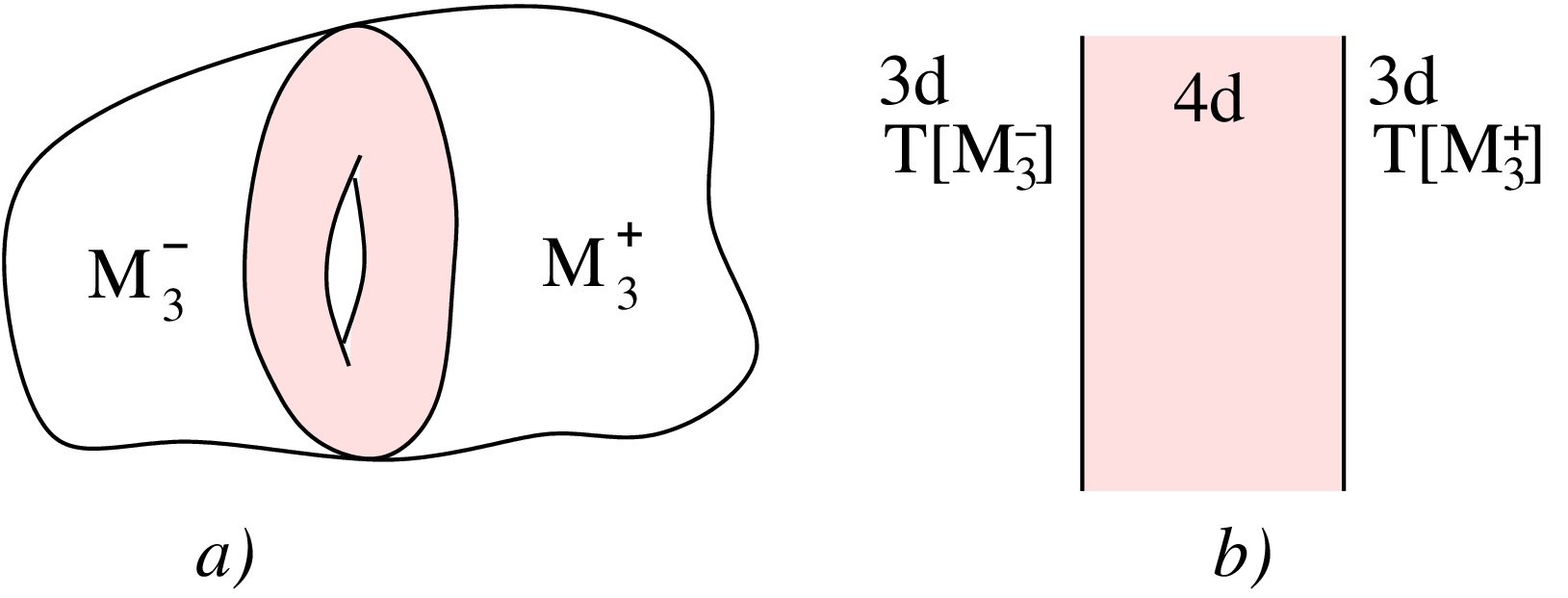}
\caption{$(a)$ A genus-1 Heegaard splitting of a 3-manifold $M_3$ becomes
a 4d $\CN=4$ super-Yang-Mills theory $(b)$ coupled to 3-dimensional $\CN=2$ theories
$T[M_3^-]$ and $T[M_3^+]$ at the boundary.}
\label{fig:M3d3d}
\end{figure}

Another approach, that also leads to \eqref{TLensU1}, is based on the Heegaard splitting of $M_3$.
Indeed, as we already mentioned earlier, $L(p,q)$ is a Dehn surgery on the unknot in $S^3$
with the coefficient $-\frac{p}{q}$. It means that $M_3 = L(p,q)$ can be glued from two copies
of the solid torus, $S^1 \times D^2$, whose boundaries are identified via non-trivial map $\phi : T^2 \to T^2$.
The latter is determined by its action on homology $H_1 (T^2; \Z) \cong \Z \oplus \Z$ which,
as usual, we represent by a $2 \times 2$ matrix
\be
\phi =
\begin{pmatrix}
p & r \\
q & s
\end{pmatrix}
\label{knottwist}
\ee
with $ps - qr = 1$.
If $(-\frac{p}{q}) = [a_1, a_2, \ldots , a_n]$ is given by the continued fraction expansion \eqref{contfract},
we can explicitly write
\be
\begin{pmatrix}
p & r \\
q & s
\end{pmatrix}
=
\begin{pmatrix}
- a_1 & -1 \\
1 & 0
\end{pmatrix}
\begin{pmatrix}
- a_2 & -1 \\
1 & 0
\end{pmatrix}
\ldots
\begin{pmatrix}
- a_n & -1 \\
1 & 0
\end{pmatrix}
\ee
This genus-1 Heegaard decomposition has a simple translation to physics, illustrated in Figure~\ref{fig:M3d3d}.
Again, let us first consider the simple case with $N=1$ and $q=1$.
Then, the 6d $(0,2)$ theory on $T^2$ gives 4d $\CN=4$ supersymmetric Maxwell theory,
in which the $SL(2,\Z)$ action \eqref{knottwist} on a torus is realized as the electric-magnetic duality transformation.
On the other hand, each copy of the solid torus defines a ``Lagrangian'' boundary condition
that imposes Dirichlet boundary condition on half of the $\CN=4$ vector multiplet
and Neumann boundary condition on the other half.
Hence, the combined system that corresponds to the Heegaard splitting of $L(p,1)$
is 4d $\CN=4$ Maxwell theory on the interval with two Lagrangian boundary conditions
that are related by an S-duality transformation $\phi = \bigl( \begin{smallmatrix}
p & -1 \\
1 & 0
\end{smallmatrix} \bigr)$
and altogether preserve $\CN=2$ supersymmetry in three non-compact dimensions.

\begin{figure}[ht]
\centering
\includegraphics[width=2.5in]{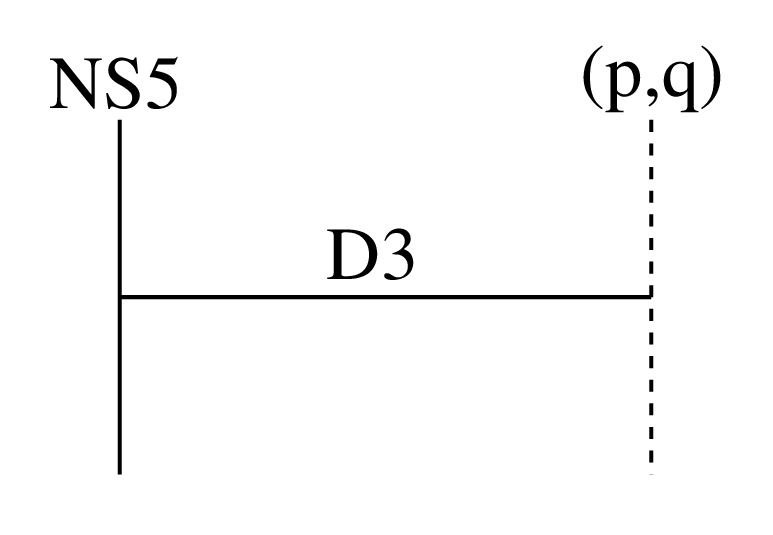}
\caption{The effective 3d $\CN=2$ theory on a D3-brane stretched between NS5-brane and a 5-brane of type $(p,q)$
is a Chern-Simons theory at level $k = - \frac{p}{q}$. We describe it as a ``quiver Chern-Simons theory''
with {\it integer} levels $a_i$ given by the continued fraction $- \frac{p}{q} = [a_1, \ldots, a_n]$.}
\label{fig:IIBbranes}
\end{figure}

Following the standard techniques \cite{HananyW,Gaiotto:2008sa},
this theory can be realized on the world-volume of a D3-brane stretched between two fivebranes,
which impose suitable boundary conditions at the two ends of the interval.
If both boundary conditions were the same, we could take both fivebranes to be NS5-branes.
However, since in this brane approach the S-duality of $\CN=4$ gauge theory is realized
as S-duality of type IIB string theory, it means that the two fivebranes on which D3-brane ends
are related by a transformation \eqref{knottwist}.
In particular, if we choose one of the fivebranes to be NS5, then the second fivebrane must be
a $(p,q)$ fivebrane, with D5-brane charge $p$ and NS5-brane charge $q$, as shown in Figure~\ref{fig:IIBbranes}.
In the present case, $q=1$ and the effective theory on the D3-brane stretched between NS5-brane
and a 5-brane of type $(p,1)$ is indeed $\CN=2$ abelian Chern-Simons theory \eqref{SUSYCSaction} at level $-p$,
in agreement with \eqref{TLensU1}.

\bigskip

This approach based on Heegaard splitting and the brane construction suggests that $T[L(p,q); U(1)]$
associated to a more general gluing automorphism \eqref{knottwist}
should be a 3d $\CN=2$ theory on the D3-brane stretched between NS5-brane and a 5-brane of type $(p,q)$.
This theory on the D3-brane, shown in Figure~\ref{fig:IIBbranes},
indeed has the effective Chern-Simons coupling at level $-\frac{p}{q}$ \cite{Kitao:1998mf,BHKK,Ohta}.
However, a better way to think about this $\CN=2$ theory --- that avoids using fractional Chern-Simons levels
and that we take as a proper Lagrangian formulation of $T[L(p,q); U(1)]$ --- is based on writing the general
$SL(2,\Z)$ element \eqref{knottwist} as a word in standard $S$ and $T$ generators that obey $S^4 = (ST)^3 = \text{id}$,
\be
\phi \; = \; S \, T^{a_1} \, S \, T^{a_2} \, \cdots \, S \, T^{a_n}
\ee
and implementing it as a sequence of operations on the 3d $\CN=2$ abelian gauge theory {\it a la} \cite{Witten:2003ya}.
Specifically, the $T$ element of $SL(2,\Z)$ acts by adding a level-1 Chern-Simons term,
\be
T\,:\quad \Delta \CL = \frac{1}{4\pi} \int d^4 \theta \; V \Sigma = \frac{1}{4\pi} A\wedge dA + \ldots
\label{TonT}
\ee
while the $S$ transformation introduces a new $U(1)$ gauge (super)field $\tilde A$
coupled to the ``old'' gauge (super)field $A$ via Chern-Simons term
\be
S\,:\quad \Delta \CL = \frac{1}{2\pi} \int d^4 \theta \; \tilde V \Sigma = \frac{1}{2 \pi} \tilde A \wedge dA + \ldots
\label{SonT}
\ee
Equivalently, the new vector superfield containing $\tilde A$ couples to the ``topological'' current $*F = *dA$
carried by the magnetic charges for $A$.

Using this $SL(2,\Z)$ action on abelian theories in three dimensions, we propose the following candidate
for the generalization of the Lens space theory \eqref{TLensU1} to $|q| \ge 1$:
\be
T[L(p,q); U(1)] \; = \; U(1)^n \text{ theory with Chern-Simons coefficients } Q_{ij}
\label{TpqLensU1}
\ee
where the matrix $Q$ is given by \eqref{Qmatrix} and $-\frac{p}{q} = [a_1, \ldots, a_n]$
is the continued fraction expansion \eqref{contfract}.
Note, the matrix of Chern-Simons coefficients in this Lens space theory can be conveniently
represented by a quiver diagram identical to the plumbing graph in Figure~\ref{fig:Anplumbing}.
The proposal \eqref{TpqLensU1} for the Lens space theory is, in fact, a special case of \eqref{quiverCS}
and can be justified in the same way, by comparing the critical points of the twisted superpotential \eqref{WquiverCS}
with solutions to \eqref{xxvacQ}.

Both methods that we used to derive the basic 3d $\CN=2$ Lens space theory \eqref{TLensU1}
suggest a natural generalization to $G=U(N)$:
\be
T[L(p,1); U(N)] \; = \;
\left \{
\begin{array}{l}
U(N) \text{ SUSY Chern-Simons theory at level } -p \\[.1cm]
\text{with a chiral multiplet in the adjoint representation}
\end{array}
\right .
\label{TLensUN}
\ee
which corresponds to replacing a single D3-brane in the brane construction on Figure~\ref{fig:IIBbranes}
by a stack of $N$ D3-branes.
Indeed, the Witten index of $\CN=2$ Chern-Simons theory with gauge group $SU(N)$ and level $p$ (with or without super-Yang-Mills term)
is equal to the number of level $p$ representations of affine $SU(N)$, see \cite{Witten:1999ds} and also \cite{BHKK,Ohta,Smilga}:
\be
\CI_{SU(N)_p} \; = \; \frac{(N+p-1)!}{(N-1)! p!}
\ee
After multiplying by $\frac{p}{N}$ to pass from the gauge group $SU(N)$ to $U(N) = \frac{U(1) \times SU(N)}{\Z_N}$
we get the number of $SU(p)_N$ representations \eqref{Lpqvac},
which matches the number of $U(N)$ flat connections on the Lens space $L(p,1)$.
Note, that the role of the level and the rank are interchanged compared to what one might naturally expect.
An alternative UV Lagrangian for the theory \eqref{TLensUN}, that makes contact with the cohomology
of the Grassmannian \cite{Witten:1993xi,Kapustin:2013hpk}, is a $\CN=2$ $U(N)$ Chern-Simons action at level $-\frac{p}{2}$
coupled to a chiral multiplet in the adjoint representation and $p$ chiral multiplets in the anti-fundamental representation.
This theory was studied in detail in \cite{Gukov:2015sna}, where further connections to integrable systems and quantum equivariant K-theory of vortex moduli spaces were found.

\subsection*{3d $\CN=2$ theory $T{[} M_3 ; G {]}$ for general $M_3$ and $G$}

Now it is clear how to tackle the general case of $N$ fivebranes on a 4-manifold $M_4$ with boundary $M_3 = \partial M_4$.
This setup leads to a 2d $\CN=(0,2)$ theory $T[M_4;G]$ on the boundary of the half-space coupled to a 3d $\CN=2$ theory $T[M_3;G]$ in the bulk,
with the group $G$ of rank $N$ and Cartan type $A$, $D$, or $E$.

For a general class of 4-manifolds \eqref{M4KKK} considered here, the boundary 3-manifold is an integral surgery on a link $K$ in $S^3$.
As usual, we denote the link components $K_i$, $i = 1, \ldots, n$.
Therefore, the corresponding theory $T[M_3]$ can be built by ``gluing'' the 3d $\CN=2$ theory $T[S^3 \setminus K]$
assoiated to the link complement with $n$ copies of the 3d $\CN=2$ theory $T[S^1 \times D^2]$ associated to the solid torus:
\be
T[M_3] \; = \; T[S^3 \setminus K] \, \otimes \,
\underbrace{ \Big( \phi_{a_1} \circ T[S^1 \times D^2] \Big) \, \otimes \, \ldots \, \otimes \, \Big( \phi_{a_n} \circ T[S^1 \times D^2]
\Big)}_{\displaystyle{n \text{ copies}}}
\label{generalTM3}
\ee
As pointed out in the footnote \ref{footnt:abelian}, it is important that the theory $T[S^3 \setminus K]$
accounts for {\it all} flat $G_{\C}$ connections on the link complement, including the abelian ones.
Such theories are known for $G_{\C} = SL(2,\C)$ and for many simple knots and links \cite{Nawata,FGSS},
in fact, even in a more ``refined'' form that knows about categorification and necessarily incorporates all branches of flat connections.
For $G_{\C}$ of higher rank, it would be interesting to work out such $T[S^3 \setminus K]$ following \cite{Dimofte:2013iv}.
In particular, the results of \cite{Dimofte:2013iv} elucidate one virtue of 3d $\CN=2$ theories $T[M_3;G]$:
they always seem to admit a UV description with only $U(1)$ gauge fields (but possibly complicated matter content and interactions).
This will be especially important to us in section \ref{sec:2dtheory}: in order to identify a 2d $(0,2)$ theory $T[M_4]$
asociated to a 4-manifold $M_4$ bounded by $M_3$ we only need to understand boundary conditions of abelian 3d $\CN=2$ theories.

The second basic ingredient in \eqref{generalTM3} is the theory $T[S^1 \times D^2]$ associated to the solid torus.
This theory is very simple for any $N \ge 1$ and corresponds to
the Dirichlet (D5-brane) boundary condition of $\CN=4$ super-Yang-Mills theory, {\it cf.} Figure \ref{fig:M3d3d}.
To be more precise, if we denote by $\mathbb{T} \subset G$ the maximal torus of $G$,
then $G_{\C}$ flat connections on $T^2 = \partial \big( S^1 \times D^2 \big)$
are parametrized by two $\mathbb{T}_{\C}$-valued holonomies, modulo the Weyl group $W$ of $G$,
\be
(x,y) \; \in \; \left( \mathbb{T}_{\C} \times \mathbb{T}_{\C} \right) / W
\ee
Only a middle dimensional subvariety in this space corresponds to $G_{\C}$ flat connections that
can be extended to the solid torus $S^1 \times D^2$.
Namely, since one of the cycles of $T^2$ (the meridian of $K_i$) is contractible in $N(K_i) \cong S^1 \times D^2$,
the $G_{\C}$ holonomy on that cycle must be trivial, {\it i.e.}
\be
\CV_{T[S^1 \times D^2]} \; = \; \left \{ (x_i,y_i) \in \frac{\mathbb{T}_{\C} \times \mathbb{T}_{\C}}{W} \; \Big| \; x_i = 1 \right \}
\label{Tsolidtorus}
\ee
The $SL(2,\Z)$ transformation $\phi_{a_i}$ gives a slightly more interesting theory $\phi_{a_i} \circ T[S^1 \times D^2]$,
whose space of supersymetric vacua \eqref{repVar} is simply an $SL(2,\Z)$ transform of \eqref{Tsolidtorus}:
\be
\CV_{\phi_{a_i} \circ T[S^1 \times D^2]} \; = \;
\left \{ (x_i,y_i) \in \frac{\mathbb{T}_{\C} \times \mathbb{T}_{\C}}{W} \; \Big| \; x_i^{a_i} y_i = 1 \right \}
\label{TSLZsolidtorus}
\ee
See {\it e.g.} \cite{Apol} for more details on Dehn surgery in the context of complex Chern-Simons theory.

The space of vacua \eqref{TSLZsolidtorus} essentially corresponds to $\CN=2$ Chern-Simons theory at level~$a_i$.
Therefore, when performing a surgery on $K_i$, the operation of gluing back $N(K_i) \cong S^1 \times D^2$
with a twist $\phi_{a_i} \in SL(2,\Z)$ means gauging the $i$-th global symmetry of the 3d $\CN=2$ theory $T[S^3 \setminus K]$
and introducing a Chern-Simons term at level $a_i$.
Before this operation, in the theory $T[S^3 \setminus K]$ associated to the link complement,
the twisted masses and Fayet-Illiopoulos parameters $(\log x_i , \log y_i)$
are expectation values of real scalars in background vector multiplets that couple to flavor and topological currents, respectively

For instance, when $G_{\C} = SL(2,\C)$ and $K$ is a knot ({\it i.e.} a link with a single component),
the holonomy eigenvalues $x$ and $y$ are both $\C^*$-valued, and the space of vacua $\CV_{T[S^3 \setminus K]}$
is the algebraic curve $A_K (x,y)=0$, the zero locus of the $A$-polynomial.
Therefore, modulo certain technical details, the vacua of the combined theory \eqref{generalTM3}
in this case can be identified with the intersection points of the two algebraic curves, {\it cf.} \eqref{vacspaceintersect}:
\be
\CV_{T[M_3]} \; = \; \{ A_K (x,y) =0 \} \cap \{ x^{a} y = 1 \}
\label{AsurgxyTM3}
\ee
modulo $\Z_2$ action of the $SL(2,\C)$ Weyl group $(x,y) \mapsto (x^{-1},y^{-1})$.
Note, both the $A$-polynomial $A_K (x,y)$ of any knot and the equation $x^a y = 1$ are invariant under this symmetry.
In particular, if $K$ is the unknot we have $A (\text{unknot}) = y-1$ and these two conditions give an $SL(2,\C)$ analogue of \eqref{xxvacQ}.

As a simple illustration one can consider, say, a negative definite 4-manifold whose Kirby diagram consists of
the left-handed trefoil knot $K={\bf 3_1}$ with the framing coefficient $a=-1$:
\be
{-1 \atop {\,\raisebox{-.08cm}{\includegraphics[width=0.8cm]{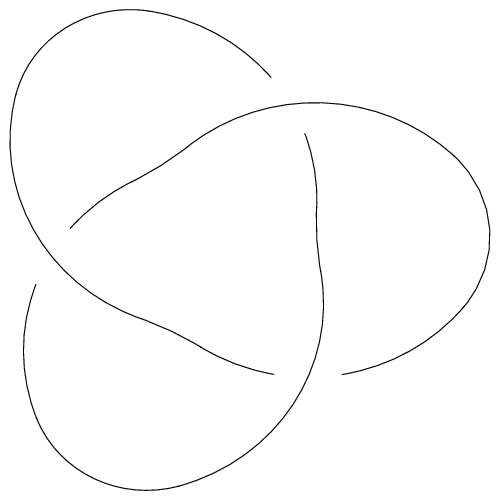}}\,}}
\ee
Using standard tools in Kirby calculus (that we review shortly), it is easy to verify that
the boundary of this 4-manifold is the Poincar\'e homology sphere $\Sigma (2,3,5)$, {\it cf.} \eqref{Brieskorn},
realized here as a $-1$ surgery on the trefoil knot in $S^3$.
Therefore, the corresponding theory $T[\Sigma (2,3,5)]$ can be constructed as in \eqref{generalTM3}.
The knot complement theory that accounts for all flat connections is well known in this case \cite{FGSS};
in fact, \cite{FGSS} gives two dual descriptions of $T[S^3 \setminus {\bf 3_1}]$.
In this theory, the twisted mass $\log x$ is the vev of the real scalar
in background vector multiplet $V$ that couples to the $U(1)_x$ flavor symmetry current.
Gauging the flavor symmetry $U(1)_x$ by adding a $\CN=2$ Chern-Simons term for $V$ at level $a = -1$
gives the desired Poincar\'e sphere theory:
\be
\CL_{T[\Sigma (2,3,5)]} \; = \; \CL_{T[S^3 \setminus {\bf 3_1}]} - \frac{1}{4\pi} \int d^4 \theta \; V \Sigma
\ee
Upon compactification on $S^1$, the field $\sigma = \log x$ is complexified and the critical points \eqref{Wcritptseqs}
of the twisted superpotential in the effective 2d $\CN=(2,2)$ theory $T[\Sigma (2,3,5)]$,
\be
\exp \, \frac{\partial}{\partial \log x}
\left( \tilde \CW_{T[S^3 \setminus K]} + \frac{a}{2} (\log x)^2 \right) \; = \; 1 \,,
\ee
automatically reproduce the equations \eqref{AsurgxyTM3} for flat $SL(2,\C)$ connections.

\subsection{Gluing along a common boundary}
\label{sec2:gluing}

Given two manifolds $M_4^+$ and $M_4^-$ which have the same boundary (component) $M_3$,
there is a natural way to build a new 4-manifold labeled by a map $\varphi : M_3 \to M_3$
that provides an identification of the two boundaries:
\be
M_4 \; = \; M_4^- \cup_{\varphi} M_4^+
\label{MMMgluing}
\ee

\begin{figure}[ht]
\centering
\includegraphics[width=5.0in]{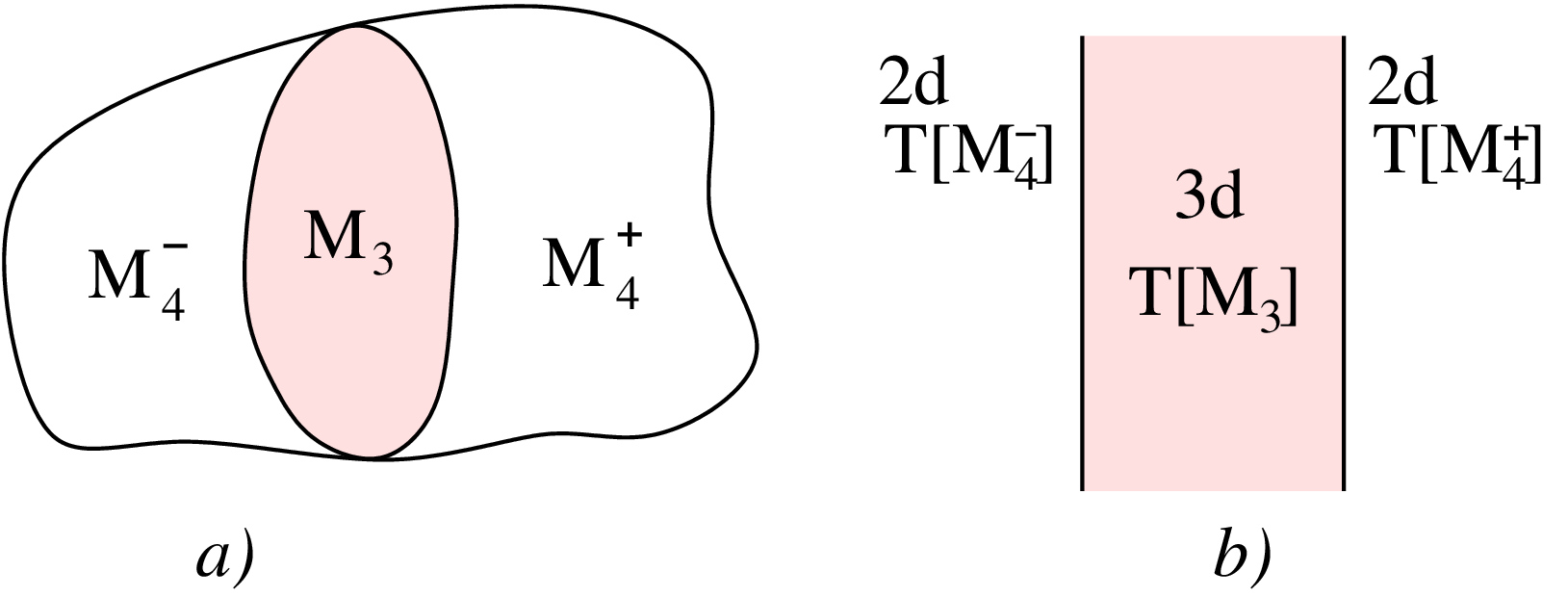}
\caption{$(a)$ Two 4-manifolds glued along a common boundary $M_3 = \pm \partial M_4^{\pm}$ correspond to $(b)$
three-dimensional $\CN=2$ theory $T[M_3]$ on the interval
coupled to two-dimensional $\CN=(0,2)$ theories $T[M_4^-]$ and $T[M_4^+]$ at the boundaries of the interval.}
\label{fig:M4d2d}
\end{figure}

For example, let $M_4^-$ be the negative $E_8$ plumbing, and let $\bar M_4^+$ be the handlebody on the left-handed trefoil
knot with the framing coefficient $a=-1$.
As we already mentioned earlier, both of these 4-manifolds are bounded by
the Poincar\'e homology sphere $\Sigma (2,3,5)$, {\it i.e.}
\be
E_8 \qquad \overset{\partial}{\approx} \qquad
{-1 \atop {\,\raisebox{-.08cm}{\includegraphics[width=0.8cm]{left31}}\,}}
\ee
Therefore, in order to glue these 4-manifolds ``back-to-back'' as illustrated in Figure \ref{fig:M4d2d},
we need to reverse the orientation of one of them, which in the language of Kirby diagrams
amounts to replacing all knots with mirror images and flipping the sign of all framing numbers:
\be
M_4 (K_1^{a_1}, \ldots, K_n^{a_n})
\quad \xrightarrow[~ \text{reversal} ~]{~ \text{orientation} ~} \quad
M_4 ( \bar K_1^{-a_1}, \ldots, \bar K_n^{- a_n})
\label{revorientation}
\ee
Thus, in our example we need to change the left-handed trefoil knot $K={\bf 3_1}$ with framing $a=-1$
to the right-handed trefoil knot $\bar K$ with framing coefficient $+1$.
The resulting 4-manifold $M_4^+$ with a single 2-handle that corresponds to this Kirby diagram
has boundary $M_3 = \partial M_4^+ = - \partial M_4^-$, so that now it can be glued to $M_4^- = E_8$ plumbing.

\bigskip

Gluing 4-manifolds along a common boundary, as in \eqref{MMMgluing}, has a nice physical interpretation.
Namely, it corresponds to the following operation on the 2d $\CN=(0,2)$ theories $T[M_4^{\pm}]$
that produces a new theory $T[M_4]$ associated to the resulting 4-manifold $M_4 = M_4^- \cup_{\varphi} M_4^+$.
As we already explained in section \ref{sec:bdrycond}, partial topological reduction of the 6d fivebrane theory
on a 4-manifold with a boundary $M_3$ leads to a coupled 2d-3d system
of 3d $\CN=2$ theory $T[M_3]$ with a B-type boundary condition determined by the 4-manifold.
(If the 4-manifold in question has other boundary components, besides $M_3$, then the reduction of the 6d fivebrane theory
leads to a wall / interface between $T[M_3]$ and other 3d $\CN=2$ theories; this more general possibility will be discussed in the next section.)

In the case at hand, we have two such 4-manifolds, $M_4^-$ and $M_4^+$,
with oppositely oriented boundaries $\partial M_4^{\pm} = \pm M_3$.
What this means is that $T[M_4^+]$ defines a B-type boundary condition --- with 2d $\CN=(0,2)$ supersymmetry on the boundary ---
in 3d $\CN=2$ theory $T[M_3]$, while $T[M_4^-]$ likewise defines a B-type boundary condition in the theory $T[-M_3]$.
Equivalently, $T[-M_3]$ can be viewed as a theory $T[M_3]$ with the reversed parity:
\be
T[-M_3] \; = \; P \circ T[M_3]
\label{TM3parity}
\ee
where $P: (x^0,x^1,x^2) \to (x^0, x^1, -x^2)$.
This operation, in particular, changes the signs of all Chern-Simons couplings in $T[M_3]$.

Therefore, thanks to \eqref{TM3parity}, we can couple $T[M_4^-]$ and $T[M_4^+]$
to the {\it same} 3d $\CN=2$ theory $T[M_3]$ considered in space-time $\R^2 \times I$, where $I$ is the interval.
In this setup, illustrated in Figure \ref{fig:M4d2d}, theories $T[M_4^{\pm}]$ define boundary
conditions at the two ends of the interval $I$.
As a result, we get a layer of 3d $\CN=2$ theory $T[M_3]$ on $\R^2 \times I$
sandwiched between $T[M_4^-]$ and $T[M_4^+]$.
Since the 3d space-time has only two non-compact directions of $\R^2$, in the infra-red
this system flows to a 2d $\CN=(0,2)$ theory, which we claim to be $T[M_4]$.

The only element that we need to explain is the map $\varphi : M_3 \to M_3$ that enters
the construction \eqref{MMMgluing} of the 4-manifold $M_4$.
If exist, non-trivial self-diffeomorphisms of $M_3$ correspond to self-equivalences (a.k.a. dualities) of the theory $T[M_3]$.
Therefore, a choice of the map $\varphi : M_3 \to M_3$ in \eqref{MMMgluing} means coupling theories $T[M_4^{\pm}]$
to different descriptions/duality frames of the 3d $\CN=2$ theory $T[M_3]$ or, equivalently,
inserting a duality wall (determined by $\varphi$) into the sandwich of $T[M_4^-]$, $T[M_3]$, and $T[M_4^+]$.
Of course, one choice of $\varphi : M_3 \to M_3$ that always exists is the identity map;
it corresponds to the most natural coupling of theories $T[M_4^{\pm}]$ to the same description of $T[M_3]$.
Since $\varphi : M_3 \to M_3$ can be viewed as a special case of a more general cobordism between two
different 3-manifolds that will be discussed in section \ref{sec:cobwalls}, when talking about
gluing 4-manifolds we assume that $\varphi = \text{id}$ unless noted otherwise.
Then, we only need to know which 4-manifolds have the same boundary.

\subsection*{3d Kirby moves}

\bigskip
\begin{figure}[ht]
\centering
\includegraphics[width=5.5in]{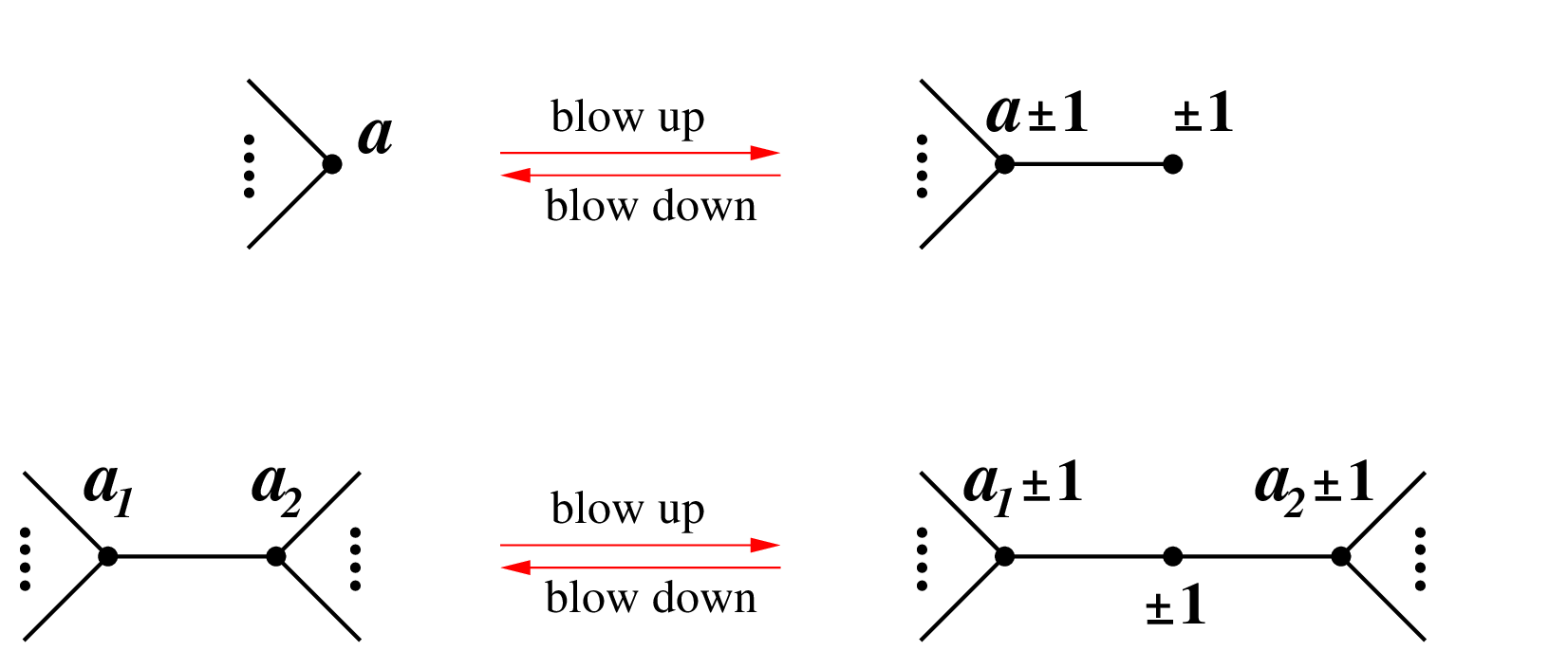}
\caption{Blowing up and blowing down does not change the boundary $M_3 = \partial M_4$.}
\label{fig:blowups}
\end{figure}

Since our list of operations includes gluing 4-manifolds along their common boundary components,
it is important to understand how $M_3 (\Upsilon)$ depends on the plumbing graph $\Upsilon$
and which 4-manifolds $M_4 (\Upsilon)$ have the same boundary (so that they can be glued together).
Not surprisingly, the set of moves that preserve the boundary
$M_3 (\Upsilon) = \partial M_4 (\Upsilon)$
is larger than the set of moves that preserve the 4-manifold $M_4 (\Upsilon)$.

\begin{figure}[ht]
\centering
\includegraphics[width=5.5in]{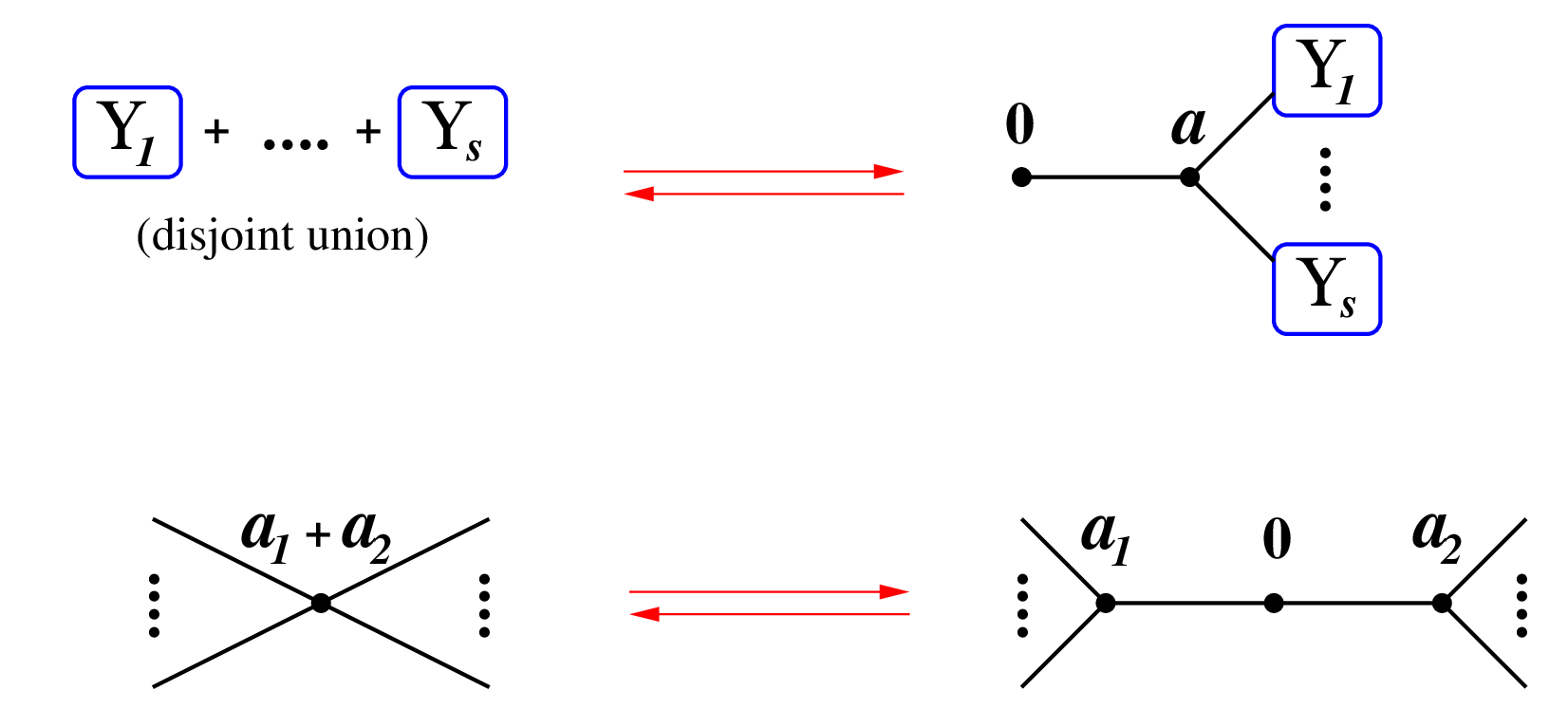}
\caption{``3d Kirby moves'' that do not change $M_3 = \partial M_4$.}
\label{fig:M3moves}
\end{figure}

Specifically, plumbing graphs $\Upsilon_1$ and $\Upsilon_2$ describe the same
3-manifold $M_3 (\Upsilon_1) \cong M_3 (\Upsilon_2)$ if and only if they can be
related by a sequence of ``blowing up'' or ``blowing down''
operations shown in Figure \ref{fig:blowups},
as well as the moves in Figure \ref{fig:M3moves}.
The blowing up (resp. blowing down) operations include adding (resp. deleting)
a component of $\Upsilon$ that consists of a single vertex with label $\pm 1$.
Such blow ups have a simple geometric interpretation as boundary connected sum
operations with very simple 4-manifolds $\cp^2 \setminus \{ \text{pt} \}$ and $\bar{\cp}^2 \setminus \{ \text{pt} \}$,
both of which have $S^3$ as a boundary and, therefore, only change $M_4$ but not $M_3 = \partial M_4$.
As will be discussed shortly, this also has a simple physical counterpart in physics of 3d $\CN=2$ theory $T[M_3]$,
where the blowup operation adds a decoupled ``trivial'' $\CN=2$ Chern-Simons term \eqref{TLensUN} at level $\pm 1$,
which carries only boundary degrees of freedom and has a single vacuum, {\it cf.} \eqref{Lpqvac}.
For this reason, blowing up and blowing down does not change $T[M_3;G]$ and only changes $T[M_4;G]$ by free Fermi multiplets,
for abelian as well as non-abelian $G$.

\bigskip

\begin{figure}[ht]
\centering
\includegraphics[width=5.5in]{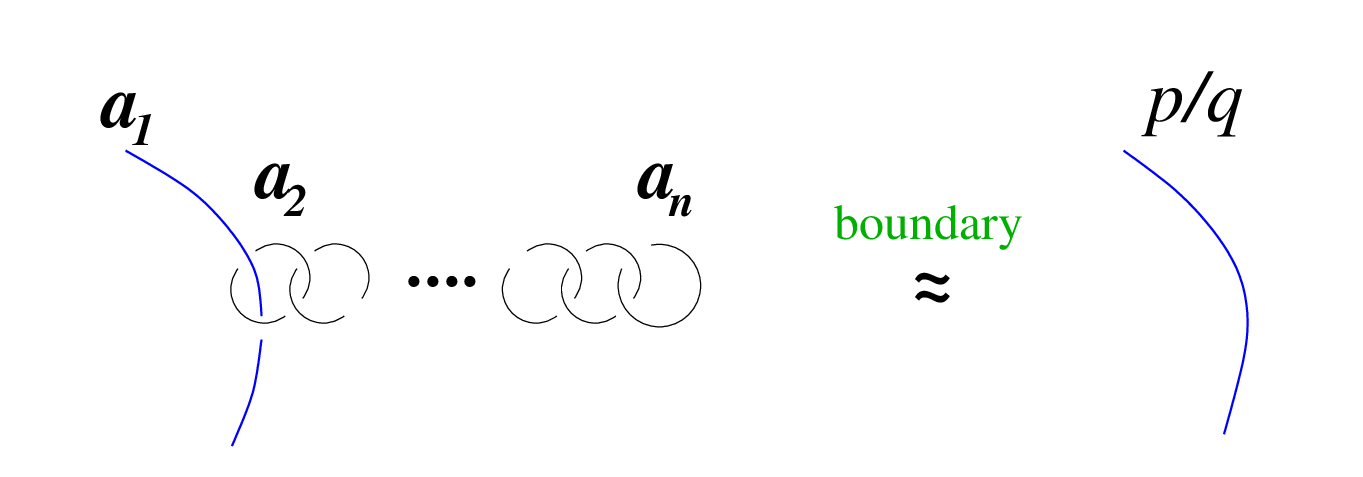}
\caption{Boundary diffeomorphisms relating integral surgery and Dehn surgery.}
\label{fig:M3diffoe}
\end{figure}

Applying these moves inductively, it is easy to derive a useful set of rules illustrated in Figure~\ref{fig:M3diffoe}
that, for purposes of describing the boundary of $M_4$, allow to collapse linear chains of sphere plumbings
with arbitrary framing coefficients $a_i$ via continued fractions
\be
\frac{p}{q} \; = \; a_1 - \cfrac{1}{a_2 - \cfrac{1}{\ddots - \cfrac{1}{a_n}}}
\label{contfractplus}
\ee
To illustrate how this works, let us demonstrate that the $A_{n-1}$ plumbing,
as in Figure \ref{fig:Anplumbing}, with $a_i = -2$
can be glued to a disc bundle with Euler number $-n$ over $S^2$ to produce
a smooth 4-manifold $(\bar{\cp}^2)^{\# n}$.
In particular, we need to show that these two 4-manifolds we are gluing naturally
have the same boundary with opposite orientation.
This is a simple exercise in Kirby calculus.

Starting with the $A_{n-1}$ linear plumbing, we can take advantage of the fact that $\pm 1$ vertices
can be added for free and consider instead
\be
\overset{\displaystyle{+1}}{\bullet}
\qquad
\overset{\displaystyle{-2}}{\bullet} \frac{\phantom{xxx}}{\phantom{xxx}}
\overset{\displaystyle{-2}}{\bullet} \frac{\phantom{xxx}}{\phantom{xxx}}
\overset{\displaystyle{-2}}{\bullet} \frac{\phantom{xxx}}{\phantom{xxx}}
\quad \cdots \quad
\frac{\phantom{xxx}}{\phantom{xxx}}
\overset{\displaystyle{-2}}{\bullet} 
\ee
Clearly, this operation (of blowing up) changes the 4-manifold, but not the boundary $M_3$.
Now, we slide the new $+1$ handle over the $-2$ handle.
According to \eqref{aaslide}, this preserves the framing $+1$ of the new handle
and changes the framing of the $-2$ handle to $-2+1 = -1$ (since they were originally unlinked), resulting in
\be
\overset{\displaystyle{+1}}{\bullet}
\frac{\phantom{xxx}}{\phantom{xxx}}
\overset{\displaystyle{-1}}{\bullet} \frac{\phantom{xxx}}{\phantom{xxx}}
\overset{\displaystyle{-2}}{\bullet} \frac{\phantom{xxx}}{\phantom{xxx}}
\overset{\displaystyle{-2}}{\bullet} \frac{\phantom{xxx}}{\phantom{xxx}}
\quad \cdots \quad
\frac{\phantom{xxx}}{\phantom{xxx}}
\overset{\displaystyle{-2}}{\bullet}
\ee
Note, this plumbing graph with $n$ vertices is a result of applying the first move in Figure \ref{fig:blowups}
to the $A_{n-1}$ linear plumbing, which we have explained ``in slow motion.''
Since we now have a vertex with weight $-1$, we can apply the second move in Figure \ref{fig:blowups}
to remove this vertex at the cost of increasing the weights of the two adjacent vertices by $+1$,
which gives
\be
\overset{\displaystyle{+2}}{\bullet}
\frac{\phantom{xxx}}{\phantom{xxx}}
\overset{\displaystyle{-1}}{\bullet} \frac{\phantom{xxx}}{\phantom{xxx}}
\overset{\displaystyle{-2}}{\bullet} \frac{\phantom{xxx}}{\phantom{xxx}}
\quad \cdots \quad
\frac{\phantom{xxx}}{\phantom{xxx}}
\overset{\displaystyle{-2}}{\bullet}
\ee
This last step made the plumbing graph shorter, of length $n-1$,
and there is a new vertex with weight $-2 + 1 = -1$
on which we can apply the blow down again.
Doing so will change the weight of the leftmost vertex from $+2$ to $+3$
and after $n-3$ more steps we end up with a plumbing graph
\be
\overset{\displaystyle{n-1}}{\bullet}
\frac{\phantom{xxx}}{\phantom{xxx}}
\overset{\displaystyle{-1}}{\bullet}
\ee
Applying the first move in Figure \ref{fig:blowups} we finally get the desired relation
\be
A_{n-1} \quad \overset{\partial}{\approx} \quad \overset{\displaystyle{+n}}{\bullet}
\label{AObdry}
\ee
Since reversing orientation on the 4-manifold is equivalent \eqref{revorientation} to replacing all knots with mirror images
and flipping the sign of all framing numbers, this shows that $A_{n-1}$ linear plumbing has the same Lens
space boundary as the disc bundle with Euler number $-n$ over $S^2$, but with opposite orientation.
In particular, it follows that these 4-manifolds with boundary can be glued along their common boundary
in a natural way. (No additional orientation reversal or other operation is needed.)

\bigskip

Following these arguments, it is easy to show a more general version of the first move in Figure \ref{fig:blowups}
called {\it slam-dunk}:
\be
\overset{\displaystyle{p/q}}{\bullet}
\frac{\phantom{xxx}}{\phantom{xxx}}
\overset{\displaystyle{a}}{\bullet}
\frac{\phantom{xxx}}{\phantom{xxx}}
\quad \cdots \quad
\qquad \overset{\partial}{\approx} \qquad
\overset{\displaystyle{a-\tfrac{q}{p}}}{\bullet}
\frac{\phantom{xxx}}{\phantom{xxx}}
\quad \cdots
\label{slamdunk}
\ee
which, of course, is just a special case of the boundary diffeomorphism in Figure~\ref{fig:M3diffoe}.
Another useful rule in 3d Kirby calculus that can be deduced by the same argument
allows to collapse a (sub)chain of $(-2)$'s:
$$
\frac{\phantom{xxx}}{\phantom{xxx}}
\overset{\displaystyle{a}}{\bullet}
\frac{\phantom{xxx}}{\phantom{xxx}}
\underbrace{
\overset{\displaystyle{-2}}{\bullet}
\frac{\phantom{xxx}}{\phantom{xxx}}
\quad \cdots \quad
\frac{\phantom{xxx}}{\phantom{xxx}}
\overset{\displaystyle{-2}}{\bullet} }_{\displaystyle{n \text{ times}}}
\frac{\phantom{xxx}}{\phantom{xxx}}
\overset{\displaystyle{b}}{\bullet}
\frac{\phantom{xxx}}{\phantom{xxx}}
\qquad \overset{\partial}{\approx} \qquad
\frac{\phantom{xxx}}{\phantom{xxx}}
\overset{\displaystyle{a+1}}{\bullet}
\frac{\phantom{xxx}}{\phantom{xxx}}
\overset{\displaystyle{n+1}}{\bullet}
\frac{\phantom{xxx}}{\phantom{xxx}}
\overset{\displaystyle{b+1}}{\bullet}
\frac{\phantom{xxx}}{\phantom{xxx}}
$$
which is a generalization of \eqref{AObdry}.

\subsection*{Physical interpretation of 3d Kirby moves}

All these moves that preserve the boundary 3-manifold $M_3 (\Upsilon) = \partial M_4 (\Upsilon)$
have an elegant and simple interpretation as equivalences (dualities) of the corresponding 3d $\CN=2$ theory $T[M_3 (\Upsilon);U(N)]$.
Let us illustrate this in the basic case of $N=1$, {\it i.e.} a single fivebrane.
Then, as we explained in section \ref{sec:bdrycond}, all theories $T[M_3 (\Upsilon);U(1)]$ admit a description
as supersymmetric Chern-Simons theories, and 3d Kirby moves are precisely the equivalence relations on the matrix
of Chern-Simons coefficients in the quantum theory.

Indeed, the simplest version of blowing up (resp. blowing down) operation that adds (resp. removes) an isolated vertex with label $\pm 1$
in the theory $T[M_3 (\Upsilon);U(1)]$ correspond to changing the matrix of Chern-Simons coefficients
\be
Q \; \to \; Q \oplus \langle \pm 1 \rangle
\ee
that is, adds (resp. removes) a $U(1)$ vector multiplet $V$ with the Lagrangian
\be
\Delta \CL \; = \; \pm \frac{1}{4\pi} \int d^4 \theta \; V \Sigma \; = \; \pm \frac{1}{4\pi} A\wedge dA + \ldots
\ee
A theory defined by this Lagrangian is trivial. In particular, it has one-dimensional Hilbert space.
Therefore, tensor products with copies of this trivial theory are indeed equivalences of $T[M_3 (\Upsilon);U(1)]$.
The same is true in the non-abelian case as well, where blowups change $T[M_3;G]$ by ``trivial'' Chern-Simons
terms at level $\pm 1$ that carry only boundary degrees of freedom
(and, therefore, only affect the physics of the 2d boundary theory $T[M_4;G]$, but not the 3d bulk theory $T[M_3;G]$).

Similarly, we can consider blowing up and blowing down operations shown in Figure \ref{fig:blowups}.
If in the plumbing graph $\Upsilon$
a vertex with label $\pm 1$ is only linked by one edge to another vertex with label $a \pm 1$,
it means that the Lagrangian of the 3d $\CN=2$ theory $T[M_3 (\Upsilon);U(1)]$ has the following terms
\be
\CL \; = \; \frac{1}{4\pi} \int d^4 \theta \left( \pm V \Sigma + 2 \tilde V \Sigma + (a \pm 1) \tilde V \tilde \Sigma + \ldots \right)
\ee
where ellipses stand for terms that do not involve the vector superfield $V$ or its field strength $\Sigma$.
Since the action is Gaussian in $V$, we can integrate it out by solving the equations of motion $\pm V + \tilde V = 0$.
The resulting Lagrangian is
\be
\CL' \; = \;
\frac{1}{4\pi} \int d^4 \theta \left( \pm \tilde V \tilde \Sigma \mp 2 \tilde V \tilde \Sigma + (a \pm 1) \tilde V \tilde \Sigma + \ldots \right) =
\frac{1}{4\pi} \int d^4 \theta \left( a \tilde V \tilde \Sigma + \ldots \right)
\ee
This gives a physics realization of the blowing up and blowing down operations in the top part of Figure~\ref{fig:blowups}.
We can easily generalize it to that in the lower part of Figure~\ref{fig:blowups}.
Starting with the right side of the relation, the terms in the Lagrangian which involve
the superfield $V$ at Chern-Simons level $\pm 1$ look like
\be
\CL \; = \; \frac{1}{4\pi} \int d^4 \theta \left(
\pm V \Sigma + 2 V_1 \Sigma + (a_1 \pm 1) V_1 \Sigma_1 + 2 V_2 \Sigma + (a_2 \pm 1) V_2 \Sigma_2 + \ldots \right)
\ee
Integrating out $V$ yields $\pm V + V_1 + V_2 = 0$ and the effective Lagrangian
\be
\CL' \; = \; \frac{1}{4\pi} \int d^4 \theta \left( a_1 V_1 \Sigma_1 \mp 2 V_1 \Sigma_2 + a_2 V_2 \Sigma_2 + \ldots \right)
\ee
which, as expected, describes the left side of the relation in the lower part of Figure~\ref{fig:blowups}.
{}From this physical interpretation of the blowing up and blowing down operations in the $N=1$ case
one can draw a more general lesson:
the reason that 2-handles with framing coefficients $a = \pm 1$ are ``nice'' corresponds to the fact that
3d $\CN=2$ theory $T \big[ M_3 \big( {\pm 1 \atop \bullet} \big) \big]$ is trivial.

The physical interpretation of 3d Kirby moves in Figure~\ref{fig:M3moves} is even simpler:
2-handles with framing coefficients $a_i = 0$ correspond to superfields in 3d theory $T[M_3 (\Upsilon)]$
that serve as Lagrange multipliers.
Again, let us explain this in the basic case of a single fivebrane ($N=1$).
Let us consider the first move in Figure~\ref{fig:M3moves} and, as in the previous discussion,
denote by $V$ the $U(1)$ vector superfield associated with a 2-handle (vertex) with framing label 0.
Then, the relevnt terms in the Lagrangian of the theory $T[M_3 (\Upsilon); U(1)]$
associated to the right part of the diagram are
\be
\CL \; = \; \frac{1}{4\pi} \int d^4 \theta \left(2 V \tilde \Sigma + a \tilde V \tilde \Sigma + \ldots \right)
\ee
Note, there is no Chern-Simons term for $V$ itself, and it indeed plays the role of the Lagrange multiplier
for the condition $\tilde \Sigma = 0$. Therefore, integrating out $V$ makes $\tilde V$ pure gauge and
removes all Chern-Simons couplings involving $\tilde V$.
The resulting quiver Chern-Simons theory is precisely the one associated with the left diagram
in the upper part of Figure~\ref{fig:M3moves}.

Now, let us consider the second move in Figure~\ref{fig:M3moves}, again starting from the right-hand side.
The relevant part of the Lagrangian for $T[M_3 (\Upsilon); U(1)]$ looks like
\be
\CL \; = \; \frac{1}{4\pi} \int d^4 \theta \left(
2 V \Sigma_1 + a_1 V_1 \Sigma_1 + 2 V \Sigma_2 + a_2 V_2 \Sigma_2 + \ldots \right)
\ee
where the dependence on $V$ is again only linear. Hence, integrating it out makes
the ``diagonal'' combination $V_1 + V_2$ pure gauge, and for $V' = V_1 = - V_2$ we get
\be
\CL' \; = \; \frac{1}{4\pi} \int d^4 \theta \left( (a_1 + a_2) V' \Sigma' + \ldots \right)
\ee
which is precisely the Lagrangian of the quiver Chern-Simons theory associated to
the plumbing graph in the lower left corner of Figure~\ref{fig:M3moves}.

Finally, since all other boundary diffeomorphisms in 3d Kirby calculus follow from these basic moves,
it should not be surprising that the manipulation in Figure~\ref{fig:M3diffoe} as well as the slam-dunk move \eqref{slamdunk}
also admit an elegant physical interpretation. However, for completeness, and to practice a little more
with the dictionary between 3d Kirby calculus and equivalences of 3d $\CN=2$ theories, we present the details here.
Based on the experience with the basic moves, the reader might have (correctly) guessed that
both the boundary diffeomorphism in Figure \ref{fig:M3diffoe} and the slam-dunk move \eqref{slamdunk}
correspond to integrating out vector multiplets.

Specifically, for the plumbing graph on the left side of \eqref{slamdunk} the relevant terms in the Lagrangian
of the theory $T[M_3 (\Upsilon); U(1)]$ look like
\be
\CL \; = \; \frac{1}{4\pi} \int d^4 \theta \left(
\frac{p}{q} V \Sigma + 2 \tilde V \Sigma + a \tilde V \tilde \Sigma+ \ldots \right)
\ee
Since there are no other terms in the Lagrangian of $T[M_3 (\Upsilon); U(1)]$ that contain
the superfield $V$ or its (super)field strength $\Sigma$, we can integrate it out.
Replacing $V$ by the solution to the equation $\frac{p}{q} V + \tilde V=0$ gives
the Lagrangian for the remaining fields
\be
\CL \; = \; \frac{1}{4\pi} \int d^4 \theta \left(
\big( a - \frac{q}{p} \big) \tilde V \tilde \Sigma+ \ldots \right)
\ee
which is an equivalent description of the theory $T[M_3 (\Upsilon); U(1)]$,
in fact, the one associated with the right-hand side of the slam-dunk move \eqref{slamdunk}.
By now it should be clear what is going on.
In particular, by iterating this process and integrating in or integrating out $U(1)$ vector superfields,
it is easy to show that quiver Chern-Simons theories associated to Kirby diagrams in Figure \ref{fig:M3diffoe}
are indeed equivalent.

\subsection{Cobordisms and domain walls}
\label{sec:cobwalls}

Now, it is straightforward to generalize the discussion in previous sections to 4-manifolds with two (or more) boundary components.
The lesson we learned is that each boundary component of $M_4$ corresponds to a coupling with 3d $\CN=2$ theory labeled by that component.

\begin{figure}[ht]
\centering
\includegraphics[width=5.0in]{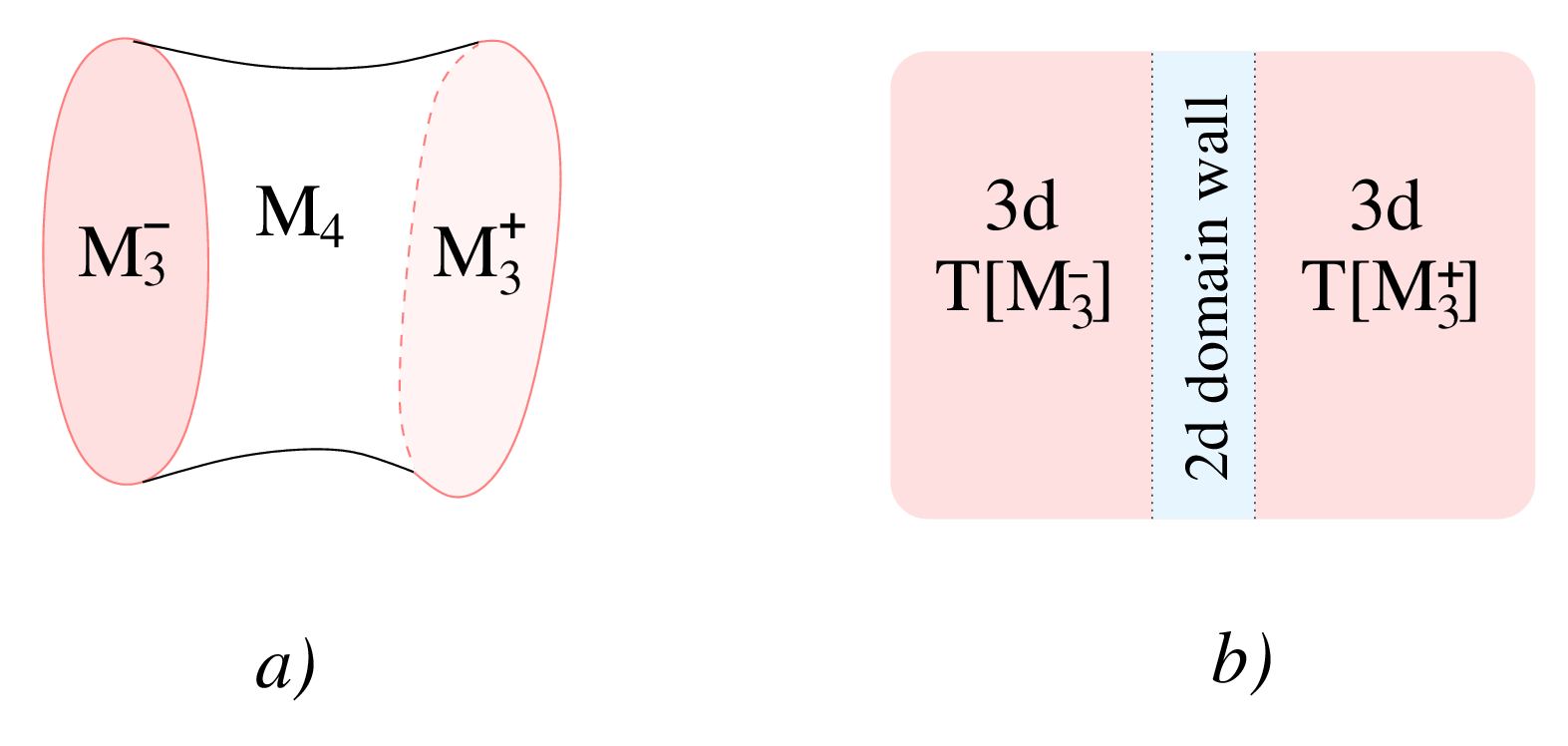}
\caption{$(a)$ A cobordism between 3-manifolds $M_3^-$ and $M_3^+$ corresponds to $(b)$
a 2d $\CN=(0,2)$ theory $T[M_4]$ on the domain wall (interface) coupled
to 3d $\CN=2$ theories $T[M_3^-]$ and $T[M_3^+]$ on both sides.}
\label{fig:M4d2dwall}
\end{figure}

In general, when a 4-manifold $M_4$ has one or more boundary components, it is convenient to view it as a (co)bordism from $M_3^-$ to $M_3^+$,
where $M_3^{\pm}$ is allowed to be empty or contain several connected components, see Figure~\ref{fig:M4d2dwall}$a$.
If $M_3^- = \emptyset$ (or $M_3^+ = \emptyset$), then the corresponding 3d $\CN=2$ theory $T[M_3^-]$ (resp. $T[M_3^+]$) is trivial.
And, when $M_3^{\pm}$ has more than one connected component, the corresponding theory $T[M_3^{\pm}]$
is simply a tensor product of 3d $\CN=2$ theories associated with those components.
(In fact, we already encountered similar situations, {\it e.g.} in \eqref{generalTM3},
when we discussed {\it 3-manifolds} with several boundary components.)

\bigskip

What kind of 2d theory $T[M_4]$ corresponds to a cobordism from $M_3^-$ to $M_3^+$?
There are several ways to look at it.
First, trying to erase any distinction between $M_3^+$ and $M_3^-$, we can view any such 4-manifold
as a cobordism from $\emptyset$ to $M_3^+ \sqcup - M_3^-$, {\it i.e.} as a 4-manifold with boundary $M_3 = M_3^+ \sqcup - M_3^-$,
thus reducing the problem to the one already considered.
Indeed, using \eqref{TM3parity}, to a 4-manifold $M_4$ with boundary $M_3^+ \sqcup - M_3^-$ we associate
a 3d $\CN=2$ theory $T[M_3^+] \otimes \big( P \circ T[M_3^-] \big)$ on a half-space $\R_+ \times \R^2$ coupled to a boundary theory $T[M_4]$.
In turn, this product 3d theory on a half-space is
equivalent --- via the so-called ``folding'' trick \cite{Wong:1994np,Oshikawa:1996dj,Bachas:2001vj} ---
to a 3d theory $T[M_3^+]$ or $T[M_3^-]$ in two regions of the full three-dimensional space $\R^3$,
separated by a 2d interface (that in 3d context might be naturally called a ``defect wall'').
This gives another, perhaps more natural way to think of 2d $\CN=(0,2)$ theory $T[M_4]$
associated to a cobordism from $M_3^-$ to $M_3^+$, as a theory trapped on the interface separating
two 3d $\CN=2$ theories $T[M_3^-]$ or $T[M_3^+]$, as illustated in Figure~\ref{fig:M4d2dwall}.

In order to understand the physics of fivebranes on 4-manifolds, it is often convenient to compactify
one more direction, {\it i.e.} consider the fivebrane world-volume to be $S^1 \times \R \times M_4$.
In the present context, it leads to an effective two-dimensional theory with $\CN=(2,2)$ supersymmetry
and a B-type defect\footnote{The converse is not true since some line defects in 2d come from
line operators in 3d.} labeled by $M_4$.
In fact, we already discussed this reduction on a circle in section \ref{sec:bdrycond},
where it was noted that the effective 2d $\CN=(2,2)$ theory
--- which, with some abuse of notations, we still denote $T[M_3]$ ---
is characterized by the twisted superpotential $\tilde \CW (x_i)$.
Therefore, following the standard description of B-type defects in $\CN=(2,2)$ Landau-Ginzburg
models \cite{Hori:2004zd,Brunner:2007qu,Brunner:2008fa,Carqueville:2010hu},
one might expect that a defect $T[M_4]$ between two theories $T[M_3^-]$ and $T[M_3^+]$
can be described as a matrix (bi-)factorization of the difference of the corresponding superpotentials
\be
\tilde W_{T[M_3^+]} (x_i) - \tilde W_{T[M_3^-]} (y_i)
\label{WWxy}
\ee
While conceptually quite helpful, this approach is less useful for practical description of the defect walls
between $T[M_3^-]$ and $T[M_3^+]$, which we typically achieve by other methods.
The reason, in part, is that superpotentials $\tilde \CW$ are non-polynomial for most theories $T[M_3]$.
We revisit this approach and make additional comments in section \ref{sec:2dtheory}.

Note, if 2d theories in question were $\CN=(2,2)$ sigma-models based on target manifolds
$X_{T[M_3^+]}$ and $X_{T[M_3^-]}$, respectively, then B-type defects between them could be
similarly represented by correspondences, or (complexes of) coherent sheaves,
or sometimes simply by holomorphic submanifolds
\be
\Delta \; \subset \; X_{T[M_3^+]} \times X_{T[M_3^-]}
\label{DXX}
\ee

Much like defect lines in 2d, defect walls in 3d can be classified according to their properties
and the symmetries they preserve: topological, conformal, reflective or transmissive,
parameter walls, (duality) transformation walls, {\it etc.}
Various examples of such walls in 3d $\CN=2$ theories were studied in \cite{Gadde:2013wq}.
For instance, parameter walls are labeled by (homotopy types of) paths on the moduli space $\CV_{T[M_3]}$
and correspond to (autoequivalence) functors acting on the category of B-type boundary conditions.
Transformation walls, on the other hand, in general change 3d $\CN=2$ theory,
{\it e.g.} by implementing the $SL(2,\Z)$ action \cite{Witten:2003ya} described in \eqref{TonT}-\eqref{SonT}.
Topological defects in abelian Chern-Simons theories --- which, according to our proposal \eqref{quiverCS},
are relevant to cobordisms between 3-manifolds --- have been studied {\it e.g.} in \cite{Kapustin:2010hk,Kapustin:2010if,Fuchs:2012dt}.
In supersymmetric theories, topological defects are quite special as they are of A-type and B-type at the same time.

\begin{figure}[ht]
\centering
\includegraphics[width=5.0in]{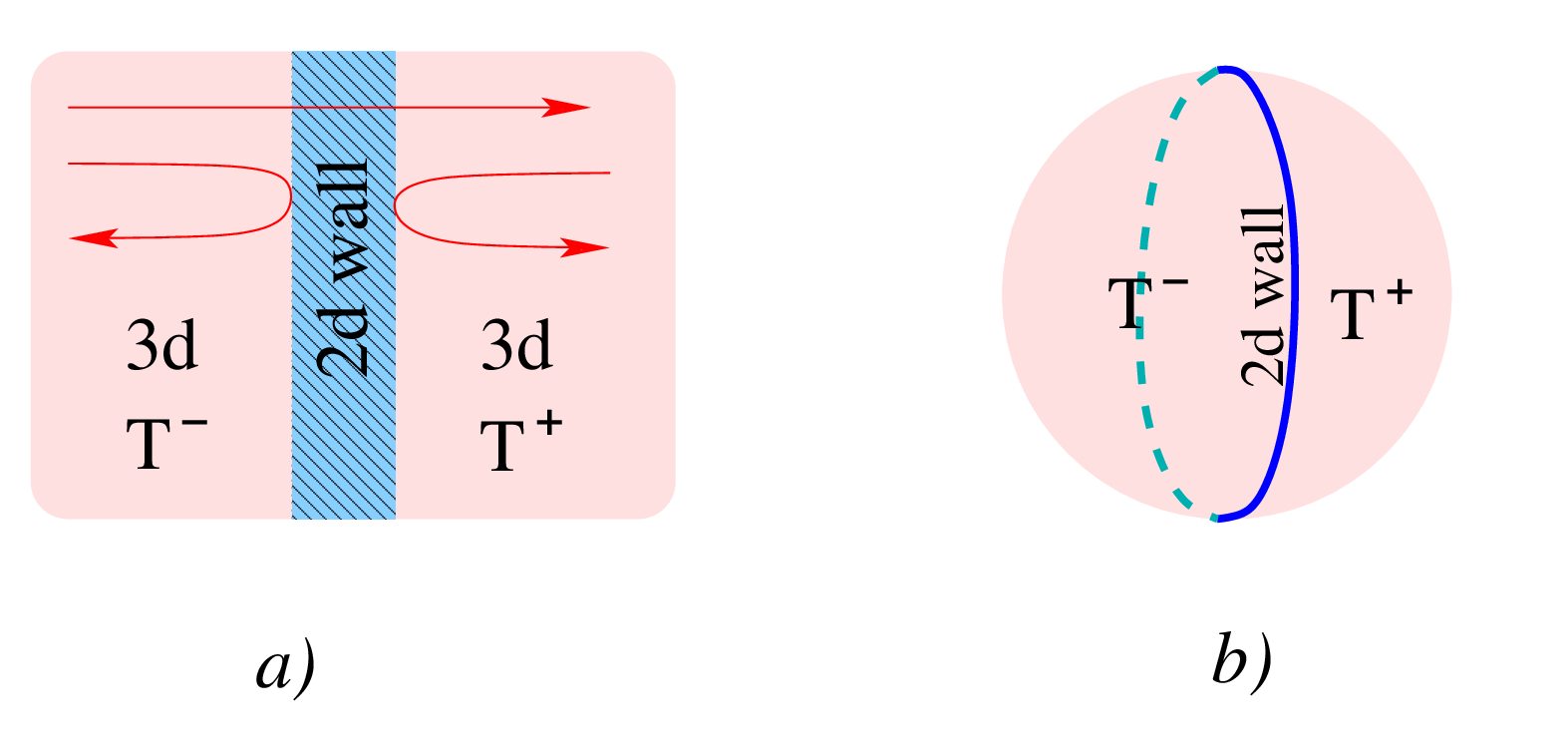}
\caption{A generic defect wall between two 3d $\CN=2$ theories $(a)$ in flat space-time
and $(b)$ the corresponding configuration on $S^1 \times S^2$.
The index of the latter system is obtained from two copies
of the ``half-index'' $\CI_{S^1 \times_q D^{\pm}} (T^{\pm}) \simeq Z_{\text{vortex}} (T^{\pm})$
convoluted via the index (flavored elliptic genus) of the defect wall supported on $S^1 \times S^1_{\text{eq}}$,
where $D^{\pm}$ is the disk covering right (resp. left) hemisphere of the $S^2$
and $S^1_{\text{eq}}:= \partial D^+ = - \partial D^-$ is the equator of the $S^2$.}
\label{fig:wallindex}
\end{figure}

The next best thing to topological defects are conformal ones, which in 2d are usually characterized
by their reflective or transmissive properties. Extending this terminology to walls in 3d, below we
consider two extreme examples, which, much like Neumann and Dirichlet boundary conditions,
provide basic ingredients for building mixed types. See Figure \ref{fig:wallindex}$a$
for an illustration of a generic defect wall (neither totally reflective nor fully transmissive).

\bigskip

\subsection*{Fully transmissive walls}

The simplest example of a totally transmisive wall (which is also conformal)
is a trivial wall between the theory $T[M_3]$ and itself.
It corresponds to the identity cobordism $M_3 \times I$ and in the language of boundary
conditions \eqref{DXX} is represented by the ``diagonal''
\be
\Delta_X \; \subset \; X \times X
\ee
and similarly for the LG models \eqref{WWxy}.

In view of \eqref{quiverCS} and \eqref{TLensUN}, more interesting examples of maximally transmissive defects
are walls between $\CN=2$ Chern-Simons theories with gauge groups $G$ and $H \subset G$ that have $H$-symmetry throughout.
Such defects can be constructed by decomposing the Lie algebra
\be
\frak{g} = (\frak{g} / \frak{h} )^{\perp} \oplus \frak{h}^{\parallel}
\ee
and imposing Dirichlet type boundary conditions on the coset degrees of freedom and
Neumann boundary conditions on degrees of freedom for $H \subset G$.
Equivalently, via the level-rank or, in the supersymmetric context, Giveon-Kutasov duality \cite{Giveon:2008zn}
equally important are level-changing defect walls in $\CN=2$ Chern-Simons theories.
See {\it e.g.} \cite{Fuchs:2012dt} for the study of defect walls with these properties in a purely bosonic theory
and \cite{Quella:2002ct,Bachas:2009mc} for various constructions in closely related WZW models one dimension lower.

\subsection*{Maximally reflective walls}

Maximally reflective domain walls between 3d theories $T[M_3^-]$ or $T[M_3^+]$ do not allow these theories to communicate at all.
Typical examples of such walls are products of boundary conditions, $\CB^-$ and $\CB^+$, for $T[M_3^-]$ and $T[M_3^+]$, respectively:
\be
T[M_4] \; = \; \CB^- \otimes \CB^+
\ee
In the correspondence between 4-manifolds and 2d $\CN=(0,2)$ theories trapped on the walls, they correspond to disjoint unions
$M_4 = M_4^- \sqcup M_4^+$, such that $\partial M_4^{\pm} = M_3^{\pm}$.

\subsection*{Fusion}

Finally, the last general aspect of domain walls labeled by cobordisms that we wish to mention is composition (or, fusion),
illustrated {\it e.g.} in Figure~\ref{fig:sequence}.
As we explain in the next section, the importance of this operation is that any 4-manifold of the form \eqref{M4KKK} and,
therefore, any 2d $\CN=(0,2)$ theory associated to it can built --- in general, in more than one way --- as a sequence of basic fusions.
Notice, while colliding general defect walls can be singular, the fusion of B-type walls on $S^1 \times \R^2$ is smooth
(since they are compatible with the topological twist along $\R^2$).

\subsection{Adding a 2-handle}
\label{sec:addinghandle}

We introduced many essential elements of the dictionary (in Table~\ref{tab:dict})
between 4-manifolds and the corresponding 2d theories $T[M_4]$, and illustrated some of them in simple examples.
Further aspects of this dictionary and more examples will be given in later sections and future work.
One crucial aspect --- which, hopefully, is already becoming clear at this stage --- is that a basic building block is a 2-handle.
Indeed, adding 2-handles one-by-one, we can build {\it any} 4-manifold of the form \eqref{M4KKK}!
And the corresponding 2d theory $T[M_4]$ can be built in exactly the same way,
following a sequence of basic steps, each of which corresponds to adding a new 2-handle.

\begin{figure}[ht]
\centering
\includegraphics[width=5.0in]{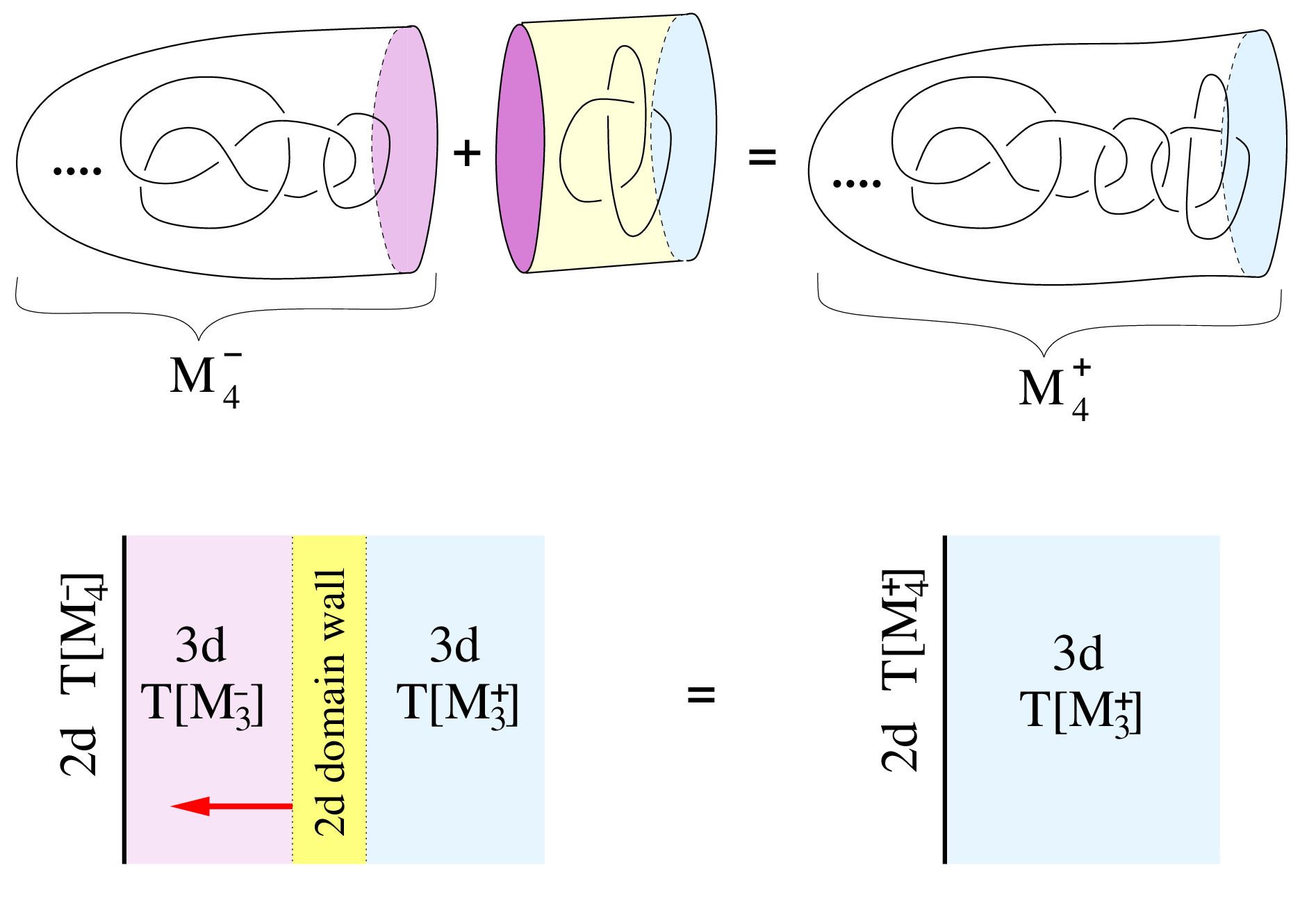}
\caption{The operation of attaching a 2-handle to $M_4^-$ can be represented by a cobordism, namely
the closure of $M_4^+ \setminus M_4^-$. This operation corresponds to fusing a 2d wall (interface)
determined by the cobordism with a boundary theory $T[M_4^-]$ to produce a new boundary theory $T[M_4^+]$.
Equivalently, the system on the left --- with a domain wall sandwiched between
3d $\CN=2$ theories $T[M_3^-]$ and $T[M_3^+]$ --- flows in the infra-red to a new boundary condition determined by $T[M_4^+]$.}
\label{fig:2hndlcobordism}
\end{figure}

\begin{figure}[ht]
\centering
\includegraphics[width=5.0in]{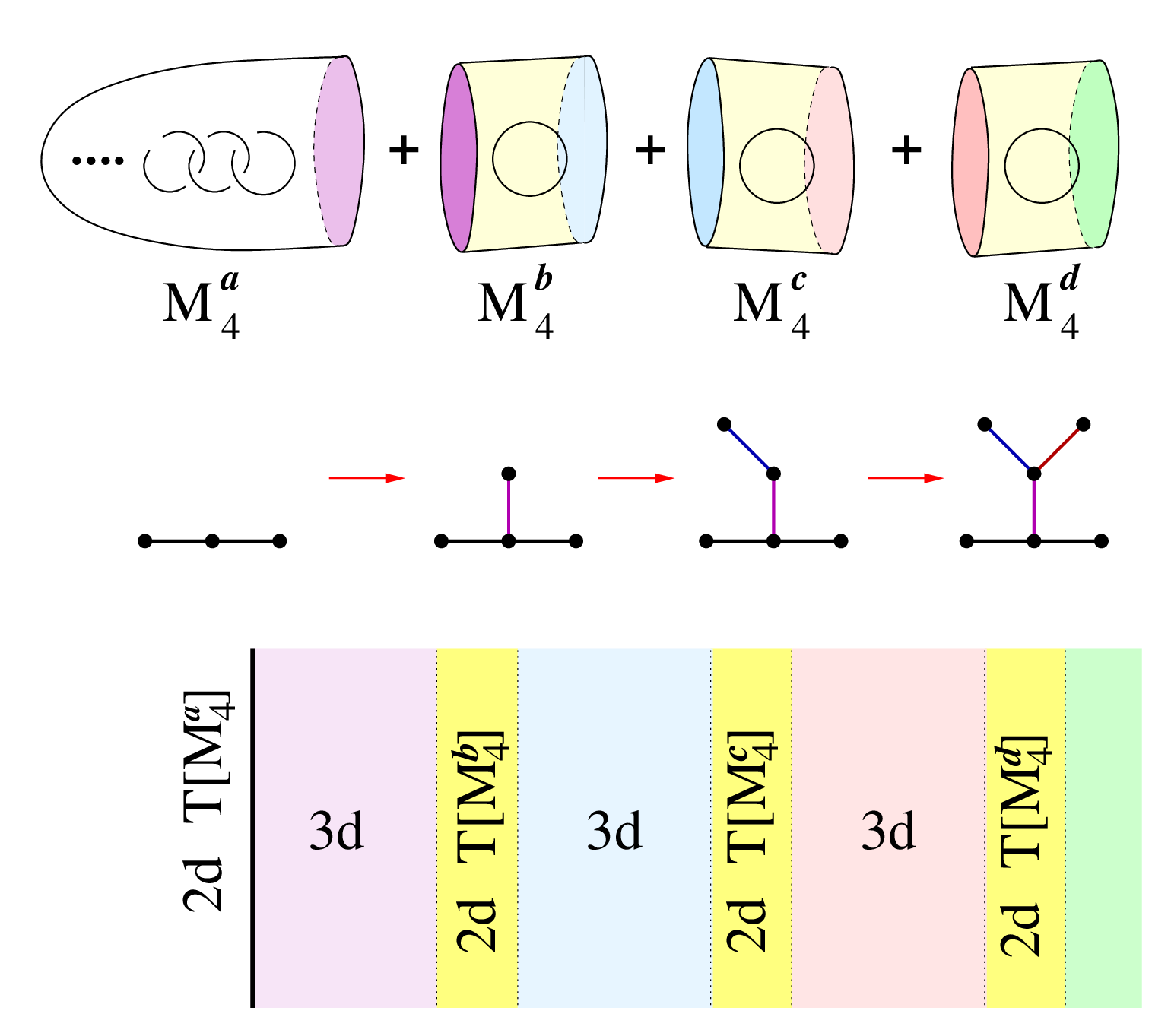}
\caption{The process of building a 4-manifold $M_4$ labeled by a plumbing tree can be represented by
a sequence of basic cobordisms with $b_2=1$, where each step adds a new 2-handle.
Each cobordism corresponds to a 2d wall (interface), and the process of building $M_4$
corresponds to defining $T[M_4]$ as the IR limit of the layered system of
3d theories trapped between walls shown on the lower part of the figure.
Note, in general, there are many equivalent ways of building the same 4-manifold $M_4$
by attaching 2-handles in a different order; they correspond to equivalent descriptions
(dualities) of the same 2d $(0,2)$ theory $T[M_4]$.}
\label{fig:sequence}
\end{figure}

In this section, we shall look into details of this basic operation and, in particular,
explain that adding a new 2-handle at any part of the Kirby diagram can be represented by a cobordism.
Then, using the dictionary between cobordisms and walls (interfaces) in 3d,
that we already explained in section \ref{sec:cobwalls}, we learn that
the operation of adding a 2-handle can be described by a fusion with the corresponding wall,
as illustrated in Figures \ref{fig:2hndlcobordism} and \ref{fig:sequence}.

This interpretation of adding 2-handles is very convenient and very powerful,
especially for practical ways of building theories $T[M_4]$.
For instance, it can be used to turn a small sample of concrete examples into a large factory for producing many new ones.
Indeed, suppose one has a good understanding of a (possibly rather small) family of 4-manifolds
that can be obtained from one another by adding 2-handles.
Then, by extracting\footnote{Explaining how to do this is precisely the goal of the present section.}
the ``difference'' one gets a key to a much larger class of 4-manifolds and the corresponding theories $T[M_4]$
that can be constructed by composing the basic steps (of adding 2-handles) in a variety of new ways,
thus, potentially taking us well outside of the original family.
A good starting point for implementing this algorithm and deducing
the set of basic cobordisms (resp. the 2d $(0,2)$ domain wall theories)
can be a class of ADE sphere plumbings, as in Figures \ref{fig:Anplumbing} and \ref{fig:E8},
for which the Vafa-Witten partition function is known to be
the level $N$ character of the corresponding WZW model \cite{Nakajima,VafaWitten}.
We pursue this approach in section \ref{sec:VW}
and identify the corresponding basic operations of adding 2-handles with certain coset models.

\bigskip

Suppose our starting point is a 4-manifold $M_4^-$ with boundary
\be
\partial M_4^- = M_3^{-}
\ee
Attaching to it an extra 2-handle we obtain a new 4-manifold $M_4^+$ with a new boundary
\be
\partial M_4^+ = M_3^+
\ee
A convenient way to describe this operation --- which admits various generalizations
and a direct translation into operations on $T[M_4^-]$ --- is to think of (the closure of)
$M_4^+ \setminus M_4^-$ as a (co)bordism, $B$, from $M_3^-$ to $M_3^+$.
In other words, we can think of $M_4^+$ as a 4-manifolds obtained by gluing $M_4^-$ to a cobordism $B$ with boundary
\be
\partial B = - M_3^- \cup M_3^+
\ee
Therefore,
\be
M_4^+ \; = \; M_4^- \cup_{\varphi} B
\label{MMpmandB}
\ee
where $\varphi : M_3 \to M_3$ is assumed to be the identity map, unless noted otherwise.

\bigskip

We have $H_3 (M_4^+, B) \cong H_3 (M_4^-,M_3^-)\cong H^1(M_4^-)$ by Poincar\'e duality. The latter is trivial, $H^1 (M_4^-) = 0$.
Then, comparing the exact sequence for the pair $(M_4^+,B)$ with the exact sequence for the triple $(M_4^+,B,M_3^+)$
we get the following diagram
\be
\begin{array}{ccccccccc}
0 & \to & H_2 (B) & \to & H_2 (M_4^+) & \to & H_2 (M_4^+ , B) & & \\
\parallel & & \downarrow & & \downarrow & & \parallel & & \\
0 & \to & H_2 (B,M_3^+) & \to & H_2 (M_4^+ , M_3^+) & \to & H_2 (M_4^+ , B) & \to & H_1 (B, M_3^+) = 0 \\
& & & & \phantom{P.D.} \wr \!\! \parallel \mathtt{P.D.} & &  \wr \!\! \parallel & & \\
& & & & H^2 (M_4^+) & &  H_2 (M_4^- , M_3^-) & & \\
& & & & \wr \!\! \parallel & & \phantom{P.D.} \wr \!\! \parallel \mathtt{P.D.} & & \\
& & & & H_2 (M_4^+)^* & \xrightarrow[~]{~\iota^*~} &  H_2 (M_4^-)^* & &
\end{array}
\ee
In this diagram, the map from $H_2 (M_4^+)$ to its dual $H_2 (M_4^+)^* \cong H^2 (M_4^+)$ is given
by the intersection form $Q^+ \equiv Q_{M_4^+}$. 
Therefore, we get
\be
0 \to H_2 (B) \to H_2 (M_4^+) \xrightarrow[~]{~ Q^+ ~} H_2 (M_4^+)^* \xrightarrow[~]{~\iota^*~}  H_2 (M_4^-)^*
\ee
Since the second map, from $H_2 (B)$ to $H_2 (M_4^+)$, is injective, it follows that
\be
H_2 (B) \; = \; \text{ker} \left( \iota^* \circ Q^+ \right)
\label{H2cob}
\ee
This useful result can tell us everything we want to know about the cobordism $B$ from the data of $M_4^-$ and $M_4^+$.

In particular, when both $M_4^+$ and $M_4^-$ are sphere plumbings, and the plumbing tree of the former
is obtained by adding a new vertex (with an edge) to the plumbing tree of the latter, as in Figure~\ref{fig:sequence},
the second homology of the cobordism $B$ is one-dimensional,
\be
b_2 (X) \; = \; 1 \,,
\ee
and, therefore, its intersection form is determined by the self-intersection of a single generator $s \in H^2 (B)$.
Thus, introducing a natural basis $\{ s_i \}$ for $H_2 (M_4^+)$, such that the intersection pairing
\be
Q^+ (s_i, s_i) = Q^+_{ij}
\label{QQplus}
\ee
is determined by the (weighted) plumbing tree, the generator $s \in H^2 (B)$ can be expressed
as a linear combination
\be
s = \sum_{i=1}^{b_2 (M_4^+)} k_i s_i
\label{cobgenlinexp}
\ee
where the coefficients $k_i \in \Z$ are determined by \eqref{H2cob}:
\be
Q^+ (s,x) = 0 \,, \qquad \forall x \in H_2 (M_4^-)
\label{Qorthog}
\ee
In practice, of course, it suffices to verify this orthogonality condition only on the basis elements of $H_2 (M_4^-)$.
Then, it determines the cohomology generator \eqref{cobgenlinexp} and, therefore, the self-intersection number $Q^+ (s,s)$.

As a warm-up, let us illustrate how this works in the case of a linear plumbing in Figure~\ref{fig:Anplumbing},
where for simplicity we start with the case where all Euler numbers $a_i = -2$.
Namely, if $M_4^-$ has a linear plumbing graph with $n-1$ vertices
and $M_4^+$ has a linear plumbing graph with $n$ vertices,
then the condition \eqref{Qorthog} becomes
\be
Q(s,s_i) = 0 \,, \qquad i = 1, \ldots, n-1
\ee
or, more explicitly,
\bea
-2 k_1 + k_2 & = & 0 \\
k_{i-1} - 2k_i + k_{i+1} & = & 0 \qquad i = 2, \ldots, n-1 \nonumber
\eea
Solving these equations we find the generator $s \in H^2 (B)$,
\be
s = s_1 + 2 s_2 + 3 s_3 + \ldots + n s_n
\label{slinprog}
\ee
for the cobordism $B$ that relates $A_{n-1}$ and $A_n$ linear plumbings.
Now, the self-intersection is easy to compute:
\be
Q^+ (s,s) = - n(n+1)
\label{Ann1bord}
\ee

It is easy to generalize this calculation to linear plumbings with arbitrary framing coefficients $a_i$,
as well as plumbing graphs which are not necessarily linear.
As the simplest example of the latter, let us consider a 2-handle attachment in the first step of Figure \ref{fig:sequence}
that turns a linear plumbing graph with three vertices
\be
M_4^- ~: \qquad\qquad
\overset{\displaystyle{a}}{\bullet}
\frac{\phantom{xxx}}{\phantom{xxx}}
\overset{\displaystyle{b}}{\bullet}
\frac{\phantom{xxx}}{\phantom{xxx}}
\overset{\displaystyle{c}}{\bullet}
\ee
into a non-linear plumbing graph with a trivalent vertex:
\be
M_4^+ ~: \qquad\qquad
\begin{array}{ccc}
& \overset{\displaystyle{d}}{\bullet} & \\
& \vline & \\
\overset{\displaystyle{a}}{\bullet}
\frac{\phantom{xxx}}{\phantom{xxx}}
& \underset{\displaystyle{b}}{\bullet} &
\frac{\phantom{xxx}}{\phantom{xxx}}
\overset{\displaystyle{c}}{\bullet}
\end{array}
\label{Qfortrinion}
\ee
In order to determine the cobordism $B$ that does the job we are again going to use \eqref{H2cob}
or, better yet, its more explicit version \eqref{Qorthog} suitable for arbitrary plumbing trees.
As before, denoting by $s_i$ the generators of $H_2 (M_4^+)$ with the intersection pairing \eqref{QQplus},
which is easy to read off from \eqref{Qfortrinion}, we get the system of linear equations \eqref{Qorthog}
that determines the generator \eqref{cobgenlinexp} of the cobordism $B$:
\bea
Q^+ (s,s_1) & = & a k_1 + k_2 \; = \; 0 \nonumber \\
Q^+ (s,s_2) & = & k_1 + b k_2 + k_3 + k_4 \; = \; 0 \\
Q^+ (s,s_3) & = & k_2 + c k_3 \; = \; 0 \nonumber
\eea
Of course, in case of negative-definite 4-manifolds $a$, $b$, $c$, and $d$ are all supposed to be negative.
Solving these equations we find the integer coefficients in \eqref{cobgenlinexp},
\be
k_1 \; = \; \frac{c}{\gcd (a,c)}
\,, \quad
k_2 \; = \; - \frac{ac}{\gcd (a,c)}
\,, \quad
k_3 \; = \; \frac{a}{\gcd (a,c)}
\,, \quad
k_4 \; = \; \frac{abc - a - c}{\gcd (a,c)}
\ee
which, in turn, determine the intersection form on $B$:
\be
Q^+ (s,s) = \frac{(abcd-ac-ad-cd)(abc-a-c)}{\gcd (a,c)^2}
\label{QBtrin}
\ee
For instance, if $a=b=c=d=-2$, we get $Q_B = \langle -4 \rangle$.


\section{Top-down approach: fivebranes and instantons}
\label{sec:VW}

In this section we approach the correspondence between 4-manifolds and 2d $\CN=(0,2)$ theories $T[M_4;G]$
by studying the (flavored) elliptic genus \eqref{indexdef} which, according to \eqref{ZVWvsindex},
should match the Vafa-Witten partition function.

In particular, we propose the ``gluing rules'' that follow operations on 4-manifolds
introduced in section \ref{sec:generalities}
and identify the set of basic cobordisms with branching functions in certain coset models.
In the non-abelian case, the key ingredient in the gluing construction is the integration measure,
which we propose to be the index of a 2d $(0,2)$ vector multiplet.
Another key ingredient, which plays an important role in \eqref{ZVWvsindex} for non-compact 4-manifolds,
is a relation between {\it discrete basis} and {\it continuous basis} introduced in section \ref{sec:contbasis}.

\subsection{Vafa-Witten theory}
\label{sec:VWpf}

In order to realize the Vafa-Witten twist of 4d $\CN=4$ super-Yang-Mills \cite{VafaWitten} in M-theory,
we start with the six-dimensional $(2,0)$ theory realized on the world-volume of $N$ fivebranes.
The R-symmetry group of the $(2,0)$ theory is $Sp(2)_r \cong SO(5)_r$
and can be viewed as a group of rotations in the five-dimensional space transverse to the fivebranes.
A $(2,0)$ tensor multiplet in six dimensions contains 5 scalars, 2 Weyl fermions and a chiral 2-form,
which under $Sp(2)_r$ transform as ${\bf 5}$, ${\bf 4}$, and ${\bf 1}$, respectively.

We are interested in the situation when the M-theory space is $S^1\times\mathbb{R}_t\times M_7\times\C$,
where $M_7$ is a 7-manifold with $G_2$ holonomy and $\R_t$ may be considered as the time direction.
We introduce a stack of $N$ fivebranes supported on the subspace $S^1\times \R_t\times M_4$,
where $M_4$ is a coassociative cycle in $M_7$.
This means that the normal bundle of $M_4$ inside $M_7$ is isomorphic to the self-dual part of $\Lambda^2 T^*M_4$:
\begin{equation}
 T_{M_7/M_4} \; \cong \; \Lambda_+^2 T^*M_4 \,.
 \label{normalbundle}
\end{equation}
Moreover, the neighborhood of $M_4$ in $M_7$ is isomorphic (as a $G_2$-manifold)
to the neighborhood of the zero section of $\Lambda_+^2 T^*M_4$.

Since both the eleven-dimensional space-time and the fivebrane world-volume in this setup have $S^1$
as a factor, we can reduce on this circle to obtain $N$ D4-branes supported on $\mathbb{R}\times M_4$ in type IIA string theory.
The D4-brane world-volume theory is maximally supersymmetric ($\CN=2$) super-Yang-Mills
in five dimensions with the following field content:
\be
\begin{array}{l@{\;}|@{\;}c@{\;}|@{\;}c}
\multicolumn{3}{c}{\text{spectrum of } 5d~\text{ super-Yang-Mills}} \\[.1cm]
& Spin(5)_E & Sp(2)_r \\\hline
\text{1-form} & {\bf 5} & {\bf 1} \\
\text{scalars} & {\bf 1} & {\bf 5} \\
\text{fermions} & {\bf 4} & {\bf 4}
\end{array}
\notag
\ee
The rotation symmetry in the the tangent bundle of $M_4$ is $Spin(4)_E \cong SU(2)_L \times SU(2)_R$
subgroup of the $Spin(5)_E$ symmetry of the Euclidean five-dimensional theory.
Five normal direction to the branes are decomposed into three directions normal to $M_4$ inside $M_7$
and two directions of $\C$-plane. This corresponds to the following decomposition of the R-symmetry group:
\be
SO(5)_r \; \to \; SO(3)_A \times SO(2)_U \cong SU(2)_A \times U(1)_U.
\ee
The fields of the 5d super-Yang-Mills transform under the resulting $SU(2)_L \times SU(2)_R \times SU(2)_A \times U(1)_U$
symmetry group as
\be \begin{array}{r@{\qquad}l}
\text{bosons}: & ({\bf 5}, {\bf 1} ) \oplus ({\bf 1}, {\bf 5}) \to
({\bf 2},{\bf 2},{\bf 1})^0 \oplus ({\bf 1},{\bf 1},{\bf 1})^0 \oplus ({\bf 1},{\bf 1},{\bf 3})^0 \oplus ({\bf 1},{\bf 1},{\bf 1})^{\pm 2} \\
\text{fermions}: & ({\bf 4}, {\bf 4} ) \to ({\bf 2},{\bf 1},{\bf 2})^{\pm 1} \oplus ({\bf 1},{\bf 2},{\bf 2})^{\pm 1}
\end{array}
\ee
Non-trivial embedding of the D4-branes in space-time with the normal bundle (\ref{normalbundle})
corresponds \cite{Bershadsky:1995qy} to identifying $SU(2)_L$ with $SU(2)_A$ and gives precisely
the topological twist introduced by Vafa in Witten \cite{VafaWitten}.
The spectrum of the resulting theory looks like:
\be \begin{array}{r@{\qquad}l}
\text{bosons}: &
({\bf 2},{\bf 2})^0 \oplus ({\bf 1},{\bf 1})^0 \oplus ({\bf 3},{\bf 1})^0 \oplus ({\bf 1},{\bf 1})^{\pm 2} \\
\text{fermions}: & ({\bf 1},{\bf 1})^{\pm 1} \oplus ({\bf 3},{\bf 1})^{\pm 1} \oplus ({\bf 2},{\bf 2})^{\pm 1}
\end{array}
\label{twistedspectrum}
\ee
where we indicate transformation under the symmetry group $SU(2)_L' \times SU(2)_R \times U(1)_U$.
Here, the subgroup $SU(2)_L' \times SU(2)_R$ is the new rotation symmetry along $M_4$,
whereas $U(1)_U$ is the R-symmetry\footnote{Note,
in \cite{VafaWitten} the symmetry group $U(1)_U$ is enhanced to the global symmetry group $SU(2)_U$
due to larger R-symmetry of the starting point.} of the effective $\CN=2$ supersymmetric quantum mechanics $T_\text{1d}[M_4]$ on $\R_t$. The $U(1)_U$ quantum number is called the ghost number.

{}From \eqref{twistedspectrum} it is clear that the resulting supersymmetric quantum mechanics $T_\text{1d}[M_4]$ has two
supercharges, which are scalar from the viewpoint of the 4-manifold $M_4$ and which carry ghost number $U=+1$ and $U=-1$, respectively. When the quantum mechanics is lifted to the 2d theory $T[M_4]$ on $S^1\times \R_t$ they become supercharges of $\CN=(0,2)$ SUSY.
Among the bosons, two states $({\bf 1},{\bf 1})^{\pm 2}$ with non-zero ghost number are scalars $\phi$ and $\bar \phi$
that are not affected by the twist, the state $({\bf 3},{\bf 1})^0$ is the self-dual 2-form field $B$,
and finally the state $({\bf 1},{\bf 1})^0$ is the scalar field $C$, all transforming in the adjoint representation of the gauge group.
The state $({\bf 2},{\bf 2})^0$ is, of course, the gauge connection on $M_4$:

\be \begin{array}{l@{\qquad}l}
({\bf 2}, {\bf 2})^0  & \text{gauge connection}~A \\
({\bf 3}, {\bf 1})^0  & \text{self-dual}~2\text{-form}~B \\
({\bf 1}, {\bf 1})^{\pm 2}  & \text{complex scalar}~\phi \\
({\bf 1}, {\bf 1})^0  & \text{real scalar}~C
\end{array}
\ee

Now let us consider a situation where the time direction is also compactified to a circle: $\R_t\rightsquigarrow S^1_t$
in a way that allows the M-theory circle $S^1$ to fiber non-trivially over $S^1_t$,
so that the twisted product $S^1 \rtimes S^1_t$ is a torus with the complex modulus $\tau$.
Then, the theory on the fivebranes can be described as a theory on D4-branes supported on $M_4$,
{\it i.e.} the four-dimensional topologically twisted $\CN=4$ SYM with coupling constant $\tau$ \cite{VafaWitten}.

The path integral of the Vafa-Witten theory localizes on the solutions to the following equations
\be
\begin{aligned}
F_A^+ - \frac{1}{2} [B \times B] + [C,B] & = 0 \\
d_A^* B - d_A C & = 0
\end{aligned}
\qquad \text{where} \qquad
\begin{aligned}
A & \in \CG_P \\
B & \in \Omega^{2,+} (M_4; \text{ad}_P) \\
C & \in \Omega^{0} (M_4; \text{ad}_P)
\end{aligned}
\ee
where $\CG_P$ denotes the space of connections of a principal bundle $P$.
Under certain conditions (see \cite{VafaWitten} for details)
the only non-trivial solutions are given by configurations with
vanishing self-dual part of the curvature
\begin{equation}
 F^+_A \; = \; 0
\end{equation}
and trivial other fields ($B=0$ and $d_A C=0$).
The partition function is then given by the generating function of the Euler numbers of instanton moduli spaces:
\begin{equation}
 Z_\VW[M_4](q) \; = \; \sum_{m}\chi(\CM_m)q^{m- \tfrac{c}{24}}
\end{equation}
where
$$
 \qquad \CM_m \; = \; \left\{A\in \CG_P\;:\; F^+_A=0,\;  \langle \mathrm{ch}, [M_4]\rangle \equiv \frac{1}{8\pi^2}\int_{M_4} \Tr F^2=m\right\} \, / \, \mathrm{Gauge} \,,
$$
$$
 q=e^{2\pi i\tau}
$$
and $c$ is a constant that depends on the topology of $M_4$. In \cite{VafaWitten} it was proposed that
\begin{equation}
 c \; = \; N \cdot \chi(M_4)
\end{equation}
where $N$ is the rank of the gauge group and $\chi(M_4)$ is the Euler characteristic\footnote{When $M_4$
is non-compact $\chi (M_4)$ should be replaced by the regularized Euler characteristic,
and when $G=U(N)$ one needs to remove by hand the zero-mode to ensure that the partition function
does not vanish identically.} of $M_4$.
The constant $c$ can be interpreted as the left-moving central charge $c_L$ of the dual 2d $(0,2)$ theory $T[M_4]$.

In general, when the manifold $M_4$ is not compact and the gauge group is $U(N)$,
anti-self-dual configurations can also be distinguished by the first Chern class $c_1$ and the boundary conditions at infinity.
In order to have finite action, the connection should be asymptotically flat:
\begin{equation}
 \left. A\right|_{\d M_4} \; = \; A_\rho,\qquad F_{A_\rho} \; = \; 0 \,.
\end{equation}
Therefore, as we already mentioned in section \ref{sec:bdrycond},
different asymptotics can be labeled by flat connections on the boundary 3-manifold $M_3 = \partial M_4$:
\begin{equation}
 \rho\in \CM_\text{flat}[M_3] \; \equiv \; \mathrm{Hom}(\pi_1(M_3),U(N)) \, / \, \text{Gauge} \,.
\label{flatgeneral}
\end{equation}
The dependence on the first Chern class can be captured by introducing the following topological
term in the action, {\it cf.} \cite{Dijkgraaf:2007sw}:
\begin{equation}
 \Delta S=\frac{1}{2\pi}\int_{\xi}\Tr F\equiv \langle c_1 , \xi\rangle
\end{equation}
where $\xi\in H_2(M_4)\otimes \C$. It is useful to define the following exponential map:
\begin{equation}
 \begin{array}{rl}
  \exp: H_2(M_4)\otimes \C & \longrightarrow (\C^*)^{b_2} \\
 \xi & \longmapsto x
 \end{array}
\end{equation}
such that $\ker (\exp)=H_2(M_4, \Z)$ and also the ``power'' operation
\begin{equation}
\begin{array}{rl}
  (\C^*)^{b_2} \times H^2(M_4) & \longrightarrow \C^* \\
  (x,h) & \longmapsto x^h\equiv e^{2\pi i\langle h,\xi \rangle}
\end{array}
\label{powermap}
\end{equation}
for some preimage $\xi$ of $x$.
The refined Vafa-Witten partition function then depends on $b_2(M_4)$ additional fugacities and is given by
\begin{equation}
 Z_\VW[M_4]_\rho(q,x) \; = \; \sum_{m,c_1}\chi(\CM_{m,c_1,\rho})\,q^{m-\tfrac{c}{24}}\,x^{c_1}
 \label{VWgeneratingfunction}
\end{equation}
where
$$
 \qquad \CM_{m,c_1,\rho} = \left\{A\in \CG_P\;:\; F^+_A=0,\; \langle \mathrm{ch}, [M_4]\rangle =m,\; [\Tr F]=2\pi c_1,\;A|_{M_3}=A_\rho \right\} / \mathrm{Gauge} .
$$

{}From the point of view of the 2d theory $T[M_4;U(N)]$, the fugacities $x$ in \eqref{VWgeneratingfunction} play the role of flavor fugacities in the elliptic genus. This tells us that $T[M_4;U(N)]$ has flavor symmetry of rank $b_2$ associated to 2-cycles of $M_4$.

In what follows, if not explicitly stated otherwise, we will consider 4-manifolds \eqref{M4KKK} with
\be
\begin{array}{c}
b_2^+(M_4)=0 \,, \qquad \pi_1(M_4)=0 \,, \qquad H_2(M_3,\mathbb{Z})=0 \,, \qquad H_1(M_3, \R)=0 \\
\Gamma \equiv H_2(M_4,\mathbb{Z})\cong \mathbb{Z}^{b_2} \,, \qquad \Gamma^* \equiv H^2({M_4},\Z)\cong \mathbb{Z}^{b_2}
\end{array}
\label{classM}
\ee
The last two conditions mean that there is no torsion in second (co)homology.
As explained in section \ref{sec:plumbing},
such manifolds are uniquely defined by the intersection form or, alternatively, by the plumbing graph.

\subsection{Gluing along 3-manifolds}
\label{sec:VW-gluing}
In this section we will describe how the Vafa-Witten partition function behaves under cutting and gluing of 4-manifolds. Suppose one can produce a 4-manifold $M_4$ by gluing $M_4^+$ and $M_4^-$ along a common boundary component $M_3$. For simplicity, in the following we actually assume that $M_3$ is the only boundary component for both $M_4^+$ and $M_4^-$ (that is, the resulting manifold $M_4$ does not have any boundary). The generalization to the case when $M_4^{\pm}$ have other boundary components (that will become boundary components of $M_4$ after the gluing) is straightforward. For the same reason we will also suppress the dependence of the moduli spaces on the first Chern class $c_1$ or, equivalently, the dependence of the Vafa-Witten partition function on the fugacities $x$ in (\ref{VWgeneratingfunction}).

\begin{figure}[ht]
\centering
\includegraphics{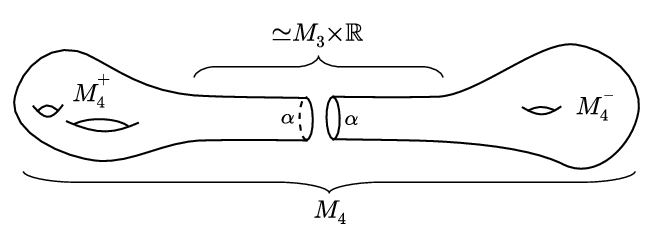}
\caption{Gluing of $M_4^+$ and $M_4^-$ along the common boundary $M_3$.}
\label{fig:VW-gluing}
\end{figure}

Since for $b_2^+ > 1$ we expect the topology of the instanton moduli spaces to be independent under smooth deformations of the 4-manifold, consider the situation where the boundary neighborhoods of $M_4^\pm$ look like long ``half-necks'' of the form $\mathbb{R}_+\times M_3$, as illustrated in Figure~\ref{fig:VW-gluing}. Very naively the Vafa-Witten partition function on $M_4$ is given by a sum of products of partition functions on $M_4^\pm$ with identified boundary conditions. However in this way we count instantons living on the long neck $M_3\times \mathbb{R}$ twice and we need to cancel out this contribution.

Let us address this issue more systematically. Let $\widetilde{\mathcal{M}}_{m}^{\alpha\beta}$ be the moduli space of $m$ instantons\footnote{Here and in what follows the instanton number is not necessarily an integer.} on $M_3\times \mathbb{R}$ with boundary conditions $\alpha,\beta\in\mathcal{M}_\text{flat}[M_3]$. One can always factor out the part of the moduli space associated to translations along $\mathbb{R}$:
\begin{equation}
 \widetilde{\mathcal{M}}_{m}^{\alpha\beta}=\bar{\mathcal{M}}_m^{\alpha\beta}\times\mathbb{R}.
\end{equation}
Let us denote the corresponding generating function for Euler characteristics as follows:
\begin{equation}
 K^{\alpha\beta}[M_3]\equiv \sum_{m}\chi(\widetilde{\mathcal{M}}_{m}^{\alpha\beta})q^m.
 \label{neckkernel}
\end{equation}

Now let $\mathcal{M}_m$ and $\mathcal{M}_{m,\alpha}^\pm$ be instanton moduli spaces for $M_4$ and $M_4^\pm$ respectively. Then
\begin{equation}
 \mathcal{M}_m=\bigcup_{\substack{\alpha \\ m_++m_-=m}} \mathcal{M}_{m_+,\alpha}^+\times \mathcal{M}_{m_-,\alpha}^-.
\end{equation}
The problem, however, is that this union is \textit{not} disjoint. Various terms have common boundary components corresponding to particular degeneration of instanton configurations. Common codimension-1 boundary components have the following form:
\begin{equation}
 \mathcal{M}^+_{m_+,\alpha}\times\bar{\mathcal{M}}_{\Delta}^{\alpha\beta}\times\mathcal{M}^-_{m_-,\beta}\subset
 \;\;\begin{array}{c}
    \partial\left(\mathcal{M}^+_{m_++\Delta,\beta}\times\mathcal{M}^-_{m_-,\beta}\right) \\
   \text{and} \\
    \partial\left(\mathcal{M}^+_{m_+,\alpha}\times\mathcal{M}^-_{\Delta+m_-,\alpha}\right).
 \end{array}
\end{equation}
The first case can be intuitively understood from a limit when we separate a localized configuration with instanton number $\Delta$ in $M_4^+$ and push it to the boundary. And in the second case we do the same for $M_4^-$. Similarly, there are common codimension-2 boundary components:
\begin{equation}
  \mathcal{M}^+_{m_+,\alpha}\times\bar{\mathcal{M}}_{\Delta_1}^{\alpha\beta}\times\bar{\mathcal{M}}_{\Delta_2}^{\beta\gamma}\times\mathcal{M}^-_{m_-,\gamma}\subset
 \;\;\begin{array}{c}
    \partial\left(\mathcal{M}^+_{m_++\Delta_1+\Delta_2,\gamma}\times\mathcal{M}^-_{m_-,\gamma}\right) \\
        \partial\left(\mathcal{M}^+_{m_++\Delta_1,\beta}\times\mathcal{M}^-_{\Delta_2+m_-,\beta}\right) \\
    \partial\left(\mathcal{M}^+_{m_+,\alpha}\times\mathcal{M}^-_{\Delta_1+\Delta_2+m_-,\alpha}\right)
 \end{array}
\end{equation}
and so on.

Then, applying inclusion-exclusion principle for Euler characteristic we get
\begin{multline}
 \chi(\mathcal{M}_m)=\sum_{\substack{\alpha \\ m_++m_-=m}} \chi\left(\mathcal{M}_{m_+,\alpha}^+\times \mathcal{M}_{m_-,\alpha}^-\right)\\
 -\sum_{\substack{\alpha,\beta;\;\;\Delta>0 \\ m_++\Delta+m_-=m}} \chi\left(
 \mathcal{M}^+_{m_+,\alpha}\times\bar{\mathcal{M}}_{\Delta}^{\alpha\beta}\times\mathcal{M}^-_{m_-,\beta}
 \right)\\
  +\sum_{\substack{\alpha,\beta,\gamma;\;\;\Delta_{1,2}>0 \\ m_++\Delta_1+\Delta_2+m_-=m}} \chi\left(
 \mathcal{M}^+_{m_+,\alpha}\times\bar{\mathcal{M}}_{\Delta_1}^{\alpha\beta}\times\bar{\mathcal{M}}_{\Delta_2}^{\beta\gamma}\times\mathcal{M}^-_{m_-,\gamma}
 \right)-\ldots
\end{multline}
which translates into the following relation for the generating functions:
\begin{multline}
 Z_\VW[M_4]=\sum_\alpha Z_\VW[M_4^+]_\alpha Z_\VW[M_4^-]_\alpha-
 \sum_{\alpha,\beta} Z_\VW[M_4^+]_\alpha (K^{\alpha\beta}[M_3]-\delta^{\alpha\beta}) Z_\VW[M_4^-]_\beta\\
 +\sum_{\alpha,\beta,\gamma }Z_\VW[M_4^+]_\alpha (K^{\alpha\beta}[M_3]-\delta^{\alpha\beta})(K^{\beta\gamma}[M_3]-\delta^{\beta\gamma}) Z_\VW[M_4^-]_\gamma-\ldots\\
 =\sum_{\alpha,\beta} Z_\VW[M_4^+]_\alpha (K^{-1}[M_3])^{\alpha\beta} Z_\VW[M_4^-]_\beta
 \label{gluingkernel}
\end{multline}
where $K^{-1}[M_3]$ denotes the matrix inverse to $K[M_3]$ defined in (\ref{neckkernel}). The relation (\ref{gluingkernel}) obviously holds when $M_4=M_4^+=M_4^-=M_3\times\mathbb{R}$. Let us note that in the case when $M_3$ is a lens space the ``gluing kernel'' $K[M_3]$ can be explicitly computed using the results of \cite{austin1990,furuta1990invariant}.

For later convenience, let us define a modified Vafa-Witten partition with an \textit{upper} index denoting the boundary condition:
\begin{equation}
 Z_\VW[M_4^-]^\alpha\equiv \sum_{\beta}(K^{-1}[M_3])^{\alpha\beta} Z_\VW[M_4^-]_\beta.
\end{equation}
Intuitively this modification can be understood as excluding instantons approaching the boundary. Then the relation between partition functions takes the following simple form:
\begin{equation}
 Z_\VW[M_4]=\sum_\alpha Z_\VW[M_4^+]_\alpha Z_\VW[M_4^-]^\alpha.
\end{equation}

\subsection{Relation to affine Lie algebras}
\label{sec:affinelie}

Before we discuss cobordisms, let us review the relation between Vafa-Witten theory on ALE spaces
and affine Lie algebras \cite{Nakajima,VafaWitten,Dijkgraaf:2007sw}, that will be our starting point
for constructing generalizations.
Namely, let $M_4$ be a hyper-K\"ahler ALE space obtained by a resolution of the quotient singularity $\C^2/H$,
where $H$ is a finite subgroup of $SU(2)$. According to the McKay correspondence,
finite subgroups of $SU(2)$ have $ADE$ classification and therefore for each $H$
there is a corresponding simple Lie algebra $\mathfrak{g}$ of the same $ADE$ type.
{}From the work of Nakajima \cite{Nakajima} it follows that the partition function
of the Vafa-Witten theory with the gauge group $U(N)$ is given by the character
of the integrable representation of the corresponding affine Lie algebra $\hat{\mathfrak{g}}$ at level $N$:
\begin{equation}
 Z_\VW^{U(N)}[M_4]_\rho(q,x) \; = \; \chi_\rho^{\hat{\mathfrak{g}}_N}(q,x) \,.
 \label{VWcharacter0}
\end{equation}
Let us explain in some detail the role of the parameters $\rho$, $q$ and $x$ on the right hand side of this formula.
First, the formula (\ref{VWcharacter0}) exploits the fact that there is a one-to-one correspondence
between $U(N)$ flat connections on $M_3 \cong S^3/H$ and integrable representations of $\hat{\mathfrak{g}}_N$.
The right hand side of (\ref{VWcharacter0}) can then be understood as a character of $\hat{\mathfrak{g}}_N$
for a given representation $\rho$. Let us consider how the identification between flat connections
and integrable representations works in a simple case when $H=\Z_p$, $M_4=A_{p-1}$ and $\mathfrak{g}=\mathfrak{su}(p)$.
The set of flat connections (\ref{flatgeneral}) in this case is given by the ordered partitions of $N$ with $p$ parts,
which are in one-to-one correspondence with Young diagrams that have at most $p-1$ rows and $N$ columns:
\begin{multline}
 \mathrm{Hom}(\Z_p,U(N))/U(N)=\left\{\left(
 \begin{array}{ccc}
    z_1 &  & 0 \\
     & \ddots & \\
     0 & & z_N
 \end{array}
\right)^p=1 \right\}/S_N=\\
\left\{\mathrm{diag}(\underbrace{1,\ldots,1}_{N_0},\underbrace{e^{\frac{2\pi i}{p}},\ldots,e^{\frac{2\pi i}{p}}}_{N_1},\ldots,
\underbrace{e^{2\pi i\frac{p-1}{p}},\ldots,e^{2\pi i\frac{p-1}{p}}}_{N_{p-1}})\right\}\cong\\
\left\{\begin{minipage}[h]{60mm}{
\includegraphics[width=60mm]{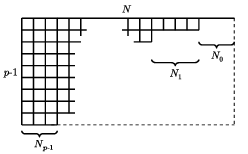}}
 \end{minipage}\right\}
\end{multline}
Young diagrams of such type indeed describe integrable representation of $\hat{\mathfrak{su}}(p)_N$. The variables $(q,x)$ in the right hand side of (\ref{VWcharacter0}) play the role of coordinates on the (complexified) torus corresponding to the Cartan subalgebra $\hat{\mathfrak{h}}$ of $\hat{\mathfrak{g}}_N$. In particular, $\tau$ is a coordinate on $\hat{\mathfrak{h}}$ in the direction of $L_0$ and $x$ can be interpreted as coordinates on the (complexified) maximal torus of the Lie group $G$ corresponding to the ordinary Lie algebra $\mathfrak{g}$. This is in agreement with the fact that the lattice $\Gamma^*$ for an ALE space of the $ADE$ type is the same as the weight lattice of the corresponding simple Lie algebra $\mathfrak{g}$ and $\xi$ in (\ref{powermap}) is then the element of the dual space. The dual lattice $\Gamma$ is the same as the root lattice of $\mathfrak{g}$ and the intersection form $Q$ plays the role of the normalized Killing form. It follows that the abelian quiver CS with coefficients $Q_{ij}$ is the same as the ordinary CS with the gauge group $G$ restricted to the Cartan subalgebra, which can be interpreted as a level-rank duality.

\begin{table}[htb]
\centering
\begin{tabular}{|c|c|}
 \hline
  {\bf Physics \& Geometry} & {\bf Algebra} \\
 \hline
 \hline
 plumbing graph & Dynkin diagram of $\mathfrak{g}$ \\
 \hline
 fugacities $x$ & maximal torus of $G$ \\
 \hline
 coupling $\tau$ & coordinate on $\hat{\mathfrak{h}}$ along $L_0$ \\
 \hline
 intersection form & normalized Killing form of $\mathfrak{g}$ \\
 \hline
 $b_2(M_4)$ & rank of $\mathfrak{g}$\\
 \hline
 $H_2(M_4)$ & root lattice of $\mathfrak{g}$\\
 \hline
 $H^2(M_4)$ & weight lattice of $\mathfrak{g}$ \\
 \hline
 boundary condition & integrable representation of $\hat{\mathfrak{g}}$\\
 \hline
 rank of the gauge group & level of $\hat{\mathfrak{g}}$ \\
 \hline
 $Z_\VW[M_4]$ & character of $\hat{\mathfrak{g}}$ \\
 \hline
  cobordism $B$: $M_4^+=B\cup M_4^-$  & embedding $\mathfrak{g}^-\subset \mathfrak{g}^+ $ \\
 \hline
  $Z_\VW[B]$ & branching functions \\
 \hline
\end{tabular}
\caption{Dictionary between Vafa-Witten theory and (affine) Lie algebras.}
\label{tabledictionary}
\end{table}

Now let us describe the gluing of 4-manifolds considered in section \ref{sec2:gluing} in the language of (affine) Lie algebras.
Suppose the manifold $M_4^+$ with boundary $M_3^+$ is defined by a plumbing graph of $ADE$ type
which can be interpreted as a Dynkin diagram of Lie algebra $\mathfrak{g}^+$ with root lattice $\Gamma_+\equiv H_2(M_4^+)$.
Let us pick up a subalgebra $\mathfrak{g}^-\subset\mathfrak{g}^+$ and consider the manifold $M_4^+$
with properties \eqref{classM} such that the lattice $\Gamma_- \equiv H_2(M_4^-)$ is the root lattice of $\mathfrak{g}^-$.
The lattice $\Gamma_-$ is a sublattice of $\Gamma_+$ and the manifold $M_4^+$ can be obtained by gluing $M_4^-$
with a certain (co)bordism $B$ such that $\d B=M_3^-\sqcup M_3^+$ along the common boundary component $M_3^-$,
{\it cf.} \eqref{MMpmandB}.
In the rest of the paper we will sometimes use the following schematic (but intuitive)
notation for the process of obtaining a manifold $M_4^+$ by gluing a cobordism $B$ to~$M_4^-$:
\begin{equation}
 M_4^-\stackrel{B}{\rightsquigarrow} M_4^+.
\end{equation}
From the gluing principle described in the previous section we have:
\begin{equation}
 Z^{U(N)}_\VW[M_3^+]_\rho(q,x) \; = \; \sum_{\lambda} \, Z^{U(N)}_\VW[B]^\lambda_\rho(q,x^\perp) \, Z^{U(N)}_\VW[M_3^-]_\lambda(q,x^\parallel)
\end{equation}
where the splitting of the parameters $x=(x^\perp,x^\parallel)$ corresponds
to the splitting\footnote{Let us note that $H_2(M_4^+)\neq H_2(B)\oplus H_2(M_4^-)$.
However, the lattice $H_2(M_4^+)$ can be obtained from the lattice $H_2(B)\oplus H_2(M_4^-)$
by the so-called gluing procedure that will be described in detail 
shortly.} of the homology groups
$H_2(M_4^+)\otimes \C=H_2(B)\otimes\C\oplus H_2(M_4^-)\otimes \C$. Using (\ref{VWcharacter0}) one has
\begin{equation}
 \chi^{\hat{\mathfrak{g}}_N^+}_\rho(q,x) \; = \; \sum_{\lambda}Z^{U(N)}_\VW[B]^\lambda_\rho(\tau,x^\perp) \; \chi_\lambda^{\hat{\mathfrak{g}}_N^-}(q,x^\parallel) \,.
\end{equation}
Therefore, $Z^{U(N)}_\VW[B]^\lambda_\rho$ are given by the branching functions of the embedding $\mathfrak{g}^-\subset\mathfrak{g}^+$,
\be
\boxed{\phantom{\int}
Z^{U(N)}_\VW[B]^{\lambda}_{\rho} \; = \; \chi^{\hat{\mathfrak{g}}_N^+ / \hat{\mathfrak{g}}_N^-}_{\lambda,\rho}
\phantom{\int}}
\ee

Let us consider a particular example: $M_4^+=A_{p}$ and $M_4^-=A_{p-1}$.
As was shown in section \ref{sec:addinghandle} via a variant of the ``Norman trick'' \cite{Norman,Quinn},
the cobordism $B$ in this case is a 4-manifold in family \eqref{classM}
with a single 2-cycle of self-intersection $-(p+1)p$ and the boundary $L(p,-1) \sqcup L(p+1,-1)$.
The partition function on $B$ is then given by the characters of $\mathfrak{su}(p+1)/\mathfrak{su}(p)$ cosets:
\begin{equation}
 Z^{U(N)}_\VW[B]^{\lambda}_{\rho} \; = \; \chi^{\hat{\mathfrak{su}}(p+1)_N/\hat{\mathfrak{su}}(p)_N}_{\lambda,\rho} \,.
\end{equation}

The relation between Vafa-Witten theory and (affine) Lie algebras is summarized
in Table~\ref{tabledictionary} and will play an important role in the following sections.
In the next section we consider in detail the case of the gauge group $U(1)$.
Then, in section \ref{sec:naVW}, we will make some proposals about the non-abelian case.


\subsection{Cobordisms and gluing in the abelian case}
\label{sec:abelianVW}

For a 4-manifold $M_4$ that satisfies \eqref{classM} one has the short exact sequence \eqref{HHHseq}:
\begin{equation}
0 \; \longrightarrow \; H_2({M_4}) \; \stackrel{Q}{\longrightarrow} H^2({M_4}) \;
\stackrel{i^*_{M_3}}{\longrightarrow} \; H^2({M_3}) \; \longrightarrow \; 0
\label{relhomexactseq}
\end{equation}
where the map $Q$ is given by the intersection matrix and $i_{M_3}$
is the inclusion map of the boundary $M_3 = \d M_4$ into ${M_4}$. Equivalently, $H^2({M_3})\cong \mathrm{coker} Q$.

In the case of abelian theory self-duality condition implies that
\begin{equation}
 dF=0 \,, \qquad d^*F=0.
 \label{harmonic}
\end{equation}
For manifolds with asymptotically cylindrical or conical ends it has been shown (under certain assumptions) \cite{harmonic1,harmonic2} that the space of $L^2$ integrable 2-forms satisfying conditions (\ref{harmonic}) coincides with the space harmonic 2-forms $\CH^2({M_4})$ and is isomorphic to the image of the natural map $H^2({M_4}, {M_3},\R)\longrightarrow H^2({M_4},\R)$. In our case this map is an isomorphism. Since $b_2^+({M_4})=0$ the space $\CH^2({M_4})$ is an eigenspace of the Hodge $*$ operator with eigenvalue $-1$.

For an ordinary $U(1)$ gauge theory the Dirac quantization condition implies that $[F/2\pi]\in H^2({M_4})\equiv \Gamma^*$.
However, since we are interested in gauge theory on the world-volume of a D4-brane in type IIA string theory setup,
we need to take into account the Freed-Witten anomaly \cite{Freed:1999vc}.
Specifically, the two-form $F=dA$ should be viewed as a curvature of the $U(1)$ part
of a connection on a $Spin^c(4)\equiv {Spin}(4)\times_{\Z_2} U(1)$
principal bundle over $M_4$ obtained by a lift of the $SO(4)$ orthonormal frame bundle.
Let us note that such a lift is possible for any 4-manifold,
{\it i.e.} any 4-manifold is $\text{Spin}^c$.
Not any 4-manifold, though, has a $\text{Spin}$ structure.
The obstruction is given by the second Stiefel-Whitney class $w_2 \in H_2(M_4,\Z_2)$.
Therefore, as in \cite{Gukov:1999ya,Gukov:2002zg} we have a shifted quantization condition
for the magnetic flux through a 2-cycle $C\subset M_4$:
\begin{equation}
 \int_C\frac{F}{2\pi}=\frac{1}{2}\int_C w_2=\frac{1}{2}Q(C,C) \;\mod \Z
\end{equation}
where the second equality is the Wu's formula. The class $[F/2\pi]$ then takes values in the shifted lattice:
\begin{equation}
 \left[\frac{F}{2\pi}\right] \in \tGamma^*\equiv\Gamma^*+\sw
\label{latshift}
\end{equation}
where $2\sw$ is a lift\footnote{Such lift exists because the manifold is $\text{Spin}^c$.} of $w_2$ with respect to the map $\Gamma^*\equiv H^2(M_4,\Z)\rightarrow H^2(M_4,\Z_2)$. From the Wu's formula it follows that $w_2=0$ or, equivalently, the manifold $M_4$ is Spin, if and only if the lattice $\Gamma$ is even.

Let us note that since $\pi_1({M_4})=0$ there are no non-trivial flat connections and therefore fixing $[F/2\pi]$ in $\tGamma^*$ completely determines the anti-self-dual gauge connection. On the boundary $F|_{{M_3}}=0$ and therefore $A|_{{M_3}}$ is a flat connection on ${M_3}$ which determines $[F/2\pi]$ modulo $H^2({M_4},{M_3})\equiv \Gamma$. It is easy to see that the coset space $\tGamma^*/\Gamma$ coincides with the space of flat connections. From (\ref{relhomexactseq}) it follows that $H_1({M_3})$ is a finite abelian group of order $|\det Q|$. All such groups are isomorphic to a direct sum of finite cyclic groups. Therefore the space of flat connections on the boundary is given by
\begin{equation}
 \mathrm{Hom}(\pi_1({M_3}),U(1))\cong \mathrm{Hom}(H_1({M_3}),U(1)) \cong H^2({M_3})\cong \Gamma^*/\Gamma \cong \tGamma^*/\Gamma
\end{equation}
where the last equality follows from (\ref{relhomexactseq}) and \eqref{latshift}.

The Vafa-Witten partition for $U(1)$ gauge group can be calculated explicitly for general 4-manifold ${M_4}$ in the family \eqref{classM}
for a prescribed boundary condition $\rho\in \tGamma^*/\Gamma$ and a fugacity $x\in H_2({M_4},\R)$, {\it cf.} \cite{Witten:1996hc,Dijkgraaf:2002ac}:
\begin{multline} Z_\VW^{U(1)}[{M_4}]_{\rho}(q,x)=\frac{1}{\eta^{\chi({M_4})}(q)}\sum_{[F/2\pi]\in \tGamma^* \atop [F/2\pi]=\rho\mod\Gamma} q^{\frac{1}{8\pi^2}\int F\wedge F} x^{\left[{F}/{2\pi}\right]}=\\
 \frac{1}{\eta^{\chi({M_4})}(q)}\sum_{[F/2\pi]\in \tGamma^* \atop [F/2\pi]=\rho\mod\Gamma} q^{-\frac{1}{2}Q^{-1}([F/2\pi],[F/2\pi])} x^{\left[{F}/{2\pi}\right]}=\\
 \frac{1}{\eta^{\chi({M_4})}(q)}\sum_{\gamma\in \Gamma} q^{-\frac{1}{2}Q^{-1}(Q\gamma+\rho,Q\gamma+\rho)} x^{Q\gamma+\rho}=\\
 \frac{1}{\eta^{\chi({M_4})}(q)}\sum_{\gamma\in \Gamma} q^{-\frac{1}{2}Q(\gamma+Q^{-1}\rho,\gamma+Q^{-1}\rho)} x^{Q\gamma+\rho}.
 \label{VWabelian0}
\end{multline}
The overall factor
\begin{equation}
 \frac{1}{\eta^{\chi({M_4})}(q)}=q^{-\frac{\chi({M_4})}{24}}\sum_{m=0}^\infty \chi (\text{Hilb}^{[m]}({M_4}))\, q^m
\end{equation}
is the contribution of point-like instantons. Let us remind that the moduli space of $m$ point-like instantons is given by the Hilbert scheme $\text{Hilb}^{[m]}({M_4})$ which can be understood as a regularization of the configuration space of $m$ points on $M_4$.

Since the quadratic form $-Q$ is positive definite one can always assume that the lattices $\Gamma$ and $\Gamma^*$ are embedded in the Euclidean space $\mathbb{R}^{b_2}$ so that
$$
\Gamma^*=\{n_i\omega_i|n_i\in \Z\}\subset \mathbb{R}^{b_2} \,.
$$
and
$$
\Gamma=\{n_i\lambda_i|n_i\in \Z\}\subset \Gamma^*\subset \mathbb{R}^{b_2}
$$
The basis vectors of these lattices are chosen so that $(\lambda_i,\lambda_j)=-Q_{ij}$ and $(\omega_i,\lambda_j)=\delta_{ij}$ where $(\cdot,\cdot)$ is the standard Euclidean scalar product. The shift due to the Freed-Witten anomaly can be represented then by the vector $\Delta=\frac{1}{2}\sum_i\|\lambda_i\|^2\omega_i$. In this setup (\ref{VWabelian0}) reads simply as:
\begin{multline}
 Z_\VW^{U(1)}[{M_4}]_{\rho}(q,x)=
 \frac{1}{\eta^{\chi({M_4})}(q)}\sum_{\gamma\in \Gamma\subset \R^{b_2}} q^{\frac{1}{2}\|\gamma+\rho+\Delta\|^2} x^{\gamma+\rho+\Delta}\\
  \equiv\frac{\theta_\Gamma^{(\rho+\Delta)}(x;q)}{\eta^{\chi({M_4})}(q)}
 ,\qquad
 \rho\in \Gamma^*/\Gamma.
 \label{VWabelian}
\end{multline}
where $\theta_\Gamma^{(\rho+\Delta)}$ stands for the theta-function of the lattice $\Gamma$ with the shift $\rho+\Delta$. The regularized Euler characteristic $\chi(M_4)$ coincides with dimension of the lattice $b_2$.

\subsection*{Number of vacua}

As in \cite{Gukov:1999ya,Gukov:2002zg}, the quantum mechanics $T_\text{1d}[M_4]$ on $\R_t$
obtained by reduction of an M5-brane on $S^1\times M_4$
is specified by a flat connection $A_\rho$ on the boundary and the flux at infinity which,
up to constant depending on the topology of $M_4\subset M_7$, is given by
\be
\Phi_{\infty} = N_{D0} - \frac{1}{8\pi^2} \int_{M_4}  F \wedge F
\label{fluxatinfinity}
\ee
Here, $N_{D0}$ is a non-negative integer denoting the number of point-like instantons.
The origin of the last term is the Wess-Zumino part of the D4-brane action:
\be
I_{WZ} = - \int_{\R \times M_4} C_* \wedge \text{ch} (F) \wedge \sqrt{\frac{\hat A (TM_4)}{\hat A (NM_4)}}.
\ee
Once we picked $\Phi_{\infty}$ and fixed the value of $[F/2\pi]$ modulo $\Gamma$ which specify the theory $T_\text{1d}[M_4]_{\rho,\Phi_\infty}$,
its supersymmetric vacua are obtained by finding $N_{D0} \ge 0$ and $[F/2\pi]$ which solve (\ref{fluxatinfinity}). Note, the effective theory is massive when $N_{D0}=0$. If $N_{D0} > 0$ there are moduli of point-like abelian on $M_4$. The number of vacua is given by the corresponding coefficient of (\ref{VWgeneratingfunction}):
\begin{equation}
 \# \{ \text{vacua of } T_\text{1d}[M_4]_{\rho,\Phi_\infty} \} \; = \; Z_\VW[M_4]_\rho(q,0)|_\text{coefficient of $q^{\Phi_\infty-\tfrac{c}{24}}$}
\end{equation}

Let us consider ${M_4}=A_{p-1}$ as an example. The lattice $\Gamma$ is even in this case and therefore $\tGamma^*=\Gamma^*$.
As was mentioned earlier, $\Gamma$ and $\Gamma^*$ can be interpreted as the root and weight lattices of $\mathfrak{su}(p)$.
These lattices can be naturally embedded into $\mathbb{R}^{p-1}$, which in turn can be considered as the subspace
of $\mathbb{R}^p$ orthogonal to the vector $(1,\ldots,1)$.
The root lattice can be generated by simple roots satisfying $\|\lambda_i\|^2=2$ and $(\lambda_i,\lambda_{i+1})=-1$.
The weight lattice can be generated by $\omega_r,\; r=1,\ldots, p-1$,
the highest weights of the fundamental representations which can be realized as $\Lambda^r \mathbb{C}^p$.
Let us also define $\omega_0\equiv 0$. In the coset $\Gamma^*/\Gamma\cong \Z_p$ one has $\omega_r\sim r\omega_1$.
For a given boundary condition $r=0,\ldots, p-1$ the flux at infinity has the following form:
\begin{equation}
 \Phi_\infty \; = \; N_{\text{D0}}+\frac{1}{2}\left\|\sum_{i=1}^{p-1}n_i\lambda_i+\omega_r\right\|^2 \,, \qquad n_i\in\Z \,.
 \label{fluxAp}
\end{equation}
The massive vacua of the theory $T_\text{1d}[A_{p-1}]_{\rho,\Phi_\infty}$ correspond
to the weights $w=\sum_{i=1}^{p-1}n_i\lambda_i+\omega_r$ that minimize (\ref{fluxAp}) when $N_\text{D0}=0$.
The set of such weights is precisely the set of weights of the fundamental
representation of $\mathfrak{su}(p)$ with the highest weight $\omega_r$.
Therefore one has:
\begin{equation}
 \# \{ \text{vacua of }T_\text{1d}[A_{p-1}]_r \} \; = \; \dim \Lambda^r \mathbb{C}^p \; = \; \frac{p!}{r!(p-r)!}.
\end{equation}
Up to a permutation, these weights have the following coordinates:
\begin{equation}
 w\underset{S_p}{\sim}(\underbrace{1-\frac{r}{p},\ldots,1-\frac{r}{p}}_{r},\underbrace{-\frac{r}{p},\ldots,-\frac{r}{p}}_{p-r}).
\end{equation}
The minimal value of the flux at infinity equals then
\begin{equation}
 \Phi_\infty \; = \; \frac{(p-r)r}{2p} \,.
\end{equation}

\subsection*{Gluing in the abelian case}

Consider two 4-manifolds (not necessarily connected) $M_4^\pm$, both satisfying \eqref{classM},
with boundaries $\d M_4^\pm=M_3^\pm$. Let us denote $\Gamma_\pm\equiv H_2(M_4^\pm)$ and $T_\pm \equiv H^2(M_3^\pm)\cong H_1(M_3^\pm)$ so that
\begin{equation}
 0\longrightarrow\Gamma_\pm \hookrightarrow \Gamma^*_\pm \stackrel{\pi_\pm}{\longrightarrow} T_\pm \longrightarrow 0.
\end{equation}
Suppose that $M_4^+$ can be obtained from $M_4^-$ by gluing to the latter a certain (co)bordism $B$ with boundary $\d B=-M_3^-\sqcup M_3^+$.

Also, let us suppose that $b_2(B)=0$ and the torsion groups in the long exact sequence (\ref{TTTlong}) for the pair $(B,\d B)$ are $T_2=0$ and $T_1\equiv T$. This means that the only non-trivial cohomology of $B$ and $\d B$ is contained in the following {\it finite} groups:
\begin{eqnarray}
 H_2(B,\d B )\cong H^2(B) & = & T \\
 H_1(B) \cong H^3(B,\d B) & = & T \\
 H_1(\d B)\cong H^2(\d B) & = & T_-\oplus T_+
\end{eqnarray}
The sequence (\ref{TTTlong}) then reduces to the following short exact sequence of finite abelian groups:
\begin{equation}
 \begin{CD}
  0 @>>> T @>{\upsilon=\upsilon_-\oplus\upsilon_+}>>  {T_-\oplus T_+} @>\psi>> T @>>> 0 \\
 \end{CD}
 \label{sesB}
\end{equation}
 Let us denote the family of all such ``basic'' cobordisms by $\basic$. From the Mayer-Vietoris sequence for the pair of manifolds $M_4^-$ and $B$ glued along $M_3^-$ one can deduce the following commutative diagram
\begin{equation}
\begin{CD}
  0 @>>> \Gamma_+^* @>>>  \Gamma_-^*\oplus T @>>> T_- @>>> 0 \\
  @. @VV{\pi_+}V @VV{(\pi_--\upsilon_-)\oplus\upsilon_+}V @VV{\mathrm{id}}V \\
  0 @>>> T_+ @>>> T_-\oplus T_+ @>>> T_- @>>> 0
\end{CD}
\end{equation}
where both horizontal lines form short exact sequences. From the snake lemma it follows that $\Gamma_+=\ker \pi_+$ can be realized as a sublattice of $\Gamma_-^*$:
\begin{multline}
 \Gamma_+=\ker(\pi_--\upsilon_-)\oplus\upsilon_+= \pi^{-1}_-\left[\text{im } \upsilon_-|_{\ker\upsilon_+}\right]=\\
\left\{\alpha\in \Gamma_-^* \;| \;\exists \rho\in T\;\;\text{s.t.}\;\;\alpha\mod \Gamma_-=\upsilon_-(\rho),\;\upsilon_+(\rho)=0\right\}.
\end{multline}

Let us now briefly review the notion of gluing of lattices described in detail in {\it e.g.} \cite{gannon1991gluing}. Consider some integer lattice $\Gamma$ embedded into a Euclidean space and a finite family of \textit{glue vectors} $g_i\in \Gamma^*$. Then one can define the \textit{glued lattice}
\begin{equation}
 \Gamma'=\{\gamma+\sum_in_ig_i\;|\;\gamma\in \Gamma,\;n_i\in \Z\}\subset \Gamma^*.
\end{equation}
The finite abelian group $J\equiv \Gamma'/\Gamma$ is called the \textit{glue group}. It is a subgroup of $\Gamma^*/\Gamma$ generated by the equivalence classes $[g_i]$. As was considered in detail in \cite{gannon1992lattices,gannon1992lattices2}, the gluing operation produces identities on the corresponding theta-functions defined as in (\ref{VWabelian}):
\begin{equation}
 \theta_{\Gamma'}^{(\rho)}=\sum_{\lambda\in J} \theta_{\Gamma}^{(\rho+\lambda)}
 \label{generalgluingthetas}
\end{equation}
One can see that in our case $\Gamma'=\Gamma_+$ is the gluing of $\Gamma=\Gamma_-$ with the glue group
\be
\text{im } \upsilon_-|_{\ker\upsilon_+} \; \subset \; \Gamma^*_-/\Gamma_-
\ee

Since $b_2(B)=0$ the only solutions of (\ref{harmonic}) are given by flat connections. The flat connections on $B$ correspond to the elements of $T= H^2(B)$, while the flat connections on $\d B=-M_3^-\sqcup M_3^+$ are in bijection with the elements of $T_-\oplus T_+$. In the case of an ordinary $U(1)$ gauge theory without Freed-Witten anomaly, the short exact sequence (\ref{sesB}) determines which flat connections on the boundary can be extended to flat connections in the bulk $B$. Namely, a flat connection on the boundary given by $(\mu,\nu)\in H^2(\d B)=T_-\oplus T_+$ originates from a flat connection in $B$ if it is in the image of the map $\upsilon$ or, equivalently, in the kernel of $\psi$. The Vafa-Witten partition function of a cobordism $B\in \basic$ in this case is simply given by
\begin{equation}
 Z_\VW^{U(1)}[B]_{\mu,\nu}=\delta_{\psi(\mu,\nu)}
\end{equation}
where
\begin{equation}
 \delta_\lambda=\left\{\begin{array}{cl}
                  1, & \quad \lambda=0 \\
                  0, & \quad \text{otherwise}
                 \end{array}\right.
\end{equation}

In the case when the $U(1)$ connection is replaced by the $U(1)$ part of the ${Spin}^c(4)$ connection one has to take into account the appropriate shift $\psi_0$:
\begin{equation}
 Z_\VW^{U(1)}[B]_{\mu,\nu}=\delta_{\psi(\mu,\nu)-\psi_0}.
\end{equation}

In the abelian case the ``gluing kernel'' defined in section \ref{sec:VW-gluing} is trivial: $K^{\alpha\beta}[M_3]=\delta^{\alpha\beta}$ (therefore there is no difference between partition functions with upper and lower indices). Then we should have the following relation between the Vafa-Witten partition function on $M_4^+$, $M_4^-$ and $B$, {\it cf.} \eqref{MMpmandB}:
\begin{equation}
 Z_\VW^{U(1)}[M_4^+]_\nu \; = \; \sum_{\mu\in T_-} \, Z_\VW^{U(1)}[B]_{\mu,\nu} \, Z_\VW^{U(1)}[M_4^-]_\mu \,.
\label{gluingthetas}
\end{equation}
Since the abelian Vafa-Witten partition function on an arbitrary four-manifold of the form \eqref{M4KKK} is given by the theta function of the corresponding lattice (\ref{intformlink}), the equation (\ref{gluingthetas}) can be viewed as the identity (\ref{generalgluingthetas}) that relates theta functions of the lattice $\Gamma_-$ to the theta-function of glued lattice $\Gamma_+$.

\subsection*{Composing cobordisms}

\begin{figure}[ht]
\centering
\includegraphics{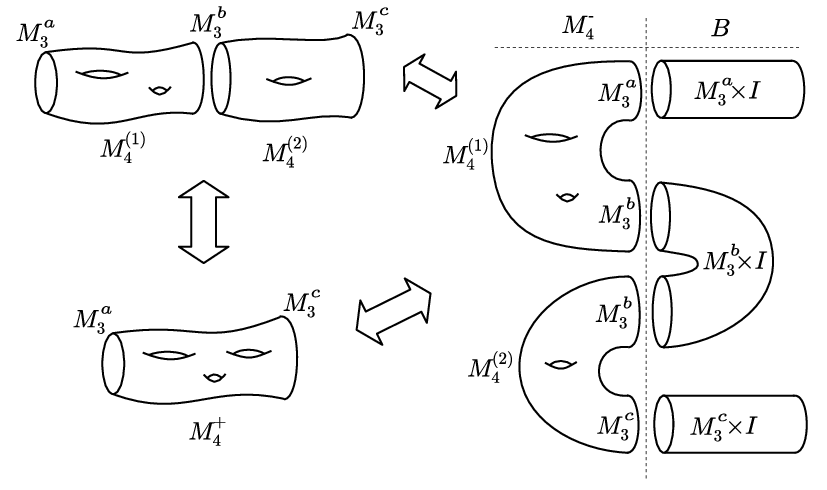}
\caption{Composition of cobordisms $M_4^{(1)}\circ M_4^{(2)} =M_4^+$ can be constructed by gluing $M_4^-=M_4^{(1)}\sqcup M_4^{(2)}$ with a basic cobordism $B\cong M_3^a \times I \sqcup M_3^b\times I \sqcup M_3^c \times I\in \basic$.}
\label{fig:cobordismcomposition}
\end{figure}

Now let us consider two four-manifolds $M_4^{(1)},\,M_4^{(2)}$, both satisfying \eqref{classM}, such that $\d M_4^{(1)}=M_3^a \sqcup M_3^b$ and $\d M_4^{(2)}=M_3^b \sqcup M_3^c$. The 3-manifold $M_3^b$ is supposed to be connected and have an opposite orientation in $M_4^{(1)}$ and $M_4^{(2)}$. The manifolds $M_3^a$ and $M_3^c$ can be empty. Then the new manifold $M^+_4=M_4^{(1)}\cup M_4^{(2)}$ obtained by gluing $M_4^{(1)}$ and $M_4^{(2)}$ along $M_3^b$ also has the properties \eqref{classM}. If we interpret $M_4^{(1)}$ as a cobordism between 3-manifolds $M_3^b$ and $M_3^a$, and $M_4^{(2)}$ as a cobordism between $M_3^c$ and $M_3^b$ then the resulting manifold $M_4^+$ is the composition of these two cobordisms. It is easy to see that this composition is a particular case of gluing described in the previous section. Namely, the manifold $M_4^+$ can be obtained by gluing $M_4^-=M_4^{(1)}\sqcup M_4^{(2)}$ with a basic cobordism, illustrated in Figure \ref{fig:cobordismcomposition},
\be
B \; \cong \; M_3^a \times I \sqcup M_3^b\times I \sqcup M_3^c \times I \; \in \; \basic
\ee
where $I$ is the interval. Let us denote $T^i=H^2(M_3^i)$, where $i=a,b,c$. Then, in the notations of the previous section, we have:
\bea
T & = & T^a\oplus T^b\oplus T^c \nonumber \\
T_- & = & T^a\oplus T^b\oplus T^b\oplus T^c \\
T_+ & = & T^a\oplus T^c \nonumber
\eea
\begin{subequations}
\begin{equation}
 \upsilon_-:\;\lambda\oplus\mu\oplus\nu ~\longmapsto~ \lambda\oplus\mu\oplus(-\mu)\oplus\nu,
\end{equation}
\begin{equation}
 \upsilon_+:\;\lambda\oplus\mu\oplus\nu ~\longmapsto~ \lambda\oplus\nu.
\end{equation}
\end{subequations}
As usual, let us denote $\Gamma_i\equiv H_2(M_4^{(i)})$ and $\Gamma_i^*\equiv H^2(M_4^{(i)})$.
Then, the lattice $\Gamma_+$ is obtained by gluing of $\Gamma_1\oplus\Gamma_2$ with the glue group
\begin{equation}
 T^b\stackrel{\text{diag}}{\hookrightarrow} \Gamma^*_1/\Gamma_1\oplus \Gamma^*_2/\Gamma_2\cong (T^a\oplus T^b)\oplus (T^b\oplus T^c).
\end{equation}
That is
\begin{equation}
 \Gamma^+=\left\{(\alpha+\mu,\beta-\mu)\;|\;\alpha\in\Gamma_1,\;\beta\in\Gamma_2,\;\mu\in T^b\right\}.
\end{equation}
The Vafa-Witten partition functions of the manifolds $M_4^{(1)}$ and $M_4^{(2)}$ are given by:
\begin{subequations}
\begin{equation}
 Z_\VW^{U(1)}[M_4^{(1)}]^\lambda_\mu(q,x) \; = \; \sum_{\alpha \in \Gamma_1}q^{\frac{1}{2}\|\alpha+\lambda+\mu\|^2}x^{\alpha+\lambda+\mu} \,,\qquad
 (\lambda,\mu)\in  T^a\oplus T^b \,,
\end{equation}
\begin{equation}
 Z_\VW^{U(1)}[M_4^{(2)}]^\mu_\nu(q,y) \; = \; \sum_{\beta \in \Gamma_2}q^{\frac{1}{2}\|\beta-\mu+\mu_0+\nu\|^2}y^{\beta+\mu+\nu} \,,\qquad
(\mu,\nu)\in  T^b\oplus T^c \,,
\end{equation}
\end{subequations}
where the boundary condition $\mu$ on the boundary component $M_3^b$ of $M_4^{(1)}$ is identified with the boundary condition $-\mu+\mu_0$ on $M_3^b\subset \d M_4^{(2)}$. The identity (\ref{gluingthetas}) in this case reads as:
\begin{multline}
 \sum_{\mu} \, Z_\VW^{U(1)}[M_4^{(1)}]^\lambda_\mu(q,x) \; Z_\VW^{U(1)}[M_4^{(2)}]^\mu_\nu(q,y) \; = \\
 =\sum_{\alpha \in \Gamma_1,\; \beta \in \Gamma_2,\; \mu } q^{\frac{1}{2}\|\alpha+\lambda+\mu+\delta_1\|^2+\frac{1}{2}\|\beta-\mu+\mu_0+\nu+\Delta_2\|^2}
 x^{\alpha+\lambda+\mu+\Delta_1}y^{\beta-\mu+\mu_0+\nu+\Delta_2}=\\
 =\sum_{\gamma \in \Gamma_+}q^{\frac{1}{2}\|\gamma+(\lambda+\Delta_1)\oplus(\nu+\Delta_2+\mu_0)\|^2}
 (x,y)^{\gamma+(\lambda+\Delta_1)\oplus(\nu+\Delta_2+\mu_0)}=\\
 =\; Z_\VW^{U(1)}[M_4^+]^\lambda_\nu(q,(x,y)) \,,\qquad (\lambda,\nu)\in T^a\oplus T^c \,.
\end{multline}
so that the new shift due to the Freed-Witten anomaly is given by $\Delta=\Delta_1\oplus(\Delta_2+\mu_0)$.

\subsection*{Examples}

Let us denote the 4-manifold associated to the Lie algebra $\mathfrak{g}$ of the $ADE$ type as $M_4(\mathfrak{g})$
and the 4-manifold with the plumbing graph $\Upsilon$ by $M_4(\Upsilon)$, as in section~\ref{sec:plumbing}. For example,
\begin{equation}
 A_{p-1}=M_4(\mathfrak{su}(p))=M_4(\underbrace{\raisebox{-0.7ex}{$\overset{-2}{\bullet}\hspace{-0.7em}-\cdots -\hspace{-0.7em}\overset{-2}{\bullet}$}}_{p-1}),
\end{equation}
\begin{equation}
 \begin{array}{c}
  \CO(-p) \\
  \downarrow \\
  \cp^1
 \end{array}\;=\; M_4(\Op{-p}),
\end{equation}
\begin{equation}
 \underbrace{\bar{\cp}^2\#\ldots\#\bar{\cp}^2}_p \setminus\{\text{pt}\}\;=\; M_4(\underbrace{\Op{-1}\ldots\Op{-1}}_p).
\end{equation}
As was previously mentioned, the lattice $\Gamma$ for the 4-manifold $M_4(\mathfrak{g})$ coincides with the root lattice of $\mathfrak{g}$, while $\Gamma^*$ is given by the corresponding weight lattice. The lattice $\Gamma$ is always even and, therefore, $M_4(\mathfrak{g})$ is Spin and $\Delta=0$. Since level-1 characters are given by theta functions on the root lattice \cite{KacPeterson},
the formula (\ref{VWcharacter0}) with $N=1$,
\begin{equation}
 Z^{U(1)}_\VW[M_4(\mathfrak{g})]_\lambda \; = \; \chi^{\hat{\mathfrak{g}}_1}_\lambda \,,
\label{VWcharacter1}
\end{equation}
also follows from (\ref{VWabelian}). The abelian Vafa-Witten partition function of the $A_{p}$ manifold was studied in detail in \cite{Dijkgraaf:2007fe}.

Let us point out that there is also the following relation between Vafa-Witten partition functions and affine characters:
\begin{equation}
 Z^{U(1)}_\VW[M_4(\Op{-p})]_\lambda(q,x) \; =\; \frac{1}{\eta(\tau)}\sum_{n\in\Z}q^{\frac{1}{2p}(pn+\lambda)^2}x^{pn+\lambda} \; \equiv \; \chi^{\hat{\mathfrak{u}}(1)_p}_\lambda \,,\qquad \lambda\in\Z_p \label{VWcharacter20}
\end{equation}
when $p$ is even. This relation is a natural generalization of (\ref{VWcharacter1}) since the one-dimensional lattice $H^2(M_4(\Op{-p}))$ can be interpreted as a weight lattice of $\hat{\mathfrak{u}}(1)_p$. Let us note that it is also consistent with the fact that $A_1=M_4(\Op{-2})$ since
\begin{equation}
 \chi^{\hat{\mathfrak{su}}(2)_1}_\lambda =\chi^{\hat{\mathfrak{u}}(1)_2}_\lambda.
\end{equation}
For general $p$ one can write
\begin{equation}
 Z^{U(1)}_\VW[M_4(\Op{-p})]_\lambda(q,x) \; = \; \frac{1}{\eta(\tau)}\sum_{n\in\Z}q^{\frac{1}{2p}(pn+\lambda+\Delta)^2}x^{pn+\lambda+\Delta} \; \equiv \; \tilde{\chi}^{\hat{\mathfrak{u}}(1)_p}_\lambda \,,\qquad \lambda\in\Z_p \label{VWcharacter2}
\end{equation}
where $\Delta=0$ if $p$ is even and $\Delta=\tfrac{1}{2}$ if $p$ is odd. Let us call $\tilde{\chi}^{\hat{\mathfrak{u}}(1)_p}$ the ``twisted'' $\hat{\mathfrak{u}}(1)_p$ character.

\begin{figure}[ht]
\centering
\includegraphics[width=2.5in]{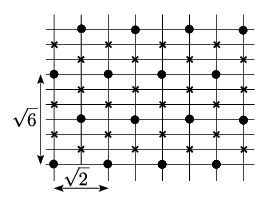}
\caption{Gluing of $A_1$ and $M_4(\Op{-6})$ gives $A_2$.}
\label{fig:su3lattice}
\end{figure}

\begin{figure}[ht]
\centering
\includegraphics[width=2.5in]{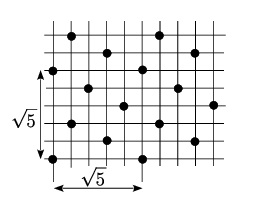}
\caption{Gluing of $M_4(\Op{-5})$ and $M_4(\Op{-5})$ gives $M_4(\Op{-1}\Op{-1})$.}
\label{fig:O5O5lattice}
\end{figure}

In Table \ref{tablegluing} we present various examples of the gluing procedure described earlier.
The corresponding gluings of lattices for many of these (and other) examples can be found in \cite{gannon1992lattices,gannon1992lattices2}. Let us note that in example 3 one can choose the gluing cobordism to be a cylinder with a hole $B=S^3/\Z_p\times I\smallsetminus \text{pt}$, {\it i.e.} one can just glue two components of $M_4^-$ along their boundaries (and then cut out a hole) in order to obtain $M_4^+$. In examples 8, 9 the cobordism $B$ is homologically equivalent to a cylinder with a hole, but not topologically, since the boundaries of $E_{8-n}$ and $A_n$ are only homologically equivalent. Consider example 2 in some detail. In general it is not posible to glue $M_4(\Op{-k})$ with $M_4(\Op{-k})$, because although the boundaries are the same, they do not have opposite orientations. However, when $k=p^2+1$ for some integer $p$ there exists an orientation {\it reversing} diffeomorphism $\varphi$ of $L(k,1)$ such that
\begin{equation}
 \begin{array}{rl}
  \varphi^*:  H^2(L(k,1)) & \longrightarrow H^2(L(k,1))\cong \Z_k \\
  \rho & \longmapsto p\rho
 \end{array}
\end{equation}
It is an automorphism of $\Z_k$ because $p$ and $k=p^2+1$ are coprime. One can also glue $A_{p^2}$ with $A_{p^2}$ using the same prescription (\textit{cf}. example 11). The gluings of lattices in examples 2 and 3 are illustrated in Figures \ref{fig:su3lattice} and~\ref{fig:O5O5lattice}.

\begin{table}[htb]
\centering
{\scriptsize
\begin{tabular}{|c|c|c|c|c|}
 \hline
  \multirow{2}{*}{} & {\bf Original 4-manifold} $M_4^-$ & {\bf End result} $M_4^+$ & \multicolumn{2}{c|}{{\bf Homological data of} $B\in\basic$ ($b_2(B)=0$)} \\
 \cline{4-5}
 & $T_-=H^2(\d M_4^-)$ & $T_+=H^2(\d M_4^+)$ & $T = H^2(B)$ & $\upsilon:T\rightarrow T_-\oplus T_+,\;\;\psi: T_-\oplus T_+ \rightarrow  T$ \\
 \hline
 \hline
 \multirow{2}{*}{1} & $M_4(\Op{-p^2})$ & $M_4(\Op{-1})$ & \multirow{2}{*}{ $\Z_p$ } & $\upsilon(\rho)=p\rho$ \\
 &  $T_-=\Z_{p^2}$ & $T_+=0$ & & $\psi(\mu)=(\mu\mod p)$ \\
 \hline
 \multirow{2}{*}{2} & $M_4(\Op{-p^2-1})\sqcup M_4(\Op{-p^2-1})$ & $M_4(\Op{-1}\Op{-1})$ & \multirow{2}{*}{ $\Z_{p^2+1}$ } & $\upsilon(\rho)=\rho\oplus p\rho$ \\
 &  $T_-=\Z_{p^2+1}$ & $T_+=0$ & & $\psi(\mu\oplus \nu)=(p\mu-\nu)$ \\
 \hline
  \multirow{2}{*}{3} & $A_{p-1}\sqcup M_4(\Op{-p})$ & $M_4(\underbrace{\Op{-1}\ldots\Op{-1}}_p)$ & \multirow{2}{*}{ $\Z_{p}$ } & $\upsilon(\rho)=\rho\oplus \rho$ \\
 &  $T_-=\Z_{p}\oplus \Z_{p}$ & $T_+=0$ & & $\psi(\mu\oplus \nu)=(\mu-\nu)$ \\
 \hline
 \multirow{2}{*}{4} & $A_{p-1}\sqcup M_4(\Op{-p(p+1)})$ & $A_{p}$ & \multirow{2}{*}{ $\Z_{p} \oplus \Z_{p+1}$ } & $\upsilon(\rho\oplus \lambda)=\rho\oplus \rho \oplus \lambda \oplus \lambda$ \\
 &  $T_-=\Z_{p}\oplus \Z_{p} \oplus \Z_{p+1}$ & $T_+=\Z_{p+1}$ & & $\psi(\mu\oplus \nu\oplus \rho \oplus \lambda)=(\mu-\nu)\oplus (\rho-\lambda)$ \\
 \hline
 \multirow{3}{*}{5} & $M_4(\raisebox{-0.7ex}{$\overset{-a_1}{\bullet}\hspace{-1em}-\cdots -\hspace{-1em}\overset{-a_n}{\bullet}$})\sqcup M_4(\Op{-p_np_{n+1}})$ & $M_4(\raisebox{-0.7ex}{$\overset{-a_1}{\bullet}\hspace{-1em}-\cdots -\hspace{-1.5em}\overset{-a_{n+1}}{\bullet}$})$ & \multirow{2}{*}{ $\Z_{p_n}\oplus \Z_{p_{n+1}}$ } & $\upsilon(\rho\oplus \lambda)=\rho\oplus \rho \oplus \lambda \oplus \lambda$ \\
 & where $p_{n+1}=a_np_n-p_{n-1}$ & $T_+=\Z_{p_{n+1}}$ & & $\psi(\mu\oplus \nu\oplus \rho \oplus \lambda)=(\mu-\nu)\oplus (\rho-\lambda)$ \\
 &  $T_-=\Z_{p_n}\oplus \Z_{p_n} \oplus \Z_{p_{n+1}}$  & & &\\
 \hline
\multirow{4}{*}{6} & $A_3\sqcup M_4(\Op{-4})$ & $D_4$ & \multirow{2}{*}{ $\Z_4\oplus \Z_2$ } & $\upsilon(\mu\oplus \nu)$ \\
 &  $T_-=\Z_4\oplus \Z_4 $ & $T_+=\Z_2\oplus\Z_2$ & & $=\mu\oplus (\mu+2\nu) \oplus (\mu\mod 2) \oplus \nu$ \\
 & & & & $\psi(\mu\oplus \nu\oplus \rho \oplus \lambda)=$ \\
 & & & & $(\nu-\mu-2\lambda)\oplus ((\mu\mod 2)-\rho)$ \\
 \hline
  \multirow{2}{*}{7} & $D_8$ & $E_8$ & \multirow{2}{*}{ $\Z_{2}$ } & $\upsilon(\rho)=\rho\oplus 0$ \\
 &  $T_-=\Z_{2}\oplus \Z_{2}$ & $T_+=0$ & & $\psi(\mu\oplus \nu)=\nu$ \\
 \hline
    \multirow{2}{*}{8} & $E_7\sqcup A_1$ & $E_8$ & \multirow{2}{*}{ $\Z_{2}$ } & $\upsilon(\rho)=\rho\oplus \rho$ \\
  &  $T_-=\Z_{2}\oplus \Z_2$ & $T_+=0$ & & $\psi(\mu\oplus \nu)=(\mu-\nu)$ \\
 \hline
    \multirow{2}{*}{9} & $E_6\sqcup A_2$ & $E_8$ & \multirow{2}{*}{ $\Z_{3}$ } & $\upsilon(\rho)=\rho\oplus \rho$ \\
  &  $T_-=\Z_{3}\oplus \Z_3$ & $T_+=0$ & & $\psi(\mu\oplus \nu)=(\mu-\nu)$ \\
 \hline
    \multirow{2}{*}{10} & $A_8$ & $E_8$ & \multirow{2}{*}{ $\Z_{3}$ } & $\upsilon(\rho)=3\rho$ \\
 &  $T_-=\Z_{9}$ & $T_+=0$ & & $\psi(\mu)=(\mu\mod 3)$ \\
 \hline
     \multirow{2}{*}{11} & $A_4\sqcup A_4$ & $E_8$ & \multirow{2}{*}{ $\Z_{5}$ } & $\upsilon(\rho)=\rho\oplus 2\rho$ \\
 &  $T_-=\Z_{5}\oplus\Z_5$ & $T_+=0$ & & $\psi(\mu\oplus \nu)=(2\mu-\nu)$ \\
 \hline
\end{tabular}
}
\caption{Examples of gluing $M_4^-\stackrel{B}{\rightsquigarrow} M_4^+$.}
\label{tablegluing}
\end{table}

Let us consider in some detail the gluing in example 3 when $p$ is even.
This example is rather interesting because both of the original 4-manifolds $A_{p-1}$ and $M_4(\Op{-p})$ are Spin,
but the resulting 4-manifold $M_4(\Op{-1}\ldots\Op{-1})$ is not Spin (since the corresponding lattice $\Z^{p}$ is not even). What is going on here?
The explanation is very instructive and reveals new aspects of the Freed-Witten anomaly in the presence of boundaries.

Each of the original ``pieces'', $A_{p-1}$ and $M_4(\Op{-p})$, admits a unique Spin structure.
However, the restrictions of these Spin structures to the boundary 3-manifold $M_3$,
along which one must glue these pieces in order to produce $M_4(\Op{-1}\ldots\Op{-1})$, are different.
To be a little more precise, as in \eqref{MMpmandB} consider the gluing map between the boundaries:
\be
\varphi: \d A_{p-1} \rightarrow \d M_4(\Op{-p})
\ee
If we introduce Spin structures on $A_{p-1}$ and $M_4(\Op{-p})$, the map $\varphi$ does not lift to a map between the restrictions of the Spin structures on the boundaries. This is why it is not possible to construct a Spin structure on $M_4(\Op{-1}\ldots\Op{-1})$ from the Spin structures on $A_{p-1}$ and $M_4(\Op{-p})$.

Nevertheless, it is possible to lift $\varphi$ to a map between the restrictions of $\text{Spin}^c$ structures on $A_{p-1}$ and $M_4(\Op{-p})$. Since ${Spin}(4)$ holonomies on the boundaries do not match, the holonomies of the $U(1)$ part of ${Spin}^c(4)$ should be identified with $-1$ factor which corresponds to the shift by $\tfrac{p}{2}$ in the $\Z_p$ space of flat connections on the boundaries. One can check that indeed
\begin{multline}
\sum_{\lambda\in \Z_p} \, Z_\VW^{U(1)}[M_4(\Op{-p})]_{\lambda+p/2}(q,x^\perp) \; Z_\VW^{U(1)}[M_4(\Op{-p})]^\lambda(q,x^\parallel) \; = \\
=\; \sum_{\lambda\in \Z_p} \, \chi^{\hat{\mathfrak{u}}(1)_p}_{\lambda+p/2}(q,x^\perp) \, \chi^{\hat{\mathfrak{su}}(p)_1}_\lambda(q,x^\parallel)
\; = \; \tilde{\chi}^{\hat{\mathfrak{u}}(p)_1}(q,x) \; \equiv \; \prod_{i=1}^p \tilde{\chi}^{\hat{\mathfrak{u}}(1)_1}(q,x_i) \; =\\
= \; Z_\VW^{U(1)}[M_4(\underbrace{\Op{-1}\cdots \Op{-1}}_{p})]
\label{upcharacter}
\end{multline}
where the splitting of parameters $x=(x^\perp, x^\parallel)$ is such that $x^\perp=(\prod_i x_i)^{1/p}$.
A version of this relation without shifts due to Freed-Witten anomaly was considered in \cite{Dijkgraaf:2007sw,Dijkgraaf:2007fe}.

In general, a gluing of the form
\begin{equation}
 M_4(\mathfrak{g}^{(1)})\sqcup\ldots \sqcup M_4(\mathfrak{g}^{(n)}) \sqcup M_4(\Op{-p_1})\sqcup\ldots \sqcup M_4(\Op{-p_m}) \quad \stackrel{B}{\rightsquigarrow} \quad M_4(\mathfrak{g})
 \label{algebraiccobordism}
\end{equation}
where all $p_i$ are even, $\mathfrak{g}^{(j)}$ and $\mathfrak{g}$ are of $ADE$ type, corresponds to the embedding of the associated algebras:
\begin{equation}
 \mathfrak{g}^{(1)}_1\oplus \ldots \oplus \mathfrak{g}^{(n)}_1 \oplus \mathfrak{u}(1)_{p_1}\oplus\ldots\oplus \mathfrak{u}(1)_{p_m} \; \subset \; \mathfrak{g}
 \label{generalembedding}
\end{equation}
where the subscripts denote the indices of the embeddings.

Let us recall that the index $\ell$ of the embedding $\mathfrak{k}_\ell\subset \mathfrak{g}$ is defined as the ratio between the normalized Killing form of $\mathfrak{g}$ restricted to the subspace $\mathfrak{k}$ and the normalized Killing form of $\mathfrak{k}$. In other words, the root lattice of $\mathfrak{k}$ is rescaled by the factor of $\sqrt{\ell}$ when embedded into the root lattice of $\mathfrak{g}$. For the corresponding affine Lie algebras, representations of $\hat{\mathfrak{g}}$ at level $k$ decompose into representations of $\hat{\mathfrak{k}}$ at level $\ell k$:
\begin{equation}
 \chi^{\hat{\mathfrak{g}}_k}_\lambda \; = \; \sum_{\mu}\,b_\lambda^\mu \,\chi^{\hat{\mathfrak{k}}_{\ell k}}_\mu \,.
\end{equation}
The coefficients $b_\lambda^\mu$ are called branching functions of the embedding $\mathfrak{k}_\ell\subset \mathfrak{g}$.

If $B\in\basic$, that is $b_2(B)=0$, the total rank on both sides of (\ref{generalembedding}) is the same:
\begin{equation}
 \sum_{i=1}^n\mathrm{rank}\,\mathfrak{g}^{(i)}+m=\mathrm{rank}\,\mathfrak{g}.
 \label{rankcons}
\end{equation}
Then, taking into account (\ref{VWcharacter1}) and (\ref{VWcharacter2}), the identity (\ref{gluingthetas}) can be interpreted as a decomposition of the characters:
\begin{equation}
\boxed{\phantom{\int}
 \chi^{\hat{\mathfrak{g}}_1}_\lambda \; = \; \sum_{\mu,\rho} \, Z_{\VW}^{U(1)}[B]_\lambda^{\mu_1\ldots\mu_n\rho_1\ldots\rho_m}
 \; \chi^{\hat{\mathfrak{g}}^{(1)}_1}_{\mu_1}\cdots\chi^{\hat{\mathfrak{g}}^{(n)}_1}_{\mu_n} \, \chi^{\hat{\mathfrak{u}}(1)_{p_1}}_{\rho_1}\cdots \chi^{\hat{\mathfrak{u}}(1)_{p_m}}_{\rho_m}
 \phantom{\int}}
 \label{branchinggeneral}
\end{equation}
so that the Vafa-Witten partition function of $B$ plays the role of branching functions for the embedding (\ref{generalembedding}) at level 1. As was shown earlier, 
the abelian Vafa-Witten partition function of $B\in \basic$ does not depend on $\tau$. This corresponds to the fact that the embedding (\ref{generalembedding}) is always conformal at level 1.

Now let us define $\tilde{B}$ as $B$ glued with $M_4(\Op{-p_1})\sqcup\ldots \sqcup M_4(\Op{-p_m})$ along the common boundary components. This 4-manifold $\tilde{B}$ is no longer in $\basic$ and has $b_2(\tilde{B})=m$. It can be considered as a cobordism for the following gluing:
\begin{equation}
 M_4(\mathfrak{g}^{(1)})\sqcup\ldots \sqcup M_4(\mathfrak{g}^{(n)}) \quad \stackrel{\tilde{B}}{\rightsquigarrow} \quad M_4(\mathfrak{g}) \,.
 \label{ADEcobordism}
\end{equation}
The identity (\ref{branchinggeneral}) can be rewritten as
\begin{equation}
 \chi^{\hat{\mathfrak{g}}_1}_\lambda \; = \; \sum_{\mu} \, Z_{\VW}^{U(1)}\left[\tilde{B}\right]_\lambda^{\mu_1\ldots\mu_n}
 \, \chi^{\hat{\mathfrak{g}}^{(1)}_1}_{\mu_1}\cdots\chi^{\hat{\mathfrak{g}}^{(n)}_1}_{\mu_n}
 \label{branchingADE}
\end{equation}
and, therefore, $Z_{\VW}^{U(1)}[\tilde{B}]$ plays the role of the level-1 branching functions for the embedding
\begin{equation}
\mathfrak{g}^{(1)}_1\oplus \ldots \oplus \mathfrak{g}^{(n)}_1  \; \subset \; \mathfrak{g}
 \label{ADEembedding}
\end{equation}
where all Lie algebras are of $ADE$ type.


\subsection{Non-abelian generalizations}
\label{sec:naVW}

As was already mentioned in section \ref{sec:affinelie}, the non-abelian generalization of (\ref{VWcharacter1}) is given by
\begin{equation}
 Z_\VW^{U(N)}[M_4(\mathfrak{g})]_\rho \; = \; \chi_\rho^{\hat{\mathfrak{g}}_N}
 \label{VWcharacter11}
\end{equation}
Hence, the Vafa-Witten partition function of a cobordism $\tilde{B}$ in (\ref{ADEcobordism})
should coincide with the branching functions for the embedding (\ref{ADEembedding}) at level $N$:
$$
\boxed{\phantom{\int}
Z_{\VW}^{U(N)}\left[\tilde{B}\right]_\lambda^{\mu_1\ldots\mu_n}  = \;
\text{branching function}~b_\lambda^{\mu_1\ldots\mu_n}
\phantom{\int}}
$$

Since the lattice $H^2(M_4(\Op{-p}))$ is one-dimensional
it is natural to expect that the corresponding Vafa-Witten partition function can be expressed in terms of $\hat{\mathfrak{u}}(1)$ characters.
As a non-abelian generalization of (\ref{VWcharacter2}) one can propose that
\begin{equation}
 Z^{U(N)}_\VW[M_4(\Op{-p})]_\lambda(q,x) \; = \; \sum_{\mu} \, C_\lambda^\mu(q) \, \tilde{\chi}^{\hat{\mathfrak{u}}(1)_{pN}}_\mu(q,x)
 \label{naVWOp}
\end{equation}
with some coefficients $C_\lambda^\mu$ independent of $x$.
This is consistent with the fact that $M_4(\Op{-2})=A_1$ because the characters of $\hat{\mathfrak{su}}(2)$
can be decomposed in terms of the $\hat{\mathfrak{u}}(1)$ characters,
where $\mathfrak{u}(1)$ is embedded as a Cartan subalgebra of $\mathfrak{su}(2)$ with index 2:
\begin{equation}
 Z^{U(N)}_\VW[M_4(\Op{-2})]_\lambda(q,x)=Z^{U(N)}_\VW[A_1]_\lambda(q,x)
 =\chi^{\hat{\mathfrak{su}}(2)_N}_\lambda(q,x)=\sum_{\mu} C_\lambda^\mu(q) \chi^{\hat{\mathfrak{u}}(1)_{2N}}_\mu(q,x)
\end{equation}
Hence, in this case $C^\mu_\lambda$ are the branching functions for the embedding $\mathfrak{u}(1)_2\subset \mathfrak{su}(2)$.
The formula (\ref{naVWOp}) is also in agreement with the results of \cite{Aganagic:2004js}.

{}From (\ref{naVWOp}) and (\ref{VWcharacter11}) it follows that $Z_\VW^{U(N)}[B]$ for the cobordism $B$ in (\ref{algebraiccobordism})
is given, up to coefficients $C$, by level-$N$ characters of the coset for the embedding (\ref{generalembedding}):
\begin{equation}
 \frac{G}{G^{(1)}\times\ldots \times G^{(n)}\times \underbrace{U(1)\times\ldots \times U(1)}_m}.
\end{equation}
Note, such coset spaces are K\"ahler manifolds because of the property (\ref{rankcons}).
This suggests that the corresponding 2d theories $T[B]$ may have a realization
in terms of $(0,2)$ gauged WZW theories studied in \cite{Johnson:1994kv,BJKZ}.

Now let us discuss various consequences and consistency checks of the proposed relation between cobordisms and branching functions.
In \cite{VafaWitten} it was argued that under the blow up of $M_4$ (that is taking the connected sum with $\bar{\cp}^2$)
the $SU(N)$ partition function on $M_4$ is multiplied by the character of $\hat{\mathfrak{su}}(N)_1$:
\begin{equation}
 Z_\VW^{SU(N)}[M_4\#\bar{\cp}^2] \; = \; Z_\VW^{SU(N)}[M_4]\,\chi^{\hat{\mathfrak{su}}(N)_1} \,.
\end{equation}
Based on our experience with abelian theory discussed in the previous section,
it is then natural to propose the following generalization to the case of $U(N)$ gauge group and non-compact 4-manifolds:
\begin{equation}
 Z_\VW^{U(N)}\left[M_4\natural\left(\bar{\cp}^2\setminus\{\text{pt}\}\right)\right](\tau,x)
\; = \; Z_\VW^{U(N)}[M_4](\tau,x^\parallel) \, \tilde{\chi}^{\hat{\mathfrak{u}}(N)_1}(\tau,x^\perp)
 \label{VWblowup}
\end{equation}
where $\natural$ denotes the boundary connected sum, $x=(x^\parallel,x_\perp)$, $x^\parallel\in \exp (H_2(M_4)\otimes\C)$,
and $x^\perp\in\exp (H_2(\bar{\cp}^2\setminus\{\text{pt}\})\otimes\C)\cong \C^*$.
The ``twisted'' $\hat{\mathfrak{u}}(N)_1$ character $\tilde{\chi}^{\hat{\mathfrak{u}}(N)_1}$ is defined as in (\ref{upcharacter}).
The parameter $x\in\C^*$ plays the role of the coordinate along the diagonal $\mathfrak{u}(1)$ of $\mathfrak{u}(N)$,
and the coordinates in the other directions of the Cartan subalgebra are set to zero.
If the manifold $M_4$ is constructed by the plumbing graph $\Upsilon$, the relation (\ref{VWblowup}) looks like
\begin{equation}
 Z_\VW^{U(N)}[M_4(\Upsilon\sqcup\Op{-1})] \; = \; Z_\VW^{U(N)}[M_4(\Upsilon)] \,\tilde{\chi}^{\hat{\mathfrak{u}}(N)_1}.
 \label{VWblowup1}
\end{equation}
In particular:
\begin{equation}
Z_\VW^{U(N)}[M_4(\underbrace{\Op{-1}\cdots \Op{-1}}_{p})] \; = \; \prod_{i=1}^p \tilde{\chi}^{\hat{\mathfrak{u}}(N)_1}(q,x_i).
\label{VWpdots}
 \end{equation}

 Let us note that the ``twisted'' $\hat{\mathfrak{u}}(N)_1$ character is given by
the product of $N$ standard theta-functions with odd characteristics:
\begin{equation}
 \tilde{\chi}^{\hat{\mathfrak{u}}(N)_1}(q,z) \; = \; \prod_{j=1}^N \, \frac{1}{\eta(q)} \, \sum_{n_j\in\Z}q^{\frac{(n_j+1/2)^2}{2}}\,z^{n_j+1/2} \; \equiv \;
 \prod_{j=1}^N\frac{\theta_2(q,z_j)}{\eta(q)} \,.
\end{equation}
Therefore, (\ref{VWpdots}) can be rewritten as
\begin{equation}
Z_\VW^{U(N)}[M_4(\underbrace{\Op{-1}\cdots \Op{-1}}_{p})](q,x) \; = \;
\prod_{i=1}^p \prod_{j=1}^N\frac{\theta_2(q,x_i)}{\eta(q)} \; = \;
 \tilde{\chi}^{\hat{\mathfrak{u}}(Np)_1}(q,x)
 \end{equation}
where the components $x_i$ play the role of the coordinates
in the diagonal directions of $p$ copies of the $\mathfrak{u}(N)$ subalgebra in $\mathfrak{u}(Np)$.
In \cite{Dijkgraaf:2007sw} it was shown that the embedding (which is conformal at level 1)
\begin{equation}
 \mathfrak{su}(N)_p\oplus \mathfrak{u}(1)_{pN}\oplus \mathfrak{su}(p)_N \; \subset \; \mathfrak{u}(Np) \,,
\end{equation}
leads to the following relation between the ``untwisted'' characters:
\begin{multline}
 \prod_{i=1}^p \prod_{j=1}^N\frac{\theta_3(q,x_iy_j)}{\eta(q)} \; \equiv \; \chi^{\hat{\mathfrak{u}}(Np)_1}(q,\{x,y\})
\;  =\\
= \; \sum_{[\lambda]}\sum_{a=1}^{N}\sum_{b=1}^{p}
 \chi^{\hat{\mathfrak{su}}(N)_p}_{\sigma^a_N(\lambda)}(q,y^\parallel)
 \chi^{\hat{\mathfrak{u}}(1)_{Np}}_{|\lambda|+ap+bN}(x^\perp y^\perp)
 \chi^{\hat{\mathfrak{su}}(p)_N}_{\sigma^b_p(\lambda^t)}(q,x^\parallel)
\label{UNpdecomp}
\end{multline}
where $x^\perp=(\prod_ix_i)^N$, $y^\perp=(\prod_jx_j)^p$, $\sigma_N$ and $\sigma_p$ denote the generators of outer automorphisms groups $\Z_N$ and $\Z_p$ of $\hat{\mathfrak{su}}(N)$ and $\hat{\mathfrak{su}}(p)$, respectively, $\lambda$ denotes an integrable representation of $\hat{\mathfrak{su}}(p)_N$ associated to a Young diagram, and $\lambda^t$ denotes an integrable representation of $\hat{\mathfrak{su}}(N)_p$ associate to the transposed Young diagram. The first sum on the right-hand side of this expression is performed over the orbits $[\lambda]$ of $\lambda$ with respect to the action of the outer automorphism group. Finally, $|\lambda|$ stands for the number of boxes in the Young diagram associated to $\lambda$. See \cite{Dijkgraaf:2007sw} for the details.

When $p=1$ and $y=0$, it follows from (\ref{UNpdecomp}) that
\begin{equation}
 Z_\VW^{U(N)}[M_4(\Op{-1})] \; = \; \chi^{\hat{\mathfrak{u}}(N)_1}(q,x) \; = \; \sum_{\lambda}\chi^{\hat{\mathfrak{su}}(N)_1}_\lambda(q,0)
\, \chi^{\hat{\mathfrak{u}}(1)_N}_\lambda(q,x)
\end{equation}
and, therefore, the coefficients $C$ in (\ref{naVWOp}) in the case $p=1$ are given by the characters of $\hat{\mathfrak{su}}(N)_1$.

Now let us consider the example 3 from Table \ref{tablegluing}:
\begin{equation}
 A_{p-1}\sqcup M_4(\Op{-p})\stackrel{B}{\rightsquigarrow} M_4(\underbrace{\Op{-1}\cdots \Op{-1}}_{p}).
\end{equation}
As was mentioned earlier, $B$ is topologically a cylinder with a hole: $B\cong L(p,1)\times I \setminus \{\text{pt}\}$.  One can expect the following identify for the corresponding non-abelian Vafa-Witten partition functions:
\begin{multline}
 Z_\VW^{U(N)}[M_4(\underbrace{\Op{-1}\cdots \Op{-1}}_{p})](q,x) \;= \\
= \; \sum_{\lambda,\mu} \; Z_\VW^{U(N)}[M_4(\Op{-p})]_\lambda(q,x^\perp) \; Z_\VW^{U(N)}[B]^{\lambda,\mu}(q) \; Z_\VW^{U(N)}[A_{p-1}]_{\mu}(q,x^\parallel) \,.
 \label{VWgluingAp}
\end{multline}
Taking into account
\begin{equation}
 Z_\VW^{U(N)}[A_{p-1}]_{\mu}(q,x^\parallel) \; = \; \chi^{\hat{\mathfrak{su}}(p)_N}_\mu(q,x^\parallel)
\end{equation}
combined with (\ref{VWpdots}) and (\ref{naVWOp}), one can interpret (\ref{VWgluingAp}) as
the ``twisted'' version of the identity (\ref{UNpdecomp}) in the case where $y$ is set to zero.



\subsection{Linear plumbings and quiver structure}
\label{sec:contbasis}

{}From example 5 in Table \ref{tablegluing} it follows that one can build
the plumbing $\raisebox{-0.7ex}{$\overset{a_1}{\bullet}\hspace{-0.6em}-\cdots -\hspace{-0.7em}\overset{a_n}{\bullet}$}$ step by step,
attaching one node at a time.
Moreover, as we explained in section \ref{sec:bdrycond}, the boundary 3-manifold is the Lens space,
$M_3(\raisebox{-0.7ex}{$\overset{a_1}{\bullet}\hspace{-0.6em}-\cdots -\hspace{-0.7em}\overset{a_n}{\bullet}$})=L(p_n,q_n)$,
where $p_n/q_n$ is given by the continued fraction \eqref{contfract} associated to the string of integers $(a_1,\ldots,a_n)$.
Therefore, the gluing discussed in sections \ref{sec:addinghandle} and \ref{sec:abelianVW}
\begin{equation}
 M_4(\raisebox{-0.7ex}{$\overset{a_1}{\bullet}\hspace{-0.6em}-\cdots -\hspace{-0.7em}\overset{a_n}{\bullet}$})
 \quad {\rightsquigarrow} \quad
 M_4(\raisebox{-0.7ex}{$\overset{a_1}{\bullet}\hspace{-0.6em}-\cdots -\hspace{-0.7em}\overset{a_n}{\bullet}\hspace{-0.5em}\frac{\hspace{1.7em}}{}\hspace{-0.8em}\overset{a_{n+1}}{\bullet}$})
\end{equation}
can be achieved with a certain cobordism $B^{p_{n+1},q_{n+1}}_{p_n,q_n}$ from the family (\ref{classM}),
which is uniquely determined by the properties
\bea
\d B^{p_{n+1},q_{n+1}}_{p_n,q_n} & = & -L(p_n,q_n) \sqcup L(p_{n+1},q_{n+1}) \\
b_2(B^{p_{n+1},q_{n+1}}_{p_n,q_n}) & = & 1 \nonumber
\eea
The cobordism $B^{p_{n+1},q_{n+1}}_{p_n,q_n}$ can be obtained by joining the cobordism $B$ in example 5 of Table \ref{tablegluing}
with $M_4(\Op{-p_np_{n+1}})$. Let us note that the Lens spaces $L(p,q)$ are homologically equivalent for
different values of $q$ and have $H_1 (L(p,q)) =\Z_p$.
A manifestation of this fact is that the abelian Vafa-Witten partition function
of the cobordism $B^{p',q'}_{p,q}$ depends only on $p$ and $p'$, and is given by
\begin{equation}
 Z_\VW^{U(1)}[B^{p',q'}_{p,q}]^{j'}_j \; = \; \sum_{n\in\mathbb{Z}}q^{\frac{pp'}{2}\left(n-\frac{j}{p}+\frac{j'}{p'}\right)^2}x^{pp'n-p'{j}+p{j'}} \,, \qquad j\in\Z_p,\;j'\in\Z_{p'}
\end{equation}
when $p$ and $p'$ are even.

This gluing procedure can be formally encoded in a quiver diagram where every vertex is labeled by pair of integers.
This quiver can be interpreted as a quiver description of the corresponding 2d theory $T[M_4]$.
A four manifold with $L({p,q})$ boundary has a ``flavor symmetry vertex'' $\boxed{p,q} \,.$
When the cobordism $B^{p',q'}_{p,q}$ is glued to it to produce the $L({p',q'})$ boundary,
we ``gauge'' the $\boxed{p,q}$ vertex with the $\boxed{p,q}$ vertex of the ``bifundamental'' $\boxed{\phantom{!}\!p,q}\frac{\quad}{}\boxed{p',q'}\,.$

Let us illustrate this gluing procedure with an example.
Consider the plumbing $\raisebox{-0.7ex}{$\overset{a_1}{\bullet}\hspace{-0.5em}\frac{\hspace{1.7em}}{}\hspace{-0.5em}\overset{a_{2}}{\bullet}$}$.
 We start with the node $\raisebox{-0.7ex}{$\overset{a_{1}}{\bullet}$}$.
The corresponding manifold $M_4(\raisebox{-0.7ex}{$\overset{a_{1}}{\bullet}$})$ can be considered as
a cobordism from the empty space to $L(a_{1},1)$. Therefore, the quiver associated to it looks like
\begin{equation}
\boxed{\phantom{\int}\phantom{1,1}}\frac{\quad\quad}{}
\boxed{\phantom{\int} a_{1}, ~1 \phantom{\int}}
\end{equation}
The boundary of the space after adding the plumbing
node $\raisebox{-0.7ex}{$\overset{a_{2}}{\bullet}$}$ is another Lens space $L({a_{1}a_{2}-1,a_{2}})$.
This space is obtained by gluing $M_4(\raisebox{-0.7ex}{$\overset{a_{1}}{\bullet}$})$ with $B^{a_{1},1}_{a_{1}a_{2}-1,a_{2}}$.
After ``gauging'' the node $\boxed{a_{1},1}$ we get the quiver
\begin{equation}
\boxed{\phantom{\int}\phantom{1,1}}\frac{\quad\quad}{}
\text{\ovalbox{$\phantom{\int} a_{1}, ~1 \phantom{\int}$}}
\frac{\quad\quad}{}\boxed{\phantom{\int} a_{1}a_{2}-1, ~a_2 \phantom{\int}}
\end{equation}

Clearly, the associated quiver in general depends on the plumbing sequence.
We expect each quiver to give a 2d $\CN=(0,2)$ theory and theories associated to the same plumbing to be dual to each other.
For the purposes of computing $Z_{\VW}$, the ``flavor symmetry node'' stands for a boundary condition label.
``Gauging'' this node means summing over all such labels.

Let us consider in more detail how this works in the case when all $a_i=-2$.
The 4-manifold constructed by the plumbing with $n$ nodes is then $A_n$,
and adding one extra node ({\it cf.} example 4 in Table \ref{tablegluing})
can be realized by the cobordism $B^{n+2,n+1}_{n+1,n}$.
As was explained in section \ref{sec:affinelie}, the relevant ingredients have the form:
\begin{equation}
 Z^{U(N)}_\VW[A_{n+1}]_\rho(q,x) \; = \; \sum_{\lambda} \, Z^{U(N)}_\VW[B^{n+2,n+1}_{n+1,n}]_{\rho}^{\lambda}(q,x^\perp) \, Z^{U(N)}_\VW[A_n]_\lambda(q,x^\parallel) \,,
 \label{Anrec}
\end{equation}
\begin{equation}
 Z^{U(N)}_\VW[A_n]_\lambda \; = \; \chi^{{\hat{\mathfrak{su}}(n+1)_N}}_{\lambda} \,,
 \label{Andiscrete}
\end{equation}
\begin{equation}
 Z^{U(N)}_\VW[B^{n+2,n+1}_{n+1,n}]^{\lambda}_{\rho} \; = \; \chi^{{\hat{\mathfrak{su}}(n+2)_N}/{\hat{\mathfrak{su}}(n+1)_N}}_{\lambda,\rho} \,.
\end{equation}
This suggests that $T[B^{n+2,n+1}_{n+1,n}]$ may have a realization in terms of ${\hat{\mathfrak{su}}(n+2)_N}/\hat{\mathfrak{su}}(n+1)_N$ coset WZW. Direct realization in terms of $(0,2)$ WZW models considered in \cite{Johnson:1994kv,BJKZ} is difficult because the coset space does not have a complex structure. However, as we will show below, it is easy to interpret the Vafa-Witten partition function on $B^{n+2,n+1}_{n+1,n}$ if we make a certain transformation changing {\it discrete} labels associated with boundary conditions to {\it continuous} variables. This transformation can be interpreted as a change of basis in TQFT Hilbert spaces associated to boundaries. Namely, let us define the Vafa-Witten partition function on $A_n$ in the continuous basis as
\begin{equation}
 Z^{U(N)}_\VW[A_{n-1}](q,x|z) \; := \; \sum_{\rho} \, \chi^{\hat{\mathfrak{u}}(N)_{n}}_{\tilde{\rho}}(q,z) \, Z^{U(N)}_\VW[A_{n-1}]_\rho(q,x)
 \label{VWcont}
\end{equation}
where we used that, due to the level-rank duality,
there is a one-to-one correspondence $\rho\leftrightarrow \tilde{\rho}$ between
integrable representations of $\hat{\mathfrak{su}}(n)_N$ and $\hat{\mathfrak{u}}(N)_n$
realized by transposing the corresponding Young diagrams. Namely,
\begin{equation}
 \chi^{\hat{\mathfrak{u}}(N)_{n}}_{\tilde{\rho}}(q,z) \; = \;
  \sum_{a=1}^{N} \,
  \chi^{\hat{\mathfrak{u}}(1)_{Nn}}_{|\rho|+an}(q,z^\perp) \,
 \chi^{\hat{\mathfrak{su}}(N)_n}_{\sigma^a_N(\rho^t)}(q,z^\parallel)
\end{equation}
in the notations of the formula (\ref{UNpdecomp}).

The fugacities $z$ in (\ref{VWcont}) can be interpreted as fugacities for flavor symmetry of $T[M_4]$ associated to the boundary $M_3=\partial M_4$. This symmetry is the gauge symmetry of $T[M_3]$. Gluing two 4-manifolds with along the common boundary $M_3$ corresponds to integrating over $z$, that is gauging the common flavor symmetry associated to $z$. Naively, the fugacities $x$ have different nature since they are assiciated to 2-cycles, not 3-dimensional boundaries. However, one can expect a relation between them since one can always produce a 3-dimensional boundary by excising a tabular neighborhood of a 2-cycle.

It is convenient to introduce the $q$-theta function defined as:
\begin{equation}
 \theta(w;q) \; := \; \prod_{r=0}^\infty (1-q^rw)(1-q^{r+1}/w) \; = \; (w;q)_\infty (q/w;q)_\infty
\end{equation}
where
\begin{equation}
 (w;q)_s \; := \; \prod_{r=0}^{s-1}(1-wq^r)
\end{equation}
is the $q$-Pochhammer symbol. From (\ref{UNpdecomp}) it follows then that in the continuous basis the Vafa-Witten partition
function takes a remarkably simple form:
\begin{equation}
 Z^{U(N)}_\VW[A_{n-1}](q,x|z) \; = \; q^{-\frac{nN}{24}}\prod_{i=1}^{n}\prod_{j=1}^{N}\theta(-q^{\frac{1}{2}}x_{i}z_{j};q)
 \label{Ancont}
\end{equation}
where the fugacities $x$ are represented by $x_i\in\mathbb{C}^*,\;i=1\ldots n$ satisfying $\prod_{i=1}^nx_i=1$.

Now, in the {\it continuous basis}, the right hand side of (\ref{Ancont}) can be interpreted
as the flavored elliptic genus \eqref{indexdef} of $nN$ Fermi multiplets,
possibly with a superpotential (to account for the $q$ shift in the argument).
In \cite{Dijkgraaf:2007sw} the transition from the $\hat{\mathfrak{u}}(Nn)_1$ character
in the right hand side of (\ref{Ancont}) to the $\hat{\mathfrak{su}}(n)_N$ character
in the right hand side of (\ref{Andiscrete}) was interpreted as gauging degerees of freedom
of D4-branes obtained by a compactification of M5-branes.

As we show explicitly in appendix \ref{sec:orthchar} for $N=2$ and conjecture for general $N$,
the characters satisfy the following orthogonality condition:
\begin{equation}
 \oint \frac{dz}{2\pi iz}\; \CI^{U(N)}_V(q,z)\,
 \chi^{\hat{\mathfrak{u}}(N)_{n}}_{{\lambda}}(q,z)\,
 \chi^{\hat{\mathfrak{u}}(N)_{n}}_{{\lambda'}}(q,z) \; = \; C_\lambda(q)\delta_{\lambda,\lambda'}
 \label{UNorth}
\end{equation}
where
\begin{equation}
 \CI^{U(N)}_V(q,z) \; = \; (q;q)_\infty^{2N} \prod_{i\neq j}\theta(z_i/z_j;q)
\end{equation}
is precisely the index \eqref{indexdef} of a 2d $\CN=(0,2)$ vector multiplet for the gauge group $G=U(N)$.
Let us note that the transormation between the continuous basis and the discrete basis
is similar to the transformation considered in \cite{Gadde:2011ik} where ordinary, non-affine characters were used.

If the Vafa-Witten partition function for the cobordism in the continuous basis is defined as
\begin{multline}
 Z^{U(N)}_\VW[B^{n+2,n+1}_{n+1,n}](q,y|z',z) \; =\\= \;
 \sum_{\lambda,\rho}\chi^{\hat{\mathfrak{u}}(N)_{n+2}}_\lambda(q,z') \cdot Z^{U(N)}_\VW[B^{n+2,n+1}_{n+1,n}]^{\lambda}_{\rho}(q,y)
 \cdot \chi^{\hat{\mathfrak{u}}(N)_{n+1}}_\rho(q,z) \cdot C^{-1}_\rho(q)
\end{multline}
the relation (\ref{Anrec}) in the continuous basis should translate into the following property:
\begin{multline}
 Z^{U(N)}_\VW[A_{n+1}](q,\{y^{n+1},x_1/y,\ldots,x_{n+1}/y\}|z') \; =\\
 = \; \oint\prod_{j=1}^N\frac{dz_{j}}{2\pi iz_{j}} \; {\CI}_{V}^{U(N)}(q,z)
 \; Z^{U(N)}_\VW[B^{n+2,n+1}_{n+1,n}](q,y|z',z)
\; Z^{U(N)}_\VW[A_{n}](q,\{x_1,\ldots,x_{n+1}\}|z)
\end{multline}
or, explicitly,
\begin{multline}
 \prod_{j=1}^{N}\Big(\theta(-q^{\frac{1}{2}}y^{n+1}z_{j}';q)\prod_{i=1}^{n+1}\theta(-q^{\frac{1}{2}}x_{i}z_{j}'/y;q)\Big) \; = \\= \;
 \oint\prod_{j=1}^N\frac{dz_{j}}{2\pi iz_{j}}(q;q)_\infty ^{2N} \prod_{i=1}^{n+1}\theta(-q^{\frac{1}{2}}x_{i}z_{j};q)
 \prod_{i\neq j}\theta(z_{i}/z_{j};q) \;
 Z^{U(N)}_\VW[B^{n+2,n+1}_{n+1,n}](q,y|z',z).
 \label{Anreccont}
\end{multline}
The contour prescription is important and we take it to mean as evaluating
the residue of the leading pole. If this is the case, then the following
ansatz for $Z^{U(N)}_\VW[B^{n+2,n+1}_{n+1,n}]$ solves the equation (\ref{Anreccont}):
\begin{equation}
 Z^{U(N)}_\VW[B^{n+2,n+1}_{n+1,n}](q,y|z',z) \; = \; \prod_{j=1}^{N}\theta(-q^{\frac{1}{2}}y^{n+1}z_{j}';q)\prod_{i,j=1}^{N}\frac{1}{\theta(z_{i}'/(z_{j}y);q)} \,.
\end{equation}
The poles of the integral come from the denominator. They are at $z_{i}= {z_{\sigma(i)}'}/{y}$ for some permutation $\sigma$.
After summing over all poles we end up with the desired result. From
the form of the partition function we see that the cobordism corresponds
to the theory of bifundamental chiral multiplets along with a fundamental Fermi multiplet.
The Fermi multiplet itself can be associated to the 2-cycle in the cobordism which increases the second Betti number $b_2$ by $1$.

Following the same reasoning one can deduce the partition function
of the cobordism $B$ transforming $A_{n_{1}-1}\sqcup\ldots\sqcup A_{n_{s}-1}\rightsquigarrow A_{n_{1}+\ldots+n_{s}-1}$.
Consider $s=2$ for simplicity. Then, $Z_\VW^{U(N)}[B]$ must satisfy
\begin{multline}
Z^{U(N)}_\VW[A_{k+l-1}](q,\{y^{l}x_{1},\ldots,x^{l}x_{k},y^{-k}w_{1},\ldots,y^{-k}w_{l}\}|z') \; =\\= \;
\oint\prod_{j=1}^N\frac{dz_{j}}{2\pi iz_{j}}\frac{d\tilde{z}_{j}}{2\pi i\tilde{z}_{j}}
\; {\cal I}_{V}^{U(N)}(q,{z}) \; {\cal I}^{U(N)}_{V}(q,\tilde{z})
\; Z_\VW^{U(N)}[B](q,y|z',z,\tilde{z}) \; \times \\
  \times \; Z_\VW^{U(N)}[A_{k-1}](q,\{x_{1},\ldots,x_{k}\}|z) \; Z_\VW^{U(N)}[A_{l-1}](q,\{w_{1},\ldots,w_{l}\}|z)
\end{multline}
\begin{multline}
\prod_{j=1}^{N}\prod_{i=1}^{k}\theta(-q^{\frac{1}{2}}y^{l}x_{i}z_{j}';q)\prod_{i=1}^{l}\theta(-q^{\frac{1}{2}}x^{-k}w_{i}z_{j}';q) = \\
=\oint\prod_{j=1}^N\frac{dz_{j}}{2\pi iz_{j}}
(q;q)^{2N}_\infty\prod_{i\neq j}\theta(z_{i}/z_{j};q)\prod_{j=1}^{N}\prod_{i=1}^{k}\theta(-q^{\frac{1}{2}}x_{i}z_{j};q)\times\\
\times \oint\frac{d\tilde{z}_{j}}{2\pi i\tilde{z}_{j}}
(q;q)^{2N}_\infty\prod_{i\neq j}\theta(\tilde{z}_{i}/\tilde{z}_{j};q)\prod_{j=1}^{N}\prod_{i=1}^{l}\theta(-q^{\frac{1}{2}}w_{i}\tilde{z}_{j};q)\times \\
\times Z_\VW^{U(N)}[B](q,y|z',z,\tilde{z})
\end{multline}
In this case, the following ansatz solves the equation:
\begin{equation}
 Z_\VW^{U(N)}[B](q,y|z',z,\tilde{z})=\prod_{i,j}\frac{1}{\theta(y^{l}z_{i}'/z_{j};q)}\prod_{i,j}\frac{1}{\theta(y^{-k}z_{i}'/\tilde{z}_{j};q)}.
\end{equation}
As we can see, this is the index of two sets of bifundamental chiral multiplets, {\it cf.} \cite{Gadde:2013wq}.
For a general cobordism $A_{n_{1}-1}\sqcup\ldots\sqcup A_{n_{s}-1}\rightsquigarrow A_{n_{1}+\ldots+n_{s}-1}$,
the corresponding 2d $\CN=(0,2)$ theory is that of $s$ sets of bifundamental chiral multiplets.

\subsection{Handle slides}

Another source of identities on the partition functions is handle slide moves described in section \ref{sec:generalities}.
Consider the following simple example. First, let us note that since $L(p,p-1)\cong L(p,1)$ the cobordism $B$ for
\begin{equation}
 M_4(\Op{-p}) \quad \stackrel{B}{\rightsquigarrow} \quad
M_4(\raisebox{-0.7ex}{$\overset{-p}{\bullet}\hspace{-0.5em}\frac{\hspace{1.7em}}{}\hspace{-0.5em}\overset{-1}{\bullet}$})
\end{equation}
is the same (although we glue along the different component of $\d B$) as for
\begin{equation}
 A_{p-2} \quad \stackrel{B}{\rightsquigarrow} \quad A_{p-1}
\end{equation}
Therefore,
\begin{equation}
  Z^{U(N)}_\VW[B]^{\lambda}_{\rho} \; = \; \chi^{\hat{\mathfrak{su}}(p)_N/\hat{\mathfrak{su}}(p-1)_N}_{\lambda,\rho} \,.
\end{equation}
as we argued in section \ref{sec:affinelie}.
On the other hand, sliding a 2-handle gives the following relation, {\it cf.} \eqref{aaslide}:
\begin{equation}
 M_4(\raisebox{-0.7ex}{$\overset{-p}{\bullet}\hspace{-0.5em}\frac{\hspace{1.7em}}{}\hspace{-0.5em}\overset{-1}{\bullet}$})
\; \cong \; M_4(\Op{-(p-1)}\,\Op{-1}) \,.
\end{equation}
Taking into account (\ref{VWblowup1}) one can expect that
\begin{equation}
\sum_{\rho} \, \chi^{\hat{\mathfrak{su}}(p)_N/\hat{\mathfrak{su}}(p-1)_N}_{\lambda,\rho}
\; Z_\VW^{U(N)}[M_4(\Op{-p})]_\rho
\; = \; \tilde{\chi}^{\hat{\mathfrak{u}}(N)_1} \; Z_\VW^{U(N)}[M_4(\Op{-(p-1)})]_\lambda \,.
\end{equation}

One can consider more complicated handle slides, for example:
\begin{equation}
\overset{-p}{\bullet}\hspace{-0.5em}\frac{\quad}{}\hspace{-0.5em}\overset{-1}{\bullet}\quad
\longrightarrow
\quad\overset{-p}{\bullet}\hspace{-0.5em}\frac{\quad_{-(p-1)}\quad\quad}{}\hspace{-1.5em}\overset{-(p-1)}{\bullet}\quad
\longrightarrow
\quad\overset{-4p+3}{\bullet}\hspace{-1.5em}\frac{\quad\quad_{-2(p-1)}\quad\quad}{}\hspace{-1.5em}\overset{-(p-1)}{\bullet}
\end{equation}
which gives the equation
\begin{eqnarray*}
\sum_{\rho} \; Z_\VW^{U(N)}[B^{p-1,1}_{4p-3,1}]^{\lambda}_{\rho}
\; Z_\VW^{U(N)}[M_4(\Op{-4p+3})]_\rho
\; = \; \tilde{\chi}^{\hat{\mathfrak{u}}(N)_1} \; Z_\VW^{U(N)}[M_4(\Op{-(p-1)})]_\lambda \,.
\end{eqnarray*}


\section{Bottom-up approach: from 2d $(0,2)$ theories to 4-manifolds}
\label{sec:2dtheory}

As explained in section \ref{sec:generalities}, a 4-manifold $M_4$ with boundary $M_3 = \partial M_4$
defines a half-BPS (B-type) boundary condition in a 3d $\CN=2$ theory $T[M_3]$,
such that the boundary degrees of freedom are described by a 2d $\CN=(0,2)$ theory $T[M_4]$.
Similarly, a cobordism between $M_3^-$ and $M_3^+$ corresponds to a wall between
3d $\CN=2$ theories $T[M_3^-]$ and $T[M_3^+]$ or, equivalently (via the ``folding trick''),
to a B-type boundary condition in the theory $T[M_3^+] \times T[-M_3^-]$,~{\it etc.}

Therefore, one natural way to approach the correspondence between 4-manifolds and 2d $(0,2)$ theories $T[M_4]$
is by studying half-BPS boundary conditions in 3d $\CN=2$ theories.
For this, one needs to develop sufficient technology for constructing such boundary conditions,
which will be the goal of the present section.

\subsection{Chiral multiplets and 3d lift of the Warner problem}

The basic building blocks of 3d $\CN=2$ theories, at least those needed for building theories $T[M_3]$,
are matter multiplets (chiral superfields) and gauge multiplets (vector superfields) with various
interaction terms: superpotential terms, Fayet-Illiopoulos terms, Chern-Simons couplings, {\it etc.}

Therefore, we start by describing B-type boundary conditions in a theory of $n$ chiral multiplets
that parametrize a K\"ahler target manifold $X$.
Examples of such boundary conditions were recently studied in \cite{Okazaki:2013kaa}
and will be a useful starting point for our analysis here.
After reformulating these boundary conditions in a more geometric language,
we generalize this analysis in a number of directions by including gauge fields and various interaction terms.

\begin{table}[htb]
\centering
\renewcommand{\arraystretch}{1.3}
\begin{tabular}{|@{\quad}c@{\quad}|@{\quad}c@{\quad}| }
\hline  $\CN=(2,2)$ {\bf supersymmetry} & $\CN=(0,2)$ {\bf supersymmetry}
\\
\hline
\hline vector superfield & Fermi + adjoint chiral \\
(twisted chiral superfield) & $(\Lambda,\Sigma)$ \\
\hline chiral superfield & chiral + Fermi \\
 & $(\Phi,\Psi)$ \\
\hline superpotential & $(0,2)$ superpotential \\
$\CW (\Phi)$ & $J = \frac{\partial \CW}{\partial \Phi}$ \\
\hline charge $q_{\Phi}$ & $E = i \sqrt{2} \, q_{\Phi} \, \Sigma \, \Phi$ \\
\hline
\end{tabular}
\caption{Decomposition of $\CN=(2,2)$ superfields and couplings into $(0,2)$ superfields and couplings.}
\label{tab:SUSY}
\end{table}

In order to describe boundary conditions that preserve $\CN=(0,2)$ supersymmetry on the boundary
it is convenient to decompose 3d $\CN=2$ multiplets into multiplets of 2d $\CN=(0,2)$ supersymmetry algebra, see {\it e.g.} \cite{Wphases}.
Thus, each 3d $\CN=2$ chiral multiplet decomposes into a bosonic 2d $(0,2)$ chiral multiplet $\Phi$
and a fermionic chiral multiplet $\Psi$, as illustrated in Table~\ref{tab:SUSY}.
Then, there are two obvious choices of boundary conditions that either impose Neumann conditions on $\Phi$
and Dirichlet conditions on $\Psi$, or vice versa.
In the first case, the surviving $(0,2)$ multiplet parametrizes a certain holomorphic submanifold $Y \subset X$,
whereas the second choice leads to left-moving fermions that furnish a holomorphic bundle $\CE$ over $Y$.
Put differently, a choice of a K\"ahler submanifold $Y \subset X$ determines a B-type boundary condition
in a 3d $\CN=2$ sigma-model on $X$, such that 2d boundary theory is a $(0,2)$ sigma-model with the target space $Y$
and a holomorphic bundle $\CE = T_{X/Y}$, the normal bundle to $Y$ in $X$:
\be
\left .
\begin{array}{l}
\Phi_i ~:~ \text{Neumann} \\[.1cm]
\Psi_i ~:~ \text{Dirichlet}
\end{array}
\right \}
\quad \Rightarrow \quad
Y \subset X
\ee
\be
\left .
\begin{array}{l}
\Phi_i ~:~ \text{Dirichlet} \\[.1cm]
\Psi_i ~:~ \text{Neumann}
\end{array}
\right \}
\quad \Rightarrow \quad
\CE = T_{X/Y}
\ee
Now let us include superpotential interactions.

\subsection*{3d matrix factorizations}

In general, there are three types of holomorphic couplings in 2d $(0,2)$ theories that play the role of a superpotential.
The first type already appears in the conditions that define bosonic and fermionic chiral multiplets:
\be
\bar D_+ \Phi_i = 0
\quad , \quad
\bar D_+ \Psi_j = \sqrt{2} E_j (\Phi)
\label{Eterms}
\ee
Here, $E_j (\Phi)$ are holomorphic functions of chiral superfields $\Phi_i$.
The second type of holomorphic couplings $J^i (\Phi)$ can be introduced by the following terms in the action
\be
S_J \; = \; \int d^2 x d \theta^+ \, \Psi_i J^i (\Phi) + c.c.
\label{Jterms}
\ee
where, as in the familiar superpotential terms, the integral is over half of the superspace.
In a purely two-dimensional $(0,2)$ theory, supersymmetry requires
\be
\sum_i E_i J^i \; = \; 0
\ee
However, if a 2d $(0,2)$ theory is realized on the boundary of a 3d $\CN=2$ theory that has a superpotential $\CW (\Phi)$,
then the orthogonality condition $E \cdot J = 0$ is modified to
\be
E (\Phi) \cdot J (\Phi) \; = \; \CW (\Phi)
\label{EJW}
\ee
This modification comes from a three-dimensional analog of the ``Warner problem'' \cite{Warner:1995ay},
and reduces to it upon compactification on a circle.
It also leads to a nice class of boundary conditions that are labeled by factorizations
(or, ``matrix factorizations'') of the superpotential $\CW(\Phi)$ and preserve $\CN=(0,2)$ supersymmetry.
For example, a 3d $\CN=2$ theory with a single chiral superfield and a superpotential $\CW = \phi^{k}$
has $k+1$ basic boundary conditions, with $(0,2)$ superpotential terms
\be
J(\phi) = \phi^m \qquad,\qquad E(\phi) = \phi^{k-m} \qquad,\qquad m = 0, \ldots , k
\ee

To introduce the last type of holomorphic ``superpotential'' couplings in $(0,2)$ theories,
we note that in 2d theories with $(2,2)$ supersymmetry there are two types of F-terms:
the superpotential $\CW$ and the twisted superpotential $\tilde \CW$.
In a dimensional reduction from 3d, the latter comes from Chern-Simons couplings.
The distinction between these two types of F-terms is absent in theories with only $(0,2)$ supersymmetry.
In particular, they both correspond to couplings of the form \eqref{Jterms}
with $J = \frac{\partial \CW}{\partial \Phi}$ or $\tilde J = \frac{\partial \tilde \CW}{\partial \Sigma}$,
except in the latter case one really deals with the field-dependent Fayet-Illiopoulos (FI) terms:
\be
S_{FI} \; = \; \int d^2 x d \theta^+ \, \Lambda_{i} \tilde J^{i} (\Sigma,\Phi) + c.c.
\label{FIterms}
\ee
where the Fermi multiplet $\Lambda_{i}$ is the gauge field strength of the $i$-th vector superfield.
The possibility of such holomorphic couplings is very natural from the (mirror) symmetry
between the superpotential and twisted superpotential in $(2,2)$ models.
However, the importance of such terms and, in particular, the fact that they can depend
on {\it charged} chiral fields was emphasized only recently \cite{Melnikov:2012nm}.
The novelty of these models is that classically they are not gauge invariant,
but nevertheless can be saved by quantum effects.
This brings us to our next topic.

\subsection{Anomaly Inflow}
\label{sec:inflow}

Now we wish to explain that not only the coupling of a 2d $\CN=(0,2)$ theory $T[M_4]$
to a 3d $\CN=2$ theory $T[M_3]$ on a half-space is convenient, but in many cases it is also necessary.
In other words, by itself a 2d theory $T[M_4]$ associated to a 4-manifold with boundary may be anomalous.
Such theories, however, do appear as building blocks in our story since the anomaly can be
canceled by inflow from the 3d space-time where $T[M_3]$ lives \cite{Callan:1984sa}.

In this mechanism, the one-loop gauge anomaly generated by fermions in the 2d $(0,2)$ theory $T[M_4]$
is typically balanced against the boundary term picked up
by anomalous gauge variation of the classical Chern-Simons action in 3d $\CN=2$ theory $T[M_3]$.
Essentially the same anomaly cancellation mechanism --- with Chern-Simons action in extra dimensions replaced by a WZW model ---
was used in a wide variety of hybrid $(0,2)$ models \cite{GPS,Johnson:1994kv,BJKZ,DistlerSharpe,AdamsG},
where the chiral fermion anomaly and the classical anomaly of the gauged WZW model were set to cancel each other out.
In particular, our combined 2d-3d system of theories $T[M_4]$ and $T[M_3]$ provides a natural home
to the ``fibered WZW models'' of \cite{DistlerSharpe}, where the holomorphic WZW component is now
interpreted as Chern-Simons theory in extra dimension.

The simplest example --- already considered in this context in \cite{Gadde:2013wq} ---
is an abelian 3d $\CN=2$ Chern-Simons theory at level $k$.
In the presence of a boundary, it has $k$ units of anomaly inflow which must be canceled by
coupling to an ``anomalous heterotic theory''
\be
\partial_{\mu} J^{\mu} \; = \; \frac{\CA_R - \CA_L}{2\pi} \alpha \epsilon^{\mu \nu} F_{\mu \nu}
\ee
whose left-moving and right-moving anomaly coefficients are out of balance by $k$ units:
\be
\CA_R - \CA_L \; = \; k
\label{ALRk}
\ee

\subsection*{Boundary conditions for $\CN=2$ Chern-Simons theories}

In general, there can be several contributions to the anomaly coefficients $\CA_{L,R}$ and, correspondingly,
different ways of meeting the anomaly cancellation condition like \eqref{ALRk}.
In the case of a single $U(1)$ gauge symmetry, there is, of course, a familiar contribiution from
fermions transforming in chiral representations of the gauge group,
\begin{subequations}\label{AAqqlr}
\be
\CA_R \; = \; \sum_{r : \text{chiral}} \tilde q_r^2
\ee
\be
\CA_L \; = \; \sum_{\ell : \text{Fermi}} q_{\ell}^2
\ee
\end{subequations}
where $\tilde q_r$ and $q_{\ell}$ are the charges of $(0,2)$ chiral and Fermi multiplets, respectively.

Besides the chiral anomaly generated by charged Weyl fermions, there can be an additional contribution
to \eqref{ALRk} from field-dependent Fayet-Illiopoulos couplings \eqref{FIterms},
such as ``charged log interactions'':
\be
\tilde J \; = \; \frac{i}{8 \pi} \sum_r N_r \log \left( \Phi_r \right)
\ee
which spoils gauge invariance at the classical level.
As explained in \cite{Melnikov:2012nm} such terms contribute to the anomaly
\be
\Delta \CA_R \; = \; - \sum_{r : \text{chiral}} \tilde q_r N_r
\label{loganomaly}
\ee
and arise from integrating out massive pairs of $(0,2)$ multiplets with unequal charges.
Note the sign difference in (\ref{AAqqlr}a) compared to \eqref{loganomaly}.

This can be easily generalized to a 2d-3d coupled system with gauge symmetry $U(1)^n$.
Namely, let us suppose that 3d $\CN=2$ theory in this combined system contains Chern-Simons interactions
with a matrix of ``level'' coefficients $k_{ij}$,
much like our quiver Chern-Simons theory \eqref{quiverCS} associated to a plumbing graph $\Upsilon$.
And suppose that on a boundary of the 3d space-time it is coupled to some interacting system of
$(0,2)$ chiral and Fermi multipets that, respectively, carry charges $\tilde q_r^i$ and $q_{\ell}^i$
under $U(1)^n$ symmetry, $i = 1, \ldots, n$.
In addition, for the sake of generality we assume that the Lagrangian of the 2d $(0,2)$ boundary theory contains
field-dependent FI terms \eqref{FIterms} with
\be
\tilde J^i \; = \; \frac{i}{8 \pi} \sum_{r} N_r^i \log \left( \Phi_r \right)
\label{logFI}
\ee
Then, the total anomaly cancellation condition for the coupled 2d-3d system --- that combines
all types of contributions \eqref{ALRk}, \eqref{AAqqlr}, and \eqref{loganomaly} --- has the following form:
\be
\sum_{r : \text{chiral}} \tilde q_r^i \tilde q_r^j
- \sum_{\ell : \text{Fermi}} q_{\ell}^i q_{\ell}^j
- \sum_{r : \text{chiral}} \tilde q_r^{(i} N_r^{j)}
\; = \; k_{ij}
\label{UnCSanomaly}
\ee
which must be satisfied for all values of $i,j = 1, \ldots, n$.
Note, that each of the contributions on the left-hand side can be viewed as a ``matrix factorization''
of the matrix of Chern-Simons coefficients.
In particular, the term $\sum \tilde q_r^{(i} N_r^{j)}$ is simply the (symmetrized) product
of the matrix of chiral multiplet charges and the matrix of the boundary superpotential coefficients,
which altogether can be viewed as a ``twisted superpotential version'' of the condition \eqref{EJW},
with \eqref{WquiverCS} and \eqref{logFI}.

Suppose for simplicity that we have a theory of free chiral and Fermi multiplets. The elliptic genus of this theory is simply
\begin{equation}
 \CI(q,x)=\frac{\prod_{\ell : \text{Fermi}}\theta(\prod_i x^{q^i_\ell}_i;q)}{\prod_{r : \text{chiral}}\theta(\prod_i x^{\tilde{q}^i_r}_i;q)}
\end{equation}
In \cite{BDP} it was argued that the right hand side can be interpreted as the ``half-index'' of CS theory, that is the partition function on $S^1\times_q D$ which has boundary $S^1\times_q S^1\cong T^2$ with modulus $\tau$. Following \cite{Gadde:2013wq} one can argue that this theory is equivalent to the quiver CS theory with coefficients $k_{ij}$ living in the half-space on the left of 2d worldvolume. That is, the original 2d-3d system is equivalent to CS theory in the whole space. The relation
\begin{equation}
 k_{ij}=\sum_{r : \text{chiral}} \tilde q_r^i \tilde q_r^j
- \sum_{\ell : \text{Fermi}} q_{\ell}^i q_{\ell}^j
\end{equation}
can be deduced by considering the limit $q\rightarrow 1$ using that $\theta(x;q)\sim \exp\{-(\log x)^2/(2\log q)\}$

Now, one can apply this to 3d $\CN=2$ theories $T[M_3;G]$ that come from fivebranes on 3-manifolds.
Luckily, many of these theories --- even the ones coming from multiple fivebranes,
{\it i.e.} associated with non-abelian $G$ --- admit a purely abelian UV description,
for which \eqref{UnCSanomaly} should suffice.
Hence, using the tools explained here one can match 4-manifolds to specific boundary conditions
that preserve $\CN=(0,2)$ supersymmetry in two dimensions.

\subsection{From boundary conditions to 4-manifolds}
\label{sec:bdrytoM4}

Let us start with boundary conditions that can be described by free fermions.
Clearly, these will give us the simplest examples of 2d $(0,2)$ theories $T[M_4]$,
some of which have been already anticipated from the discussion in the previous sections.

In particular, we expect to find free fermion description of theories $T[M_4 (\Upsilon)]$
for certain plumbing graphs $\Upsilon$.
In the bottom-up approach of the present section, we construct such theories as
boundary conditions in 3d $\CN=2$ theories $T[M_3]$ associated with $M_3 = \partial M_4$.
Thus, aiming to produce a boundary condition for the $\CN=2$ quiver Chern-Simons theory \eqref{quiverCS},
let us associate a symmetry group $U(1)_i$ to every vertex $i \in \Upsilon$ of the plumbing graph.
Similarly, to every edge between vertices ``$i$'' and ``$j$'' we associate a Fermi multiplet
carrying charges $(+1,-1)$ under $U(1)_i \times U(1)_j$.
Then, its contribution to the gauge anomaly \eqref{UnCSanomaly}
is given by the matrix of anomaly coefficients that is non-trivial only in a $2 \times 2$ block
(that corresponds to rows and columns with labels ``$i$'' and ``$j$''):
\be
- \CA_L = \begin{pmatrix}
-1 & ~1 \\
~1 & -1
\end{pmatrix}
\label{bifundFermi}
\ee
To ensure cancellation of the total anomaly, a combination of such contributions
must be set to equal the matrix of Chern-Simons coefficients $k_{ij}$,
which for the quiver Chern-Simons theory \eqref{quiverCS} is given by the symmetric bilinear form \eqref{Qplumbing}.
Therefore, by comparing \eqref{bifundFermi} with \eqref{Qplumbing}, we immediately see that
assigning $U(1)$ factors to vertices of the plumbing graph $\Upsilon$ and ``bifundamental'' charged
Fermi multiplets to edges already accounts for all off-diagonal terms (with $i \ne j$) in the intersection form $Q$.

Also, note that contributions of charged Fermi multiplets to the diagonal elements of the anomaly matrix are always negative,
no matter what combination of contributions \eqref{bifundFermi} or more general charge assignments in \eqref{UnCSanomaly} we take.
This conclusion, of course, relies crucially on the signs in \eqref{UnCSanomaly} and has an important consequence:
only negative definite intersection forms $Q$ can be realized by free Fermi multiplets.

For example, in the case of the $A_n$ plumbing graph shown in Figure~\ref{fig:Anplumbing},
we have $M_3 = L(n+1,n)$, and the $\CN=2$ quiver Chern-Simons theory $T[L(n+1,n);U(1)]$
has matrix of Chern-Simons coefficients of the form \eqref{Qmatrix} with $a_i = -2$, $i = 1, \ldots, n$.
By combining \eqref{bifundFermi} with two extra Fermi multiplets of charges $\pm 1$ under the first and the last $U(1)$ factors,
we can realize the $A_n$ intersection form as the anomaly matrix in the following 2d $\CN=(0,2)$ theory:
\be
T[M_4(A_n);U(1)] \; = \; \text{Fermi multiplets}~\Psi_{\ell = 0, \ldots, n}
\label{TM4Anabel}
\ee
with charges
\be
q (\Psi_{\ell}) \; = \;
\begin{cases}
+1 \text{ under } U(1)_1, & \text{if } \ell=0 \\
(-1,+1) \text{ under } U(1)_{\ell} \times U(1)_{\ell +1}, & \text{if } 1 \le \ell < n \\
-1 \text{ under } U(1)_n, & \text{if } \ell=n
\end{cases}
\label{AnFermicharges}
\ee
Note, the total number of Fermi multiplets in this theory is $n+1$,
which is precisely the number of Taub-NUT centers in the ALE space of type $A_n$.

Let us briefly pause to discuss the structure of the charge matrix
$(q_{\ell}^i)_{\ell = 0, \ldots, n}^{i = 1, \ldots, n}$ in \eqref{AnFermicharges}.
First, it is easy to see that each of the $U(1)^n$ gauge symmetries is ``vector-like''
in a sense that the charges add up to zero for every $U(1)$ factor.
Also note that redefining the charges $q^n_{\ell} \mapsto q^1_{\ell} + 2q^2_{\ell} + 3 q^3_{\ell} + \ldots + nq^n_{\ell}$
for all Fermi multiplets as in \eqref{slinprog} gives a new matrix of charges that, via \eqref{UnCSanomaly},
leads to a new matrix of Chern-Simons coefficients:
\be
Q \; = \; A_{n-1} \oplus \langle -n(n+1) \rangle
\ee
which splits into a matrix of Chern-Simons coefficients for a similar $U(1)^{n-1}$ theory
and an extra $\CN=2$ Chern-Simons term at level $-n(n+1)$.
In this basis we recognize the statement --- explained in section \ref{sec:addinghandle}
through a variant of the ``Norman trick'' \cite{Norman,Quinn} ---
that a sphere plumbing with $\Upsilon = A_n$ can be built from the $A_{n-1}$ sphere plumbing by a cobordism (attaching a 2-handle)
with the intersection form $Q_B = \langle -n(n+1) \rangle$, {\it cf.} \eqref{Ann1bord}.

This observation has a nice physical interpretation in the coupled 2d-3d system
described in section \ref{sec:addinghandle} and illustrated in Figures \ref{fig:2hndlcobordism} and \ref{fig:sequence}.
Namely, the system of Fermi multiplets \eqref{TM4Anabel}--\eqref{AnFermicharges} without $\Psi_{n}$
is simply the 2d $\CN=(0,2)$ theory $T[M_4(A_{n-1});U(1)]$ that can cancel anomaly and
define a consistent boundary condition in the 3d $\CN=2$ Chern-Simons theory $T[M_3(A_{n-1});U(1)]$
associated to the plumbing graph $\Upsilon = A_{n-1}$ by the general rule \eqref{quiverCS}.
In the new basis, the extra $U(1)_{i=n}$ symmetry (which is not gauged in $T[M_3(A_{n-1});U(1)]$)
is, in fact, an axial symmetry under which all $\Psi_{\ell = 0, \ldots, n-1}$ have charge $+1$.
Gauging this symmetry and adding an extra Fermi multiplet that in the new basis has charge $-n$
under $U(1)_{i=n}$ gives precisely the 2d-3d system of 3d $\CN=2$ quiver Chern-Simons theory $T[M_3(A_{n});U(1)]$
coupled to the 2d $\CN=(0,2)$ theory $T[M_4(A_n);U(1)]$ on the boundary.
This way of building $T[M_4(A_n);U(1)]$ corresponds to a fusion of the fully transmisive domain wall that carries $\Psi_{n}$
with a boundary theory $T[M_4(A_{n-1});U(1)]$, as illustrated in Figures \ref{fig:2hndlcobordism} and~\ref{fig:sequence}.

And, last but not least,
in the matrix of charges $(q_{\ell}^i)_{\ell = 0, \ldots, n}^{i = 1, \ldots, n}$ given in \eqref{AnFermicharges}
one can recognize simple roots $\alpha_{i = 1, \ldots, n}$ of the $A_n$ root system.
This suggests immediate generalizations.
For instance, for a 4-manifold \eqref{Qfortrinion} whose plumbing graph $\Upsilon = D_4$ contains a trivalent vertex,
we propose the ``trinion theory'' $T[\perp]$ to be a theory of four Fermi multiplets with the following charge
assignments under the $U(1)^4$ flavor symmetry group:
\be
\begin{array}{ccc}
& \overset{\displaystyle{-2}}{\bullet} & \\
& \vline & \\
\overset{\displaystyle{-2}}{\bullet}
\frac{\phantom{xxx}}{\phantom{xxx}}
& \underset{\displaystyle{-2}}{\bullet} &
\frac{\phantom{xxx}}{\phantom{xxx}}
\overset{\displaystyle{-2}}{\bullet}
\end{array}
\qquad : \qquad
(q_{\ell}^i)_{\text{trinion}} \; = \;
\begin{pmatrix}
1 & -1 & 0 & 0 \\
0 & 1 & -1 & 0 \\
0 & 0 & 1 & -1 \\
0 & 0 & 1 & 1
\end{pmatrix}
\label{qqD4}
\ee
The rows of this matrix are simple roots of the $D_4$ root system associated to the plumbing graph $\Upsilon$,
while the columns are the charge vectors of the Fermi multiplets $\Psi_{\ell = 1, \ldots, 4}$.
Substituting this into \eqref{UnCSanomaly}, we conclude that this 2d trinion theory can precisely cancel the anomaly
of the 3d $\CN=2$ Chern-Simons theory with gauge group $U(1)^4$ and the matrix of Chern-Simons coefficients:
\be
(Q_{ij}) \; = \;
\begin{pmatrix}
-2 & 1 & 0 & 0 \\
1 & -2 & 1 & 1 \\
0 & 1 & -2 & 0 \\
0 & 1 & 0 & -2
\end{pmatrix}
\label{QQD4}
\ee
which equals minus the Cartan matrix of the $D_4$ root system.
This is in complete agreement with our general proposal \eqref{quiverCS}
that $T[M_4 (\Upsilon)]$ defines a consistent, non-anomalous boundary condition for the 3d $\CN=2$ theory $T[M_3(\Upsilon)]$,
which in the present case is simply the quiver Chern-Simons theory defined by the symmetric bilinear form \eqref{Qplumbing}.


In section \ref{sec2:gluing} we saw that $A_{n}$ linear plumbing can be naturally glued to a twisted $D^2$ bundle over $S^2$
with Euler number $-(n+1)$ since they share the same boundary (with opposite orientation, as required for gluing).
In particular, the latter 4-manifold is represented by the Kirby diagram \eqref{single2handle} with $p=n+1$
and has boundary $M_3 = L(n+1,1)$.

The corresponding 3d $\CN=2$ theory $T[L(n+1,1);U(1)]$ was derived in \eqref{TLensU1}:
it is a $U(1)$ Chern-Simons theory at level $-(n+1)$.
This theory can be related to the $U(1)^n$ quiver Chern-Simons theory $T[L(n+1,n);U(1)]$, {\it cf.} \eqref{TpqLensU1},
by a sequence of dualities (3d Kirby moves) described in section \ref{sec2:gluing}.
In particular, this chain of dualities shows that $T[L(n+1,n);U(1)]$ and $T[L(n+1,1);U(1)]$
are related by a parity transformation \eqref{TM3parity}:
\be
T[L(n+1,n)] \; \simeq \;  P \circ T[L(n+1,1)]
\ee
which, of course, is expected to hold for any $G$, not just $G=U(1)$.

Given the explicit description of the 3d $\CN=2$ theory $T[L(n+1,1);U(1)]$,
one can study B-type boundary conditions and try to match those with 4-manifolds bounded by $L(n+1,1)$.
The anomaly cancellation condition \eqref{UnCSanomaly} suggests several possible candidates
for the $(0,2)$ boundary theory $T[M_4]$:
\begin{itemize}

\item[$a)$] $n+1$ Fermi multiplets of charge $\pm 1$ (or, more generally, a collection of Fermi multiplets
whose charges squared add up to $n+1$);

\item[$b)$] a single $(0,2)$ chiral multiplet $\Phi$ of charge $\tilde q_{\Phi} = +1$
and charged log interaction \eqref{logFI} with $N_{\Phi} = n+2$.

\end{itemize}

\subsection*{Non-abelian generalizations and cobordisms}

It is straightforward to extend this discussion to boundary theories and theories $T[M_4;G]$ trapped on walls for non-abelian $G$.
Even if $G$ is non-abelian, theories $T[M_4;G]$ and $T[M_3;G]$ often admit (multiple) UV definitions that only involve abelian gauge fields.
In some cases, however, it is convenient to build $T[M_4;G]$ and $T[M_3;G]$ using non-abelian gauge symmetries.
For instance, the Lens space theory \eqref{TLensUN} proposed in section \ref{sec:bdrycond} is a good example.

In order to accommodate such examples, we need to discuss 2d $(0,2)$ theories with non-abelian gauge symmetries,
which by itself is a very interesting subject that does not appear to be explored in the literature on $(0,2)$ heterotic models.
Specifically, consider a general 2d theory with $(0,2)$ chiral multiplets $\Phi_r$ that transform in representations $\tilde R_r$
of the gauge group $G$ and Fermi multiplets $\Psi_{\ell}$ in representations $R_{\ell}$.
The corresponding fermions couple to the non-abelian gauge field via the usual covariant derivatives,
{\it e.g.} for left-moving fermions in Fermi multiplets we have
$$
(D_{z})_{ij} \; = \; \delta_{ij} \partial_{z} + \sum_a A_{z}^a (T^a_{R_{\ell}})_{ij}
$$
and similarly for chiral multiplets.
Here, $T^a_{R}$ are matrices of size $\dim (R) \times \dim (R)$ that obey the same commutation relations
as the generators $T^a$ of the Lie algebra $\text{Lie} \, (G)$. (The latter correspond to the fundamental representation.)
Then, the anomaly cancellation condition in such a theory has the form, {\it cf.} \eqref{UnCSanomaly},
\be
\sum_{r : \text{chiral}} \Tr [T^a_{\tilde R_r} T^b_{\tilde R_r}]
- \sum_{\ell : \text{Fermi}} \Tr [T^a_{R_{\ell}} T^b_{R_{\ell}}]
\; = \; (k^+ - k^-) \cdot \Tr [T^a T^b]
\label{nonabCSanomaly}
\ee
where, in order to diversify our applications, we now assumed that the inflow from three dimensions
has two contributions, from Chern-Simons couplings at levels $k^+$ and $k^-$, respectively.
This more general form of the anomaly inflow is realized in a 2d $(0,2)$ theory trapped
on a domain wall between 3d $\CN=2$ theories $T[M_3^+]$ and $T[M_3^-]$.

The anomaly cancellation condition \eqref{nonabCSanomaly} can be written more succinctly
by using the index $C(R)$ of a representation $R$ defined via
$\Tr \left( T^a_R T^b_R \right) = C(R) \delta^{ab}$.
For example, for the fundamental and adjoint representations of $G=SU(N)$ we have
$C(\text{fund}) = \frac{1}{2}$ and $C(\text{Adj}) = N$, respectively.
In general,
\be
C(R) \; = \; h_R \frac{\dim (R)}{\dim (\text{Adj})}
\ee
where $h_R$ is the quadratic Casimir of the representation $R$.

Now we can apply \eqref{nonabCSanomaly}, say, to the Lens space theory \eqref{TLensUN}.
We conclude that a domain wall that carries a Fermi multiplet $\Psi$ in the fundamental
representation of $G=U(N)$ changes the level of the $\CN=2$ Chern-Simons theory by one unit,
\be
k^+ - k^- \; = \; -1
\ee
This is consistent with our proposal, based on matching the Vafa-Witten partition function with the superconformal index,
that the cobordism $B$ that relates $A_p$ and $A_{p+1}$ sphere plumbings corresponds to a domain wall which carries 2d $(0,2)$ theory
\be
T[B;U(N)] \; = \; \text{Fermi multiplet $\Psi$ in the fundamental representation}
\ee
The fusion of such domain walls is clearly non-singular and gives
$$
T[M_4(A_p);U(N)] \; = \; p+1~\text{Fermi multiplets $\Psi_{\ell = 0, \ldots, p}$ in $N$-dimn'l representation}
$$
In fact, the wall in this example is fully transmissive.
Notice, as in \eqref{TM4Anabel}, the total number of Fermi multiplets in this theory is greater (by one)
than the number of 2-handles in $M_4$ and equals the number of Taub-NUT centers in the ALE space of type $A_p$.


\section{Future directions}
\label{sec:conclusions}

There are many avenues along which one can continue studying 2d $\CN=(0,2)$ theories $T[M_4]$
labeled by 4-manifolds. The most obvious and/or iteresting items on the list include:

\begin{itemize}

\item
{\bf Examples:}
While focusing on the general structure,
we presented a number of concrete (abeliean and non-abelian) examples of:
$a)$ theories labeled by 4-manifolds and 3-manifolds,
$b)$ dualities that correspond to Kirby moves,
$c)$ relations between cosets and Vafa-Witten partition functions,
and $d)$ B-type walls and boundary conditions in 3d $\CN=2$ theories.
Needless to say, it would certainly be interesting to extend our list of examples in each case.

In particular, it would be interesting to study 2d $\CN=(0,2)$ theories $T[M_4]$
associated with 4-manifolds that are not definite or not simply-connected.
Such examples clearly exist ({\it e.g.} for $M_4 = T^2 \times \Sigma_g$ or $M_4 = K3$,
possibly with ``frozen singularities'' \cite{Witten:1997bs,deBoer:2001px}),
but still remain rather isolated and beg for a more systematic understanding,
similar to theories labeled by a large class of negative definite simply-connected 4-manifolds \eqref{M4KKK} considered in this paper.
Thus, in section~\ref{sec:generalities}
we briefly discussed a natural generalization 
to plumbings of twisted $D^2$ bundles over genus-$g$ Riemann surfaces.
It would be interesting to see what happens to the corresponding theories $T[M_4]$
when Riemann surfaces have boundaries / punctures and to make contact with \cite{Gadde:2011ik}.

\item
{\bf 4-manifolds with corners:}
Closely related to the last remark is the study of 4-manifolds with corners.
Although such situations were encountered at the intermediate stages in section \ref{sec:bdrycond},
we quickly tried to get rid of 3-manifolds with boundaries performing Dehn fillings.
It would be interesting to study whether Vafa-Witten theory admits the structure of extended TQFT
and, if it does, pursue the connection with gluing discussed in section \ref{sec:bdrycond}.

\item
{\bf Smooth structures:}
As was already pointed out in the introduction, it would be interesting to understand what
the existence of a smooth structure on $M_4$ means for the corresponding 2d $\CN=(0,2)$ theory $T[M_4]$.
We plan to tackle this problem by studying surface operators in the fivebrane theory.

\item
{\bf Large $N$ limit:}
It would be interesting to study the large $N$ behavior of the Vafa-Witten partition function on
plumbing 4-manifolds and make contact with holographic duals.

\item
{\bf Non-abelian $(0,2)$ models:}
It appears that not much is known about non-abelian 2d $(0,2)$ gauge dynamics.
While in general abelian (gauge) symmetries suffice for buidling theories $T[M_4]$ and $T[M_3]$,
in sections \ref{sec:bdrycond} and \ref{sec:bdrytoM4} we saw some examples where using non-abelian symmetries is convenient.

\item
{\bf Defect junctions:}
One important property of defect lines and walls is that they can form complicated networks
and foam-like structures. Following the hints from sections \ref{sec:bdrycond}--\ref{sec:cobwalls}
it would be interesting to understand if these play any role in the correspondence between
4-manifolds and 2d $(0,2)$ theories.

\item
{\bf Triangulations:}
Since a basic $d$-dimensional simplex has $d+1$ vertices, the Pachner moves in $d$ dimensions
involve adding one more vertex and then subdiving the resulting $(d+2)$-gon into basic simplices.
In particular, for $d=4$ such subdivisions always give a total of 6 simplices,
resulting in $3-3$ and $2-4$ Pachner moves for 4-manifolds \cite{Mackaay}.
It would be interesting to find a special function
(analogous to the quantum dilogarithm for $2-3$ Pachner moves in case of 3-manifolds)
that enjoys such identities.
Pursuing this approach, however, one should keep in mind that not every 4-manifold can be triangulated.
Examples of non-triangulable 4-manifolds include some natural cases
(such as Freedman's $E_8$ manifold mentioned in the Introduction)
on which the fivebrane theory is expected to be well defined and interesting.

\end{itemize}


\acknowledgments{We thank F.~Quinn, D.~Roggenkamp, C.~Schweigert, A.~Stipsicz and P.~Teichner
for patient and extremely helpful explanations.
We also thank
T.~Dimofte, Y.~Eliashberg, A.~Kapustin, T.~Mrowka, W.~Neumann, T.~Okazaki, E.~Sharpe, C.~Vafa, J.~Walcher and E.~Witten,
among others, for a wide variety of helpful comments. 
The work of A.G. is supported in part by the John A. McCone fellowship and by DOE Grant DE-FG02-92-ER40701.
The work of S.G. is supported in part by DOE Grant DE-FG03-92-ER40701FG-02 and in part by NSF Grant PHY-0757647.
The work of P.P. is supported in part by the Sherman Fairchild scholarship and by NSF Grant  PHY-1050729.
Opinions and conclusions expressed here are those of the authors and do not necessarily reflect the views of funding agencies.}


\appendix

\section{M5-branes on calibrated submanifolds and topological twists}
\label{sec:survey}

\begin{table}[thb]
\renewcommand{\arraystretch}{1.3}
\begin{centering}
{\scriptsize }%
\begin{tabular}{|c|c|c|c|c|c|}
\hline
 & {\scriptsize R-symmetry $SO(5)\supset$} & {\scriptsize Embedding of $M_4$} & {\scriptsize SUSY} & {\scriptsize Solution} & {\scriptsize Metric on $M_4$} \tabularnewline
\hline
\hline
{\scriptsize $a)$} & {\scriptsize $SO(4)\supset SU(2)\times\boxed{SU(2)}$} & {\scriptsize Cayley in $Spin(7)$} & {\scriptsize $(0,1)$} & {\scriptsize $AdS_{3}\times M_4$} & {\scriptsize Conf. half-flat}\tabularnewline
\hline
{\scriptsize $b)$} & {\scriptsize $\boxed{SO(4)}$} & {\scriptsize Lagrangian in $CY_{4}$} & {\scriptsize $(1,1)$} & {\scriptsize $AdS_{3}\times M_4$} & {\scriptsize Const. curvature}\tabularnewline
\hline
{\scriptsize $c)$} & {\scriptsize $SO(2)\times\boxed{SO(3)}$} & {\scriptsize Coassociative in $G_{2}$} & {\scriptsize $(0,2)$} & {\scriptsize $AdS_{3}\times M_4$} & {\scriptsize Conf. half-flat}\tabularnewline
\hline
{\scriptsize $d)$} & {\scriptsize $\boxed{SO(2)}\times\boxed{SO(2)}$} & {\scriptsize K\"ahler in $CY_{4}$} & {\scriptsize $(0,2)$} & {\scriptsize $AdS_{3}\times M_4$} & {\scriptsize K\"ahler-Einstein} \tabularnewline
\hline
{\scriptsize $e)$} & {\scriptsize $SO(4)\supset U(2)\supset\boxed{U(1)}$} & {\scriptsize K\"ahler in $CY_{3}$} & {\scriptsize $(0,4)$} & {\scriptsize $AdS_{3}\times S^{2} \times CY_{3}$} & {\scriptsize K\"ahler-Einstein}\tabularnewline
\hline
{\scriptsize $f)$} & {\scriptsize $SO(4)\supset\boxed{U(2)}$} & {\scriptsize Complex Lagrangian in } & {\scriptsize $(1,2)$} & {\scriptsize $AdS_{3}\times M_4$} & {\scriptsize K\"ahler-Einstein w/}\tabularnewline
 &  & {\scriptsize $d=8$ hyper-K\"ahler} &  &  & {\scriptsize Const. hol. sec. curv.}\tabularnewline
\hline
{\scriptsize $g)$} & {\scriptsize $SO(4)\supset\boxed{SO(2)}\times\boxed{SO(2)}$} & {\scriptsize $(M_2 \subset CY_{2})\times(M_2^{\prime}\subset CY_{2})$} & {\scriptsize $(2,2)$} & {\scriptsize $AdS_{3}\times M_2 \times M_2^{\prime}$} & {\scriptsize Const. curvature}\tabularnewline
\hline
\end{tabular}
\par\end{centering}{\scriptsize \par}

\caption{Supersymmetric M5 brane compactifications on a negatively curved 4-manifold $M_4$.
In the first column we box the subgroup of $SO(5)$ R-symmetry of the M5 brane theory
that is used to twist away the holonomy (or its subgroup) on $M_4$.
Except in the case $e)$, all the $AdS_{3}$ solutions are already found in 7d supergravity
and can be lifted to 11d by fibering $S^{4}$ over $M_4$, see {\it e.g.} \cite{Gauntlett:2000ng,Gauntlett:2001jj,Benini:2013cda}.
In the case $e)$, the solution is found only in 11d supergravity.
For manifolds $M_4$ with general holonomy (but still some restrictions on the metric),
only the compactifications $a)$, $b)$, and $c)$ are allowed.
In this paper, we focus on the case $c)$ as it produces $(0,2)$ superconformal theory in two dimensions.
In this case, $M_4$ is conformally half-flat; see {\it e.g.} \cite{Itohmoduli} for moduli of conformally half-flat structures.}
\label{tab:calibrations}
\end{table}

We study the twisted compactification of 6d $(2,0)$ theory on a four-manifold $M_4$.
In each of the cases listed in Table \ref{tab:calibrations},
such compactification produces a superconformal theory $T[M_4]$ in the two non-compact dimensions.
Via the computation of the $T^2$ partition function explained in the main text,
the cases $a)$, $b)$, and $c)$ correspond to previously studied topological twists of ${\cal N}=4$ super-Yang-Mills
which, in turn, are summarized in Table~\ref{tab:toptwists}.

Specifically,
in the first case $a)$ the ${\cal N}=4$ SYM is thought of as an ${\cal N}=2$ gauge theory
with an extra adjoint multiplet and the Donaldson-Witten twist \cite{Witten:1988ze}.
Its path integral localizes on solutions to the non-abelian monopole equations.
The untwisted rotation group of the DW theory is then twisted by the remaining
$SU(2)$ symmetry to obtain the case $b)$.
This twist (a.k.a. GL twist) was first considered by Marcus \cite{Marcus:1995mq}
and related to the geometric Langlands program in \cite{KW}.
The last case $c)$ is of most interest to us as it corresponds to $(0,2)$ SCFT in 2d.
On a 4-manifold $M_4$, this twist is the standard Vafa-Witten twist \cite{VafaWitten}.

\begin{table}[htb]
\renewcommand{\arraystretch}{1.3}
\begin{centering}
{\scriptsize }%
\begin{tabular}{|c|c|c|c|}
\hline
 & {\scriptsize R symmetry $SO(6)\supset$} & {\scriptsize Name} & {\scriptsize Equations} \tabularnewline
\hline
\hline
{\scriptsize $a)$} & {\scriptsize $SO(2)\times SU(2)\times\boxed{SU(2)}$} & {\scriptsize Donaldson-Witten} & {\scriptsize $F_{\alpha\beta}^{+}+[\bar{M}_{(\alpha},M_{\beta)}]=0$} \tabularnewline
 &  &  & {\scriptsize $D_{\alpha\dot{\alpha}}M^{\alpha}=0$} \tabularnewline
\hline
{\scriptsize $b)$} & {\scriptsize $SO(2)\times\boxed{SU(2)}\times\boxed{SU(2)}$} & {\scriptsize Marcus / GL} & {\scriptsize $F_{\mu\nu}^{+}-i[V_{\mu},V_{\nu}]^{+}=0$} \tabularnewline
 &  &  & {\scriptsize $(D_{[\mu}V_{\nu]})^{-}=0=D_{\mu}V^{\mu}$} \tabularnewline
\hline
{\scriptsize $c)$} & {\scriptsize $SO(3)\times\boxed{SO(3)}$} & {\scriptsize Vafa-Witten} & {\scriptsize $D_{\mu}C+\sqrt{2}D^{\nu}B_{\nu\mu}^{+}=0$} \tabularnewline
 &  &  & {\scriptsize $F_{\mu\nu}^{+}-\frac{i}{2}[B_{\mu\tau}^{+},B_{\:\:\nu}^{+\tau}]-\frac{i}{\sqrt{2}}[B_{\mu\nu}^{+},C]=0$} \tabularnewline
\hline
\end{tabular}
\par\end{centering}{\scriptsize \par}

\caption{Topological twists of ${\cal N}=4$ super-Yang-Mills.}
\label{tab:toptwists}
\end{table}

\section{Orthogonality of affine characters}
\label{sec:orthchar}

The Weyl-Kac formula for affine characters of $\hat{\mathfrak{su}}(2)_{k}$ is
\begin{equation}
 \chi_{\lambda}^{\hat{\mathfrak{su}}(2)_{k}}(q,a)=\frac{\Theta_{\lambda+1}^{(k+2)}(a;q)-\Theta_{-\lambda-1}^{(k+2)}(a;q)}{\Theta_{1}^{(2)}(a;q)-\Theta_{-1}^{(2)}(a;q)}
\end{equation}
where
\begin{equation}
 \Theta_{\lambda}^{(k)}(a;q):=e^{-2\pi ikt}\sum_{n\in\mathbb{Z}+\lambda/2k}q^{kn^{2}}a^{kn}=e^{-2\pi ikt}q^{\frac{\lambda^{2}}{4k}}\sum_{n}q^{kn^{2}+\lambda n}a^{kn+\lambda}
\end{equation}
Using the Weyl-Kac denominator formula the character can be rewritten as
\begin{equation}
 \chi_{\lambda}^{\hat{\mathfrak{su}}(2)_{k}}(q,a)=\frac{e^{-2\pi i(k+2)t}q^{\frac{(\lambda+1)^{2}}{4(k+2)}}\sum_{n}q^{(k+2)n^{2}}a^{(k+2)n}(q^{(\lambda+1)n}a^{(\lambda+1)}-q^{-(\lambda+1)n}a^{-(\lambda+1)})}{a^{-1}(q;q)\theta(a^{2};q)}.
\end{equation}
Consider the integral
\begin{multline}
 \oint\frac{da}{2\pi ia}(q;q)_\infty^{2}\theta(a^{2};q)\theta(a^{-2};q)\chi_{\lambda}^{\hat{\mathfrak{su}}(2)_{k}}(q,a)\chi_{\lambda'}^{\hat{\mathfrak{su}}(2)_{k}}(q,a)\\
  =  e^{-2\pi i(k+2)t}q^{\frac{(\lambda+1)^{2}}{4(k+2)}+\frac{(\lambda'+1)^{2}}{4(k+2)}}\times\\
  \times \sum_{n,m}\left[
    q^{(k+2)(n^{2}+m^{2})+(\lambda+1)n+(\lambda'+1)m}\oint\frac{da}{2\pi ia}a^{(k+2)(n+m)+(\lambda+1)+(\lambda'+1)}\right. \\
  -  q^{(k+2)(n^{2}+m^{2})+(\lambda+1)n-(\lambda'+1)m}\oint\frac{da}{2\pi ia}a^{(k+2)(n-m)+(\lambda+1)-(\lambda'+1)}\\
  -  q^{(k+2)(n^{2}+m^{2})-(\lambda+1)n+(\lambda'+1)m}\oint\frac{da}{2\pi ia}a^{(k+2)(-n+m)-(\lambda+1)+(\lambda'+1)}\\
  +  q^{(k+2)(n^{2}+m^{2})-(\lambda+1)n-(\lambda'+1)m}\oint\frac{da}{2\pi ia}a^{(k+2)(-n-m)-(\lambda+1)-(\lambda'+1)}
  \propto  \delta_{\lambda,\lambda'}
\end{multline}
This shows that $\hat{\mathfrak{su}}(2)_{k}$ characters are orthogonal with respect
to the measure
\begin{equation}
(q;q)_\infty^{2}\theta(a^{2};q)\theta(a^{-2};q)
\end{equation}
but this
measure is exactly the index of $SU(2)$ $(0,2)$ vector multiplet. The orthogonality of $\hat{\mathfrak{u}}(1)_k$ characters can be verified in a similar way. We conjecture that $\hat{\mathfrak{su}}(N)_{k}$ ($\hat{\mathfrak{u}}(N)_{k}$) characters are orthogonal with respect to $SU(N)$ ($U(N)$) vector multiplet measure as well.


\newpage

\bibliographystyle{JHEP_TD}
\bibliography{draft1}

\providecommand{\href}[2]{#2}\begingroup\raggedright\begin{thebibliography}{10%
0}

\bibitem{GompfS}
R.~E. Gompf and A.~I. Stipsicz, {\em {$4$}-manifolds and {K}irby calculus},
  vol.~20 of {\em Graduate Studies in Mathematics}.
\newblock American Mathematical Society, Providence, RI, 1999.

\bibitem{Gaiotto:2008cd}
D.~Gaiotto, G.~W. Moore, and A.~Neitzke, {\it {Four-dimensional wall-crossing
  via three-dimensional field theory}},  {\em Commun.Math.Phys.} {\bf 299}
  (2010) 163--224, [\href{http://xxx.lanl.gov/abs/0807.4723}{{\tt
  arXiv:0807.4723}}].

\bibitem{Gaiotto:2009we}
D.~Gaiotto, {\it {N=2 dualities}},  {\em JHEP} {\bf 1208} (2012) 034,
  [\href{http://xxx.lanl.gov/abs/0904.2715}{{\tt arXiv:0904.2715}}].

\bibitem{Alday:2009aq}
L.~F. Alday, D.~Gaiotto, and Y.~Tachikawa, {\it {Liouville Correlation
  Functions from Four-dimensional Gauge Theories}},  {\em Lett.Math.Phys.} {\bf
  91} (2010) 167--197, [\href{http://xxx.lanl.gov/abs/0906.3219}{{\tt
  arXiv:0906.3219}}].

\bibitem{DGH}
T.~Dimofte, S.~Gukov, and L.~Hollands, {\it {Vortex Counting and Lagrangian
  3-manifolds}},  {\em Lett.Math.Phys.} {\bf 98} (2011) 225--287,
  [\href{http://xxx.lanl.gov/abs/1006.0977}{{\tt arXiv:1006.0977}}].

\bibitem{DGG}
T.~Dimofte, D.~Gaiotto, and S.~Gukov, {\it {Gauge Theories Labelled by
  Three-Manifolds}},  \href{http://xxx.lanl.gov/abs/1108.4389}{{\tt
  arXiv:1108.4389}}.

\bibitem{CCV}
S.~Cecotti, C.~Cordova, and C.~Vafa, {\it Braids, Walls, and Mirrors},
  \href{http://xxx.lanl.gov/abs/1110.2115}{{\tt arXiv:1110.2115}}.

\bibitem{Freedman}
M.~Freedman, {\it {The topology of four dimensional manifolds}},  {\em J. Diff.
  Geom.} {\bf 17} (1982) 357--453.

\bibitem{Green:1987sp}
M.~B. Green, J.~Schwarz, and E.~Witten, {\it {SUPERSTRING THEORY. VOL. 1:
  INTRODUCTION}}, .

\bibitem{Estrings}
J.~Minahan, D.~Nemeschansky, C.~Vafa, and N.~Warner, {\it {E strings and N=4
  topological Yang-Mills theories}},  {\em Nucl.Phys.} {\bf B527} (1998)
  581--623, [\href{http://xxx.lanl.gov/abs/hep-th/9802168}{{\tt
  hep-th/9802168}}].

\bibitem{Donaldson}
S.~K. Donaldson, {\it {An application of gauge theory to four-dimensional
  topology}},  {\em J. Diff. Geom.} {\bf 18} (1983) 279--315.

\bibitem{FQuinn}
F.~Quinn, {\it Ends of maps. {III}. {D}imensions {$4$} and {$5$}},  {\em J.
  Differential Geom.} {\bf 17} (1982), no.~3 503--521.

\bibitem{VafaWitten}
C.~Vafa and E.~Witten, {\it {A Strong coupling test of S duality}},  {\em
  Nucl.Phys.} {\bf B431} (1994) 3--77,
  [\href{http://xxx.lanl.gov/abs/hep-th/9408074}{{\tt hep-th/9408074}}].

\bibitem{Witten}
E.~Witten, {\it {ELLIPTIC GENERA AND QUANTUM FIELD THEORY}},  {\em
  Commun.Math.Phys.} {\bf 109} (1987) 525.

\bibitem{Gadde:2013wq}
A.~Gadde, S.~Gukov, and P.~Putrov, {\it {Walls, Lines, and Spectral Dualities
  in 3d Gauge Theories}},  \href{http://xxx.lanl.gov/abs/1302.0015}{{\tt
  arXiv:1302.0015}}.

\bibitem{Benini:2013nda}
F.~Benini, R.~Eager, K.~Hori, and Y.~Tachikawa, {\it {Elliptic genera of
  two-dimensional N=2 gauge theories with rank-one gauge groups}},
  \href{http://xxx.lanl.gov/abs/1305.0533}{{\tt arXiv:1305.0533}}.

\bibitem{Freed:1999vc}
D.~S. Freed and E.~Witten, {\it {Anomalies in string theory with D-branes}},
  {\em Asian J.Math} {\bf 3} (1999) 819,
  [\href{http://xxx.lanl.gov/abs/hep-th/9907189}{{\tt hep-th/9907189}}].

\bibitem{Gadde:2011ik}
A.~Gadde, L.~Rastelli, S.~S. Razamat, and W.~Yan, {\it {The 4d Superconformal
  Index from q-deformed 2d Yang-Mills}},  {\em Phys.Rev.Lett.} {\bf 106} (2011)
  241602, [\href{http://xxx.lanl.gov/abs/1104.3850}{{\tt arXiv:1104.3850}}].

\bibitem{Bershadsky:1995qy}
M.~Bershadsky, C.~Vafa, and V.~Sadov, {\it {D-branes and topological field
  theories}},  {\em Nucl.Phys.} {\bf B463} (1996) 420--434,
  [\href{http://xxx.lanl.gov/abs/hep-th/9511222}{{\tt hep-th/9511222}}].

\bibitem{Blau:1996bx}
M.~Blau and G.~Thompson, {\it {Aspects of $N(T) \ge 2$ topological gauge
  theories and D-branes}},  {\em Nucl.Phys.} {\bf B492} (1997) 545--590,
  [\href{http://xxx.lanl.gov/abs/hep-th/9612143}{{\tt hep-th/9612143}}].

\bibitem{Acharya:2004qe}
B.~S. Acharya and S.~Gukov, {\it {M theory and singularities of exceptional
  holonomy manifolds}},  {\em Phys.Rept.} {\bf 392} (2004) 121--189,
  [\href{http://xxx.lanl.gov/abs/hep-th/0409191}{{\tt hep-th/0409191}}].

\bibitem{Gauntlett:2000ng}
J.~P. Gauntlett, N.~Kim, and D.~Waldram, {\it {M Five-branes wrapped on
  supersymmetric cycles}},  {\em Phys.Rev.} {\bf D63} (2001) 126001,
  [\href{http://xxx.lanl.gov/abs/hep-th/0012195}{{\tt hep-th/0012195}}].

\bibitem{Gauntlett:2001jj}
J.~P. Gauntlett and N.~Kim, {\it {M five-branes wrapped on supersymmetric
  cycles. 2.}},  {\em Phys.Rev.} {\bf D65} (2002) 086003,
  [\href{http://xxx.lanl.gov/abs/hep-th/0109039}{{\tt hep-th/0109039}}].

\bibitem{Benini:2013cda}
F.~Benini and N.~Bobev, {\it {Two-dimensional SCFTs from wrapped branes and
  c-extremization}},  {\em JHEP} {\bf 1306} (2013) 005,
  [\href{http://xxx.lanl.gov/abs/1302.4451}{{\tt arXiv:1302.4451}}].

\bibitem{AtiyahWard}
M.~F. Atiyah and R.~S. Ward, {\it Instantons and algebraic geometry},  {\em
  Comm. Math. Phys.} {\bf 55} (1977), no.~2 117--124.

\bibitem{Alday:2009qq}
L.~F. Alday, F.~Benini, and Y.~Tachikawa, {\it {Liouville/Toda central charges
  from M5-branes}},  {\em Phys.Rev.Lett.} {\bf 105} (2010) 141601,
  [\href{http://xxx.lanl.gov/abs/0909.4776}{{\tt arXiv:0909.4776}}].

\bibitem{Ganor:1996xg}
O.~J. Ganor, {\it {Compactification of tensionless string theories}},
  \href{http://xxx.lanl.gov/abs/hep-th/9607092}{{\tt hep-th/9607092}}.

\bibitem{Seiberg:1994rs}
N.~Seiberg and E.~Witten, {\it {Electric - magnetic duality, monopole
  condensation, and confinement in N=2 supersymmetric Yang-Mills theory}},
  {\em Nucl.Phys.} {\bf B426} (1994) 19--52,
  [\href{http://xxx.lanl.gov/abs/hep-th/9407087}{{\tt hep-th/9407087}}].

\bibitem{Witten:1994cg}
E.~Witten, {\it {Monopoles and four manifolds}},  {\em Math.Res.Lett.} {\bf 1}
  (1994) 769--796, [\href{http://xxx.lanl.gov/abs/hep-th/9411102}{{\tt
  hep-th/9411102}}].

\bibitem{Rohm:1988yz}
R.~Rohm, {\it {TOPOLOGICAL DEFECTS AND DIFFERENTIAL STRUCTURES}},  {\em Annals
  Phys.} {\bf 189} (1989) 223.

\bibitem{Asselmeyer:1996bh}
T.~Asselmeyer, {\it {Generation of source terms in general relativity by
  differential structures}},  {\em Class.Quant.Grav.} {\bf 14} (1997) 749--758,
  [\href{http://xxx.lanl.gov/abs/gr-qc/9610009}{{\tt gr-qc/9610009}}].

\bibitem{Pfeiffer:2004pe}
H.~Pfeiffer, {\it {Quantum general relativity and the classification of smooth
  manifolds}},  \href{http://xxx.lanl.gov/abs/gr-qc/0404088}{{\tt
  gr-qc/0404088}}.

\bibitem{Sladkowski:2009dc}
J.~Sladkowski, {\it {Exotic Smoothness and Astrophysics}},  {\em Acta
  Phys.Polon.} {\bf B40} (2009) 3157--3163,
  [\href{http://xxx.lanl.gov/abs/0910.2828}{{\tt arXiv:0910.2828}}].

\bibitem{Akbulut}
S.~Akbulut, {\em 4-Manifolds}.
\newblock 2012.

\bibitem{LaudenbachP}
F.~Laudenbach and V.~Po{\'e}naru, {\it A note on {$4$}-dimensional
  handlebodies},  {\em Bull. Soc. Math. France} {\bf 100} (1972) 337--344.

\bibitem{Harer}
J.~L. Harer, {\em P{ENCILS} {OF} {CURVES} {ON} 4-{MANIFOLDS}}.
\newblock ProQuest LLC, Ann Arbor, MI, 1979.
\newblock Thesis (Ph.D.)--University of California, Berkeley.

\bibitem{DGGindex}
T.~Dimofte, D.~Gaiotto, and S.~Gukov, {\it {3-Manifolds and 3d Indices}},
  \href{http://xxx.lanl.gov/abs/1112.5179}{{\tt arXiv:1112.5179}}.

\bibitem{FGS}
H.~Fuji, S.~Gukov, P.~Sulkowski, and H.~Awata, {\it {Volume Conjecture: Refined
  and Categorified}},  \href{http://xxx.lanl.gov/abs/1203.2182}{{\tt
  arXiv:1203.2182}}.

\bibitem{FGP}
H.~Fuji, S.~Gukov, and P.~Sulkowski, {\it {Super-A-polynomial for knots and BPS
  states}},  {\em Nucl.Phys.} {\bf B867} (2013) 506--546,
  [\href{http://xxx.lanl.gov/abs/1205.1515}{{\tt arXiv:1205.1515}}].

\bibitem{KW}
A.~Kapustin and E.~Witten, {\it {Electric-Magnetic Duality And The Geometric
  Langlands Program}},  {\em Commun.Num.Theor.Phys.} {\bf 1} (2007) 1--236,
  [\href{http://xxx.lanl.gov/abs/hep-th/0604151}{{\tt hep-th/0604151}}].

\bibitem{Blau:1997pp}
M.~Blau and G.~Thompson, {\it {Euclidean SYM theories by time reduction and
  special holonomy manifolds}},  {\em Phys.Lett.} {\bf B415} (1997) 242--252,
  [\href{http://xxx.lanl.gov/abs/hep-th/9706225}{{\tt hep-th/9706225}}].

\bibitem{Apol}
S.~Gukov, {\it Three-Dimensional Quantum Gravity, Chern-Simons Theory, and the
  A-Polynomial},  {\em Commun. Math. Phys.} {\bf 255} (2005), no.~3 577--627,
  [\href{http://xxx.lanl.gov/abs/hep-th/0306165}{{\tt hep-th/0306165}}].

\bibitem{GukovRTN}
S.~Gukov, {\it {Gauge theory and knot homologies}},  {\em Fortsch.Phys.} {\bf
  55} (2007) 473--490, [\href{http://xxx.lanl.gov/abs/0706.2369}{{\tt
  arXiv:0706.2369}}].

\bibitem{Saveliev}
N.~Saveliev, {\it Fukumoto-{F}uruta invariants of plumbed homology 3-spheres},
  {\em Pacific J. Math.} {\bf 205} (2002), no.~2 465--490.

\bibitem{Kapustin:2010hk}
A.~Kapustin and N.~Saulina, {\it {Topological boundary conditions in abelian
  Chern-Simons theory}},  {\em Nucl.Phys.} {\bf B845} (2011) 393--435,
  [\href{http://xxx.lanl.gov/abs/1008.0654}{{\tt arXiv:1008.0654}}].

\bibitem{Nakajima}
H.~Nakajima, {\it Instantons on ALE spaces, quiver varieties, and Kac-Moody
  Algebras},  {\em Duke Math.} {\bf 76} (1994) 365--416.

\bibitem{Acharya:2001dz}
B.~S. Acharya and C.~Vafa, {\it {On domain walls of N=1 supersymmetric
  Yang-Mills in four-dimensions}},
  \href{http://xxx.lanl.gov/abs/hep-th/0103011}{{\tt hep-th/0103011}}.

\bibitem{HananyW}
A.~Hanany and E.~Witten, {\it {Type IIB superstrings, BPS monopoles, and
  three-dimensional gauge dynamics}},  {\em Nucl.Phys.} {\bf B492} (1997)
  152--190, [\href{http://xxx.lanl.gov/abs/hep-th/9611230}{{\tt
  hep-th/9611230}}].

\bibitem{Gaiotto:2008sa}
D.~Gaiotto and E.~Witten, {\it {Supersymmetric Boundary Conditions in N=4 Super
  Yang-Mills Theory}},  {\em J.Statist.Phys.} {\bf 135} (2009) 789--855,
  [\href{http://xxx.lanl.gov/abs/0804.2902}{{\tt arXiv:0804.2902}}].

\bibitem{Kitao:1998mf}
T.~Kitao, K.~Ohta, and N.~Ohta, {\it {Three-dimensional gauge dynamics from
  brane configurations with (p,q) - five-brane}},  {\em Nucl.Phys.} {\bf B539}
  (1999) 79--106, [\href{http://xxx.lanl.gov/abs/hep-th/9808111}{{\tt
  hep-th/9808111}}].

\bibitem{BHKK}
O.~Bergman, A.~Hanany, A.~Karch, and B.~Kol, {\it {Branes and supersymmetry
  breaking in three-dimensional gauge theories}},  {\em JHEP} {\bf 9910} (1999)
  036, [\href{http://xxx.lanl.gov/abs/hep-th/9908075}{{\tt hep-th/9908075}}].

\bibitem{Ohta}
K.~Ohta, {\it {Supersymmetric index and s rule for type IIB branes}},  {\em
  JHEP} {\bf 9910} (1999) 006,
  [\href{http://xxx.lanl.gov/abs/hep-th/9908120}{{\tt hep-th/9908120}}].

\bibitem{Witten:2003ya}
E.~Witten, {\it {SL(2,Z) action on three-dimensional conformal field theories
  with Abelian symmetry}},  \href{http://xxx.lanl.gov/abs/hep-th/0307041}{{\tt
  hep-th/0307041}}.

\bibitem{Witten:1999ds}
E.~Witten, {\it {Supersymmetric index of three-dimensional gauge theory}},
  \href{http://xxx.lanl.gov/abs/hep-th/9903005}{{\tt hep-th/9903005}}.

\bibitem{Smilga}
A.~Smilga, {\it {Witten index in supersymmetric 3d theories revisited}},  {\em
  JHEP} {\bf 1001} (2010) 086, [\href{http://xxx.lanl.gov/abs/0910.0803}{{\tt
  arXiv:0910.0803}}].

\bibitem{Witten:1993xi}
E.~Witten, {\it {The Verlinde algebra and the cohomology of the Grassmannian}},
   \href{http://xxx.lanl.gov/abs/hep-th/9312104}{{\tt hep-th/9312104}}.

\bibitem{Kapustin:2013hpk}
A.~Kapustin and B.~Willett, {\it {Wilson loops in supersymmetric
  Chern-Simons-matter theories and duality}},
  \href{http://xxx.lanl.gov/abs/1302.2164}{{\tt arXiv:1302.2164}}.

\bibitem{Gukov:2015sna}
S.~Gukov and D.~Pei, {\it {Equivariant Verlinde formula from fivebranes and
  vortices}},  \href{http://xxx.lanl.gov/abs/1501.0131}{{\tt arXiv:1501.0131}}.

\bibitem{Nawata}
S.~Nawata, P.~Ramadevi, Zodinmawia, and X.~Sun, {\it {Super-A-polynomials for
  Twist Knots}},  {\em JHEP} {\bf 1211} (2012) 157,
  [\href{http://xxx.lanl.gov/abs/1209.1409}{{\tt arXiv:1209.1409}}].

\bibitem{FGSS}
H.~Fuji, S.~Gukov, M.~Stosic, and P.~Sulkowski, {\it {3d analogs of
  Argyres-Douglas theories and knot homologies}},
  \href{http://xxx.lanl.gov/abs/1209.1416}{{\tt arXiv:1209.1416}}.

\bibitem{Dimofte:2013iv}
T.~Dimofte, M.~Gabella, and A.~B. Goncharov, {\it {K-Decompositions and 3d
  Gauge Theories}},  \href{http://xxx.lanl.gov/abs/1301.0192}{{\tt
  arXiv:1301.0192}}.

\bibitem{Wong:1994np}
E.~Wong and I.~Affleck, {\it {Tunneling in quantum wires: A Boundary conformal
  field theory approach}},  {\em Nucl.Phys.} {\bf B417} (1994) 403--438.

\bibitem{Oshikawa:1996dj}
M.~Oshikawa and I.~Affleck, {\it {Boundary conformal field theory approach to
  the critical two-dimensional Ising model with a defect line}},  {\em
  Nucl.Phys.} {\bf B495} (1997) 533--582,
  [\href{http://xxx.lanl.gov/abs/cond-mat/9612187}{{\tt cond-mat/9612187}}].

\bibitem{Bachas:2001vj}
C.~Bachas, J.~de~Boer, R.~Dijkgraaf, and H.~Ooguri, {\it {Permeable conformal
  walls and holography}},  {\em JHEP} {\bf 0206} (2002) 027,
  [\href{http://xxx.lanl.gov/abs/hep-th/0111210}{{\tt hep-th/0111210}}].

\bibitem{Hori:2004zd}
K.~Hori and J.~Walcher, {\it {D-branes from matrix factorizations}},  {\em
  Comptes Rendus Physique} {\bf 5} (2004) 1061--1070,
  [\href{http://xxx.lanl.gov/abs/hep-th/0409204}{{\tt hep-th/0409204}}].

\bibitem{Brunner:2007qu}
I.~Brunner and D.~Roggenkamp, {\it {B-type defects in Landau-Ginzburg models}},
   {\em JHEP} {\bf 0708} (2007) 093,
  [\href{http://xxx.lanl.gov/abs/0707.0922}{{\tt arXiv:0707.0922}}].

\bibitem{Brunner:2008fa}
I.~Brunner, H.~Jockers, and D.~Roggenkamp, {\it {Defects and D-Brane
  Monodromies}},  {\em Adv.Theor.Math.Phys.} {\bf 13} (2009) 1077--1135,
  [\href{http://xxx.lanl.gov/abs/0806.4734}{{\tt arXiv:0806.4734}}].

\bibitem{Carqueville:2010hu}
N.~Carqueville and I.~Runkel, {\it {Rigidity and defect actions in
  Landau-Ginzburg models}},  {\em Commun.Math.Phys.} {\bf 310} (2012) 135--179,
  [\href{http://xxx.lanl.gov/abs/1006.5609}{{\tt arXiv:1006.5609}}].

\bibitem{Kapustin:2010if}
A.~Kapustin and N.~Saulina, {\it {Surface operators in 3d Topological Field
  Theory and 2d Rational Conformal Field Theory}},
  \href{http://xxx.lanl.gov/abs/1012.0911}{{\tt arXiv:1012.0911}}.

\bibitem{Fuchs:2012dt}
J.~Fuchs, C.~Schweigert, and A.~Valentino, {\it {Bicategories for boundary
  conditions and for surface defects in 3-d TFT}},
  \href{http://xxx.lanl.gov/abs/1203.4568}{{\tt arXiv:1203.4568}}.

\bibitem{Giveon:2008zn}
A.~Giveon and D.~Kutasov, {\it {Seiberg Duality in Chern-Simons Theory}},  {\em
  Nucl.Phys.} {\bf B812} (2009) 1--11,
  [\href{http://xxx.lanl.gov/abs/0808.0360}{{\tt arXiv:0808.0360}}].

\bibitem{Quella:2002ct}
T.~Quella and V.~Schomerus, {\it {Symmetry breaking boundary states and defect
  lines}},  {\em JHEP} {\bf 0206} (2002) 028,
  [\href{http://xxx.lanl.gov/abs/hep-th/0203161}{{\tt hep-th/0203161}}].

\bibitem{Bachas:2009mc}
C.~Bachas and S.~Monnier, {\it {Defect loops in gauged Wess-Zumino-Witten
  models}},  {\em JHEP} {\bf 1002} (2010) 003,
  [\href{http://xxx.lanl.gov/abs/0911.1562}{{\tt arXiv:0911.1562}}].

\bibitem{Dijkgraaf:2007sw}
R.~Dijkgraaf, L.~Hollands, P.~Sulkowski, and C.~Vafa, {\it {Supersymmetric
  gauge theories, intersecting branes and free fermions}},  {\em JHEP} {\bf
  0802} (2008) 106, [\href{http://xxx.lanl.gov/abs/0709.4446}{{\tt
  arXiv:0709.4446}}].

\bibitem{austin1990}
D.~M. Austin, {\it $SO(3)$-instantons on $L(p,q)\times \mathbf{R}$},  {\em
  Journal of Differential Geometry} {\bf 32} (1990), no.~2 383--413.

\bibitem{furuta1990invariant}
M.~Furuta and Y.~Hashimoto, {\it Invariant instantons on $S^4$},  {\em Journal
  of The Faculty of Science, The University of Tokyo, Section IA, Mathematics}
  {\bf 37} (1990), no.~3 585--600.

\bibitem{Norman}
R.~A. Norman, {\it Dehn's lemma for certain {$4$}-manifolds},  {\em Invent.
  Math.} {\bf 7} (1969) 143--147.

\bibitem{Quinn}
F.~Quinn, {\it Ends of maps. {I}},  {\em Ann. of Math. (2)} {\bf 110} (1979),
  no.~2 275--331.

\bibitem{harmonic1}
M.~F. Atiyah, V.~Patodi, and I.~Singer, {\it Spectral asymmetry and Riemannian
  geometry},  {\em Bull. London Math. Soc} {\bf 5} (1973), no.~2 229--234.

\bibitem{harmonic2}
R.~Lockhart, {\it Fredholm, Hodge and Liouville theorems on noncompact
  manifolds},  {\em Transactions of the American Mathematical Society} {\bf
  301} (1987), no.~1 1--35.

\bibitem{Gukov:1999ya}
S.~Gukov, C.~Vafa, and E.~Witten, {\it {CFT's from Calabi-Yau four folds}},
  {\em Nucl.Phys.} {\bf B584} (2000) 69--108,
  [\href{http://xxx.lanl.gov/abs/hep-th/9906070}{{\tt hep-th/9906070}}].

\bibitem{Gukov:2002zg}
S.~Gukov, J.~Sparks, and D.~Tong, {\it {Conifold transitions and five-brane
  condensation in M theory on spin(7) manifolds}},  {\em Class.Quant.Grav.}
  {\bf 20} (2003) 665--706, [\href{http://xxx.lanl.gov/abs/hep-th/0207244}{{\tt
  hep-th/0207244}}].

\bibitem{Witten:1996hc}
E.~Witten, {\it {Five-brane effective action in M theory}},  {\em J.Geom.Phys.}
  {\bf 22} (1997) 103--133, [\href{http://xxx.lanl.gov/abs/hep-th/9610234}{{\tt
  hep-th/9610234}}].

\bibitem{Dijkgraaf:2002ac}
R.~Dijkgraaf, E.~P. Verlinde, and M.~Vonk, {\it {On the partition sum of the NS
  five-brane}},  \href{http://xxx.lanl.gov/abs/hep-th/0205281}{{\tt
  hep-th/0205281}}.

\bibitem{gannon1991gluing}
T.~Gannon and C.~Lam, {\it Gluing and shifting lattice constructions and
  rational equivalence},  {\em Reviews in Mathematical Physics} {\bf 3} (1991),
  no.~03 331--369.

\bibitem{gannon1992lattices}
T.~Gannon and C.~Lam, {\it Lattices and $\Theta$-function identities. I: Theta
  constants},  {\em Journal of mathematical physics} {\bf 33} (1992) 854.

\bibitem{gannon1992lattices2}
T.~Gannon and C.~Lam, {\it Lattices and $\theta$-function identities. II: Theta
  series},  {\em Journal of mathematical physics} {\bf 33} (1992) 871.

\bibitem{KacPeterson}
V.~G. Kac and D.~H. Peterson, {\it Infinite-dimensional Lie algebras, theta
  functions and modular forms}, .

\bibitem{Dijkgraaf:2007fe}
R.~Dijkgraaf and P.~Sulkowski, {\it {Instantons on ALE spaces and orbifold
  partitions}},  {\em JHEP} {\bf 0803} (2008) 013,
  [\href{http://xxx.lanl.gov/abs/0712.1427}{{\tt arXiv:0712.1427}}].

\bibitem{Aganagic:2004js}
M.~Aganagic, H.~Ooguri, N.~Saulina, and C.~Vafa, {\it {Black holes, q-deformed
  2d Yang-Mills, and non-perturbative topological strings}},  {\em Nucl.Phys.}
  {\bf B715} (2005) 304--348,
  [\href{http://xxx.lanl.gov/abs/hep-th/0411280}{{\tt hep-th/0411280}}].

\bibitem{Johnson:1994kv}
C.~V. Johnson, {\it {Heterotic Coset Models}},  {\em Mod.Phys.Lett.} {\bf A10}
  (1995) 549--560, [\href{http://xxx.lanl.gov/abs/hep-th/9409062}{{\tt
  hep-th/9409062}}].

\bibitem{BJKZ}
P.~Berglund, C.~V. Johnson, S.~Kachru, and P.~Zaugg, {\it {Heterotic coset
  models and (0,2) string vacua}},  {\em Nucl.Phys.} {\bf B460} (1996)
  252--298, [\href{http://xxx.lanl.gov/abs/hep-th/9509170}{{\tt
  hep-th/9509170}}].

\bibitem{Okazaki:2013kaa}
T.~Okazaki and S.~Yamaguchi, {\it {Supersymmetric Boundary Conditions in Three
  Dimensional N = 2 Theories}},  \href{http://xxx.lanl.gov/abs/1302.6593}{{\tt
  arXiv:1302.6593}}.

\bibitem{Wphases}
E.~Witten, {\it {Phases of N=2 theories in two-dimensions}},  {\em Nucl.Phys.}
  {\bf B403} (1993) 159--222,
  [\href{http://xxx.lanl.gov/abs/hep-th/9301042}{{\tt hep-th/9301042}}].

\bibitem{Warner:1995ay}
N.~Warner, {\it {Supersymmetry in boundary integrable models}},  {\em
  Nucl.Phys.} {\bf B450} (1995) 663--694,
  [\href{http://xxx.lanl.gov/abs/hep-th/9506064}{{\tt hep-th/9506064}}].

\bibitem{Melnikov:2012nm}
I.~V. Melnikov, C.~Quigley, S.~Sethi, and M.~Stern, {\it {Target Spaces from
  Chiral Gauge Theories}},  {\em JHEP} {\bf 1302} (2013) 111,
  [\href{http://xxx.lanl.gov/abs/1212.1212}{{\tt arXiv:1212.1212}}].

\bibitem{Callan:1984sa}
C.~G. Callan and J.~A. Harvey, {\it {Anomalies and Fermion Zero Modes on
  Strings and Domain Walls}},  {\em Nucl.Phys.} {\bf B250} (1985) 427.

\bibitem{GPS}
S.~B. Giddings, J.~Polchinski, and A.~Strominger, {\it {Four-dimensional black
  holes in string theory}},  {\em Phys.Rev.} {\bf D48} (1993) 5784--5797,
  [\href{http://xxx.lanl.gov/abs/hep-th/9305083}{{\tt hep-th/9305083}}].

\bibitem{DistlerSharpe}
J.~Distler and E.~Sharpe, {\it {Heterotic compactifications with principal
  bundles for general groups and general levels}},  {\em Adv.Theor.Math.Phys.}
  {\bf 14} (2010) 335--398, [\href{http://xxx.lanl.gov/abs/hep-th/0701244}{{\tt
  hep-th/0701244}}].

\bibitem{AdamsG}
A.~Adams and D.~Guarrera, {\it {Heterotic Flux Vacua from Hybrid Linear
  Models}},  \href{http://xxx.lanl.gov/abs/0902.4440}{{\tt arXiv:0902.4440}}.

\bibitem{BDP}
C.~Beem, T.~Dimofte, and S.~Pasquetti, {\it {Holomorphic Blocks in Three
  Dimensions}},  \href{http://xxx.lanl.gov/abs/1211.1986}{{\tt
  arXiv:1211.1986}}.

\bibitem{Witten:1997bs}
E.~Witten, {\it {Toroidal compactification without vector structure}},  {\em
  JHEP} {\bf 9802} (1998) 006,
  [\href{http://xxx.lanl.gov/abs/hep-th/9712028}{{\tt hep-th/9712028}}].

\bibitem{deBoer:2001px}
J.~de~Boer, R.~Dijkgraaf, K.~Hori, A.~Keurentjes, J.~Morgan, {\em et.~al.},
  {\it {Triples, fluxes, and strings}},  {\em Adv.Theor.Math.Phys.} {\bf 4}
  (2002) 995--1186, [\href{http://xxx.lanl.gov/abs/hep-th/0103170}{{\tt
  hep-th/0103170}}].

\bibitem{Mackaay}
M.~Mackaay, {\it Spherical {$2$}-categories and {$4$}-manifold invariants},
  {\em Adv. Math.} {\bf 143} (1999), no.~2 288--348.

\bibitem{Itohmoduli}
M.~Itoh, {\it Moduli of half conformally flat structures},  {\em Math. Ann.}
  {\bf 296} (1993), no.~4 687--708.

\bibitem{Witten:1988ze}
E.~Witten, {\it {Topological Quantum Field Theory}},  {\em Commun.Math.Phys.}
  {\bf 117} (1988) 353.

\bibitem{Marcus:1995mq}
N.~Marcus, {\it {The Other topological twisting of N=4 Yang-Mills}},  {\em
  Nucl.Phys.} {\bf B452} (1995) 331--345,
  [\href{http://xxx.lanl.gov/abs/hep-th/9506002}{{\tt hep-th/9506002}}].

\end{thebibliography}\endgroup

\end{document}